\documentclass[onecolumn]{elsart3p} 

\usepackage[xdvi,dvipdf]{color}
\usepackage{amsmath,amssymb,bbm,epsfig}
\usepackage{longtable}

\numberwithin{equation}{section}  
\numberwithin{figure}{section}  
\numberwithin{table}{section}  

\newcommand{\bigintlim}{\int}
\newcommand{\tmmathbf}{\mathbf}

\newcommand{\tmop}{\mathrm}

\newcommand{\mathd}{\mathrm{d}}
\newcommand{\mathe}{\mathrm{e}}

\newcommand{\assign}{:=}
\newcommand{\bignone}{\,}
\newcommand{\tmem}[1]{{\em #1\/}}
\newcommand{\tmsamp}[1]{\textsf{#1}}
\newcommand{\tmscript}[1]{\scriptstyle{#1}}
\newcommand{\emdash}{---}
\newcommand{\longrightarrowlim}{\mathop{\longrightarrow}\limits}

\begin{document}
\begin{frontmatter}

\title{Quantum linear Boltzmann equation}
\author[milano]{Bassano Vacchini},
\author[munich]{Klaus Hornberger}
\address[milano]{Universit{\`a} degli Studi di Milano, Dipartimento di Fisica
and INFN Sezione di Milano,
Via Celoria 16, 20133 Milano, Italy}
\address[munich]{Arnold Sommerfeld Center for Theoretical Physics, 
Ludwig-Maximilians-Universit{\"a}t M{\"u}nchen, Theresienstra{\ss}e 37,\\
80333 Munich, Germany}

\begin{abstract} 
We review the quantum version of the linear Boltzmann equation, which
describes in a non-perturbative fashion, by means of scattering
theory, how the quantum motion of a single test particle is affected
by collisions with an ideal background gas. A heuristic derivation of
this Lindblad master equation is presented, based on the requirement of
translation-covariance and on the relation to the classical linear
Boltzmann equation. After analyzing its general symmetry properties
and the associated relaxation dynamics, we discuss a quantum Monte
Carlo method for its numerical solution. We then review important
limiting forms of the quantum linear Boltzmann equation, such as the
case of quantum Brownian motion and pure collisional decoherence, as well as
the application to matter wave optics. Finally, we point to the
incorporation of quantum degeneracies and self-interactions in the gas
by relating the equation to the dynamic structure factor of the
ambient medium, and we provide an extension of the equation to include internal
degrees of freedom.
\end{abstract}

\end{frontmatter}


\addtocontents{toc}{\protect\setcounter{tocdepth}{3}}


published in: Physics Reports 478 (2009) 71-120
\hfill
doi:10.1016/j.physrep.2009.06.001
\hspace{1.5em}

\vspace*{2\baselineskip}
{\tableofcontents}

\newpage
\section{Introduction}\label{sec:intro}

The Boltzmann equation, as the basic equation of kinetic gas theory, belongs
to the foundations of classical statistical mechanics. Apart from its great
relevance for concrete applications, it is a paradigm for the description of
classical systems far from the thermodynamic equilibrium and for the
understanding of irreversibility and equilibration
{\cite{Boltzmann1898a,Cercignani1975a,Spohn1991,Harris2004a,Balian2007b}}. A
closely related equation is used if one wants to describe how the motion of a
single distinguished test particle is affected by elastic collisions with an
ideal, stationary background gas. In both situations one deals with the
evolution of a probability distribution defined on the phase space of a single
particle. In case of the Boltzmann equation this is the marginal one-particle
distribution of a dilute gas, whose self-interaction gives rise to the
nonlinear form of the integro-differential equation. There is no such
self-interaction if an individual test particle is considered, such that the
corresponding equation is linear, like the Liouville equation. Apart form
this, the linear equation shares much of its structure and many basic
properties with the original Boltzmann equation, and it is derived the same
way using the {\tmem{Stosszahlansatz}}{\emdash}it is thus aptly called the
linear Boltzmann equation {\cite{Cercignani1975a}}. This equation for the
motion of a distinguished test particle has found important applications,
especially in the study of transport phenomena
{\cite{Davison1957a,Williams1966}}.

The present review is devoted to the quantum analog of the linear Boltzmann
equation. It is therefore important not to confuse its classical counterpart
with a ``linearized Boltzmann equation''. The latter is an approximation of
the original Boltzmann equation for a self-interacting gas, which is valid
when the gas is close to equilibrium and the interactions are a weak
perturbation. In the linear Boltzmann equation, in contrast, the coupling to
the gas can be strong and the distinguished test particle may be very far from
an equilibrium state. The background gas, on the other hand, is taken to be in
a stationary state, typically at thermal equilibrium, which remains
undisturbed by the presence of the test particle.

In the quantum physics literature a large number of master equations has been
termed ``quantum Boltzmann equation''. The name is likely to appear whenever
the dynamics is amenable to a kinetic description based on interaction events
that can be described as scattering processes. A first basic result on a
quantum extension of the original Boltzmann equation was the introduction of a
correction accounting for the effects of quantum statistics in the classical
collision term {\cite{Nordheim1928,Uehling1933,Ross1954a}}. Moreover,
Boltzmann-like kinetic equations appear in a wealth of situations with the aim
of describing relaxation and transport properties in a quantum setting. Most
recently, the study of quantum kinetic equations required for the description
of degenerate Bose gases has been the object of extensive research, see, e.g.,
the series of papers {\cite{QKI,QKII,QK-PRLI,QKIII,QKIV,QK-PRLII,QKV}} and
references therein.

In the field of mathematical physics, major efforts have been undertaken in
order to rigorously derive a quantum version of the nonlinear Boltzmann
equation. Here, one typically starts out from the many-particle Hamiltonian
with a simple interaction potential and seeks to obtain a nonlinear equation
of motion for the reduced single-particle state by considering approximations
for the hierarchy of equations of motion of many-particle distribution
functions, see
{\cite{Spohn1980a,Snider1998a,Erdos2004a,Benedetto2007a,Lukkarinen2009a,Spohn2007a}}
for reviews on recent results. Similar attempts have been devoted to the
derivation of the quantum linear Boltzmann equation, using a random potential
approach, see {\cite{Castella2001a,Castella2002a,Eng2007a}} for recent
results. However, it should be emphasized that even the classical Boltzmann
equation has not yet been properly derived in the rigorous sense of
mathematical physics. The problem is substantially more difficult in the
quantum formulation, due to the need for accommodating the unitarity of the
underlying many-body dynamics and due to the quantum restrictions of the
associated phase-space representation.

The situation is more tractable if one asks for the proper quantum version of
the {\tmem{linear}} Boltzmann equation. Due to the inherent linearity of the
problem, the quantum equation of motion must be given by a linear mapping of
the statistical operator representing the motional state of the test particle.
It can be expressed in the framework of quantum dynamical semigroups
{\cite{Alicki2007,Holevo2001,Breuer2007}} since the use of a Boltzmann
equation implies that the time evolution is Markovian. It follows that the
generator of the irreversible time evolution must have the
Gorini-Kossakowski-Sudarshan-Lindblad form {\cite{Gorini1976a,Lindblad1976a}};
this guarantees complete positivity, thus ensuring that the motion of the
reduced single particle state is compatible with an underlying many particle
state that evolves in a unitary fashion.

\begin{figure}[tb]
  \begin{center}
    \resizebox{120mm}{!}{\epsfig{file=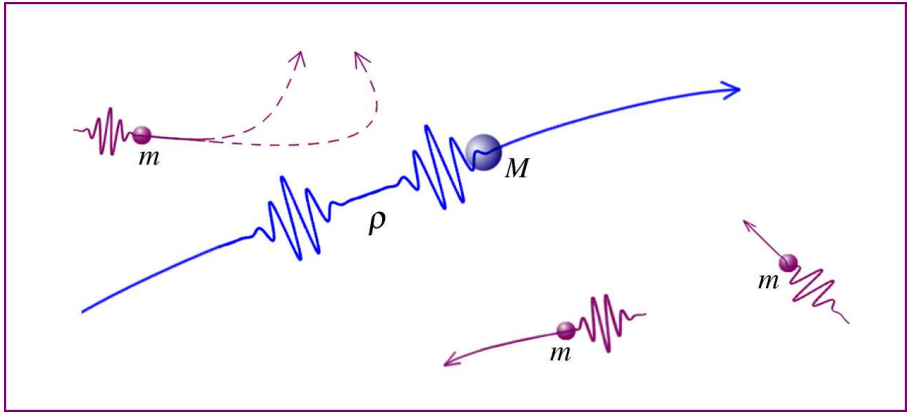}}
  \end{center}
  \caption{(Color online) The quantum linear Boltzmann equation describes the
  motion of a single, distinguished test particle in an ambient medium, taken
  to be an ideal, stationary, and uniform background gas. In the framework of
  the Markov approximation, it is valid for arbitrary initial states of
  motion, such as the non-classical spatial superpositions found in an
  interferometer. As such it describes the interplay of decoherence phenomena
  at short time scales with the dissipative effects and the thermalizing
  behavior expected in the longer-term. The interaction with the gas is
  accounted for in a non-perturbative and microscopic fashion, by means of the
  exact two-body scattering amplitudes. The equation ceases to be applicable
  if the neglect of three-body collisions or correlated scattering events
  cannot be justified, as in the case of liquids. \ \ \ \label{fig:1}}
\end{figure}

Using the framework of Lindblad master equations, the natural quantum
counterpart of the classical linear Boltzmann equation can indeed be obtained
as a unique master equation. It thus describes how the free quantum motion of
a test particle, possibly in a very non-classical, delocalized state, is
modified by the presence of a background gas, see Fig.~\ref{fig:1}. As with
the classical linear Boltzmann equation, the quantum master equation applies
provided (i) the background gas is sufficiently dilute, such that three-body
collisions can be neglected, (ii) the particle interaction is sufficiently
short-ranged, such that scattering theory can be applied, and (iii) the gas
temperature is sufficiently high, such that effects of quantum statistics can
be disregarded and the Markov assumption is justified. Both the classical and
the quantum formulation incorporate the two-body interaction between test
particle and a gas molecule in a non-perturbative fashion by means of
scattering theory. While the classical equation contains the differential
cross-section to account for the molecular collisions, the quantum master
equation involves pairs of scattering amplitudes. These objects, which feature
a complex phase, show up as operator-valued quantities if the master equation
is stated in a representation independent form. 

The main significance of the quantum linear Boltzmann equation lies in the
fact that it serves to describe the interplay of two distinct gas-induced
phenomena, typically occurring on different time scales and usually requiring
a non-perturbative description of the coupling to the gas. On one hand, this
is the collisional decoherence effect, taking place at short times when the
energy exchange between test particle and medium can be disregarded. On the
other hand, the test particle experiences friction, leading on a longer time
scale to a slowing down of the particle and its eventual thermalization. The
equation should thus predict how a test particle, starting out in a
delocalized state, gradually loses its ability to display quantum
interference, due to the dissemination of position information effected by the
gas collisions. One then expects the quantum particle, whose motional state is
increasingly hard to distinguish from a classical one, to display the
relaxation and equilibration behavior predicted by the classical linear
Boltzmann equation. It is thus an important consistency requirement that the
quantum linear Boltzmann equation indeed reduces to the classical version once
the quantum state of motion is indistinguishable from a classical state. Also
in the other relevant limiting cases the equation turns into well-established
forms, in particular in the limit of quantum Brownian motion, of pure
collisional decoherence, and of coherent forward scattering, as will be
discussed below.

In the present review we do not aim at providing a microscopic derivation of
the quantum linear Boltzmann equation; such a calculation was given recently
in some detail in {\cite{Hornberger2008a}}. Instead, we will motivate the form
of the equation in a more succinct line of reasoning by combining natural
physical and mathematical requirements. It builds on the structure of the
classical equation and on Holevo's characterization of general
translation-covariant quantum dynamical semigroups {\cite{Holevo1998a}}, by
exploiting both the required limiting expression of the equation and the
translational symmetry expected in a homogeneous medium. Apart from the
operator form, we will also discuss the equation and its limiting cases in
different representations, each admitting specific physical insights.

The physics described by the quantum linear Boltzmann equation is discussed
and applied in quite different fields of physics, ranging from mathematical
physics and statistical mechanics to matter wave optics. Traditionally,
quantum versions of the linear Boltzmann equation were introduced in order to
provide a microscopic explanation of transport phenomena, or to obtain a
derivation of quantum Brownian motion. These are situations where the relevant
quantum states are close to the state of motion of a classical particle.

Only recently it has become possible to study the influence of a gas
environment on the motion of a molecule which is very far from a classical
state, such as in a superposition of macroscopically distinct spatial
positions {\cite{Arndt2005a}}. This is a result of the experimental advances
in the interferometry of massive particles {\cite{Cronin2009a,Gerlich2007a}},
allowing one to observe the phenomenon of decoherence independently from
dissipative effects. In fact, the main experimental applications and
confirmations of the quantum linear Boltzmann equation are found in situations
where the quantum coherences described by the statistical operator play a
crucial role. This applies to the study of the refractive index for matter
waves traveling through a gas {\cite{Schmiedmayer1995a,Jacquey2007a}}, and in
particular to the analysis of collisional decoherence for massive particles
{\cite{Hornberger2003a}}. It is worth noting that the seminal paper on
decoherence by Joos and Zeh {\cite{Joos1985a}}, seeking to explain the absence
of quantum delocalization in a dust particle by the scattering of photons and
air molecules, derived and studied what the authors called a
{\tmem{Boltzmann-type master equation}}. Two decades later, the long quest for
the characterization of the phenomenon of collisional decoherence has now
reached a mature theoretical description, permitting its quantitative
experimental confirmation.

The article is organized as follows. In Sect.~\ref{sec:qlbe} we introduce the
equation both in its operator form and in the momentum representation.
Section~\ref{sec:holevo} provides an intuitive line of derivation, based on
reconciling the expression of the classical linear Boltzmann equation with the
general form of a translation-covariant generator of a quantum dynamical
semigroup. Various properties and features of the quantum linear Boltzmann
equation are then discussed in the following sections, by first addressing
phenomena with a classical counterpart, and then moving to problems which can
only be considered in a quantum setting. Specifically, Sect.~\ref{sec:relax}
considers the symmetry properties of the equation and its relaxation and
equilibration behavior, while Sect.~\ref{sec:qbm} is devoted to the limiting
case of quantum Brownian motion, obtained if the test particle is much more
massive than the gas particles. Section~\ref{sec:optics} deals with a
distinguished quantum effect, the index of refraction experienced by a matter
wave due to the presence of a background gas, which depends on the coherent
part of the quantum linear Boltzmann equation. The phenomenon of decoherence
is then addressed in Sect.~\ref{sec:decoh}, both regarding collisional
decoherence and more general forms. Section~\ref{sec:dsf} discusses an
alternative expression of the quantum linear Boltzmann equation in terms of
the dynamic structure factor of the gas, suggesting a generalization to the
case of quantum degenerate background gases. It further provides the extension
of the quantum linear Boltzmann equation to include internal degrees of
freedom. We finally present our conclusions in Sect.~\ref{sec:ceo}, where we
also point to possible future developments.

\section{The quantum linear Boltzmann equation}\label{sec:qlbe}

We start by introducing the general expression of the quantum linear Boltzmann
equation in its representation-independent operator form. In order to
understand why the scattering amplitudes and distribution functions in this
expression must be operator-valued, and why the momentum projections parallel
and perpendicular to the momentum exchange play different roles, it will be
helpful to represent it in the appropriate coordinates. We therefore proceed
by presenting the equation in the momentum basis, which will also facilitate
the comparison with the classical linear Boltzmann equation. We conclude this
section with a discussion of the Born approximation and a brief outline of
existing approaches for deriving the equation.

\subsection{The operator form }

The quantum linear Boltzmann equation is a Markovian master equation for the
statistical operator $\rho$ describing the motion of a distinguished test
particle in a gas. It has the form
\begin{eqnarray}
  \frac{\mathd}{\tmop{dt}} \rho & = & \mathcal{M} \rho \nonumber\\
  & = & \frac{1}{i \hbar} \left[ \mathsf{H}_0 + H_{\text{n}} \left(
  \mathsf{P} \right), \rho \right] + \mathcal{L} \rho,  \label{eq:lvonn}
\end{eqnarray}
where $\mathsf{H}_0 = \mathsf{P}^2 / \left( 2 M \right)$ is the kinetic energy
of the test particle and $H_{\text{n}} \left( \mathsf{P} \right)$ describes an
energy shift due to the presence of the background gas, as defined in Eq.
(\ref{eq:effectivebis}). Here and in the following we use sans-serif symbols
to denote operators in a single particle Hilbert space; variables referring to
the test particle will be set in upper-case letters.

The superoperator $\mathcal{L}$ in (\ref{eq:lvonn}) is a linear mapping which
accounts for the incoherent part of the collisional interaction with the gas.
It can be expressed in Lindblad form as {\cite{Hornberger2006b}}
\begin{eqnarray}
  \mathcal{L} \rho & = & \bigintlim \mathd \tmmathbf{Q}
  \int_{\tmmathbf{Q}^{\bot}} \mathd \tmmathbf{k}_{\bot}  \left[
  \mathe^{i\tmmathbf{Q} \cdot \mathsf{X} / \hbar} L \left(
  \tmmathbf{k}_{\bot}, \mathsf{P} ; \tmmathbf{Q} \right) \rho L^{\dag} \left(
  \tmmathbf{k}_{\bot}, \mathsf{P} ; \tmmathbf{Q} \right) \mathe^{-
  i\tmmathbf{Q} \cdot \mathsf{X} / \hbar} \right. \nonumber\\
  &  & \left. - \frac{1}{2} \left\{ \rho, L^{\dag} \left(
  \tmmathbf{k}_{\bot}, \mathsf{P} ; \tmmathbf{Q} \right) L \left(
  \tmmathbf{k}_{\bot}, \mathsf{P} ; \tmmathbf{Q} \right) \right\} \right], 
  \label{eq:qlbe}
\end{eqnarray}
where $\mathsf{X}$ and $\mathsf{P}$ are the position and the momentum operator
of the test particle, and the curly brackets denote the anti-commutator. The
integration is over all momentum transfers $\tmmathbf{Q}$, and over gas
particle momenta $\tmmathbf{p}$ from the plane perpendicular to
$\tmmathbf{Q}$, i.e., from $\tmmathbf{Q}^{\bot} \equiv \left\{ \tmmathbf{p}
\in \mathbbm{R}^3 : \tmmathbf{p} \cdot \tmmathbf{Q}= 0 \right\}$. The function
$L$, which appears operator-valued in (\ref{eq:qlbe}), contains all the
details of the collisional interaction with the gas. It is defined by
\begin{eqnarray}
  L \left( \tmmathbf{p}, \tmmathbf{P}; \tmmathbf{Q} \right) & = &
  \sqrt{\frac{n_{\tmop{gas}} m}{m_{\ast}^2 Q}} f \left( \tmop{rel} \left(
  \tmmathbf{p}_{\bot \tmmathbf{Q}}, \tmmathbf{P}_{\bot \tmmathbf{Q}} \right) -
  \frac{\tmmathbf{Q}}{2}, \tmop{rel} \left( \tmmathbf{p}_{\bot \tmmathbf{Q}},
  \tmmathbf{P}_{\bot \tmmathbf{Q}} \right) + \frac{\tmmathbf{Q}}{2} \right)
  \nonumber\\
  &  & \times \sqrt{\mu \left( \tmmathbf{p}_{\bot \tmmathbf{Q}} +
  \frac{m}{m_{\ast}} \frac{\tmmathbf{Q}}{2} + \frac{m}{M}
  \tmmathbf{P}_{\|\tmmathbf{Q}} \right)},  \label{eq:L}
\end{eqnarray}
where $m$ and $M$ are the masses of the gas particles and of the test
particle, respectively, and $m_{\ast} = mM / \left( m + M \right)$ denotes the
reduced mass. It involves also the gas number density $n_{\tmop{gas}}$, the
distribution function $\mu \left( \tmmathbf{p} \right)$ of the gas momenta,
and, most importantly, the elastic scattering amplitude $f \left(
\tmmathbf{p}_f, \tmmathbf{p}_i) \right.$ determined by the two-body
interaction between gas and test particle. The subscripts $\|\tmmathbf{Q}$ and
$\perp \tmmathbf{Q}$ denote the component of a vector $\tmop{parallel}$ and
perpendicular to the momentum transfer $\tmmathbf{Q}$, such that
$\tmmathbf{P}_{\|\tmmathbf{Q}} = \left( \tmmathbf{P} \cdot \tmmathbf{Q}
\right) \tmmathbf{Q}/ Q^2$ and $\tmmathbf{P}_{\bot \tmmathbf{Q}}
=\tmmathbf{P}-\tmmathbf{P}_{\|\tmmathbf{Q}}$ respectively. Finally, relative
momenta are stated by means of the abbreviation
\begin{eqnarray}
  \tmop{rel} \left( \tmmathbf{p}, \tmmathbf{P} \right) & \equiv &
  \frac{m_{\ast}}{m} \tmmathbf{p}- \frac{m_{\ast}}{M} \tmmathbf{P}. 
  \label{eq:reldef}
\end{eqnarray}

The Hamiltonian correction in (\ref{eq:lvonn}) is defined by the function
{\cite{Hornberger2008a}}
\begin{eqnarray}
  H_{\text{n}} \left( \tmmathbf{P} \right) & = & - 2 \pi \hbar^2
  \frac{n_{\tmop{gas}}}{m_{\ast}}  \int \mathd \tmmathbf{p} \, \mu \left(
  \tmmathbf{p} \right) \tmop{Re} \left[ f \left( \tmop{rel} \left(
  \tmmathbf{p}, \tmmathbf{P} \right), \tmop{rel} \left( \tmmathbf{p},
  \tmmathbf{P} \right) \right) \right] .  \label{eq:effectivebis}
\end{eqnarray}
As discussed in Sect.~\ref{sec:optics}, this energy shift is related to
forward scattering.

The quantum linear Boltzmann equation is valid for non-degenerate, uniform
gases with arbitrary stationary momentum distribution $\mu \left( \tmmathbf{p}
\right)$. However, for concrete calculations we will often tacitly take $\mu
\left( \tmmathbf{p} \right)$ to be the standard Maxwell-Boltzmann expression
\begin{eqnarray}
  \mu_{\beta} \left( \tmmathbf{p} \right) & = & \frac{1}{\pi^{3 / 2}
  p_{\beta}^3} \exp \left( - \frac{\tmmathbf{p}^2}{p_{\beta}^2} \right), 
  \label{eq:muMB}
\end{eqnarray}
with $p_{\beta} = \sqrt{2 m / \beta}$ the most probable momentum at
temperature $T = 1 / \left( k_{\text{B}} \beta \right)$.

\subsection{Momentum representation}\label{sec:mom}

To gain some physical insight into the equation it is convenient to express it
in the basis of improper momentum eigenstates, $\mathsf{P} |\tmmathbf{P}
\rangle =\tmmathbf{P}|\tmmathbf{P} \rangle$, which will also help in
establishing the relation to its classical counterpart. In the momentum
representation Eq.~(\ref{eq:qlbe}) can be brought into the form
\begin{eqnarray}
  \langle \tmmathbf{P}| \mathcal{L} \rho |\tmmathbf{P}' \rangle & = & \int
  \mathd \tmmathbf{Q} \, M_{\tmop{in}} \left( \tmmathbf{P}, \tmmathbf{P}' ;
  \tmmathbf{Q} \right) \langle \tmmathbf{P}-\tmmathbf{Q}| \rho |\tmmathbf{P}'
  -\tmmathbf{Q} \rangle - \frac{1}{2} \left[ M_{\tmop{out}}^{\tmop{cl}} \left(
  \tmmathbf{P} \right) + M_{\tmop{out}}^{\tmop{cl}} \left( \tmmathbf{P}'
  \right) \right] \langle \tmmathbf{P}| \rho |\tmmathbf{P}' \rangle . 
  \label{eq:poffdiag}
\end{eqnarray}
The ``gain term'', i.e., the first term on the right-hand side, is here given
by
\begin{eqnarray}
  M_{\tmop{in}} \left( \tmmathbf{P}, \tmmathbf{P}' ; \tmmathbf{Q} \right) & =
  & \int_{\tmmathbf{Q}^{\bot}} \mathd \tmmathbf{k}_{\bot} \, L \left(
  \tmmathbf{k}_{\bot}, \tmmathbf{P}-\tmmathbf{Q}; \tmmathbf{Q} \right)
  L^{\ast} \left( \tmmathbf{k}_{\bot}, \tmmathbf{P}' -\tmmathbf{Q};
  \tmmathbf{Q} \right)  \label{eq:Min3}
\end{eqnarray}
with $L$ defined in Eq.~(\ref{eq:L}). Importantly, (\ref{eq:Min3}) reduces to
the rate known from the classical linear Boltzmann equation if one approaches
the diagonals $\tmmathbf{P}=\tmmathbf{P}'$,
\begin{eqnarray}
  M_{\tmop{in}} \left( \tmmathbf{P}, \tmmathbf{P}; \tmmathbf{Q} \right) & = &
  M_{\tmop{in}}^{\tmop{cl}} \left( \tmmathbf{P}; \tmmathbf{Q} \right) . 
  \label{eq:ratein}
\end{eqnarray}
This classical gain rate, which involves the quantum mechanically defined
differential cross-section $\sigma \left( \tmmathbf{p}_f, \tmmathbf{p}_i
\right) = \left| f \left( \tmmathbf{p}_f, \tmmathbf{p}_i \right) \right|^2$,
reads as
\begin{eqnarray}
  M^{\tmop{cl}}_{\tmop{in}} \left( \tmmathbf{P}+\tmmathbf{Q}; \tmmathbf{Q}
  \right) & = & \frac{n_{\tmop{gas}} m}{m^2_{\ast} Q}
  \int_{\tmmathbf{Q}^{\bot}} \mathd \tmmathbf{k}_{\bot} \, \mu \left(
  \tmmathbf{k}_{\bot} + \frac{m}{m_{\ast}} \frac{\tmmathbf{Q}}{2} +
  \frac{m}{M} \tmmathbf{P}_{\|\tmmathbf{Q}} \right) \nonumber\\
  &  & \times \sigma \left( \tmop{rel} \left( \tmmathbf{k}_{\bot},
  \tmmathbf{P}_{\perp \tmmathbf{Q}} \right) - \frac{\tmmathbf{Q}}{2},
  \tmop{rel} \left( \tmmathbf{k}_{\bot}, \tmmathbf{P}_{\perp \tmmathbf{Q}}
  \right) + \frac{\tmmathbf{Q}}{2} \right) .  \label{eq:diretta}
\end{eqnarray}
It has a natural meaning in the sense that the transition rate
\begin{eqnarray}
  M^{\tmop{cl}} \left( \tmmathbf{P}_i \rightarrow \tmmathbf{P}_f \right) &
  \equiv & M^{\tmop{cl}}_{\tmop{in}} \left( \tmmathbf{P}_f ; \tmmathbf{P}_f
  -\tmmathbf{P}_i \right)  \label{eq:rate}
\end{eqnarray}
provides the rate of collisions which change the test particle momentum from
$\tmmathbf{P}_i$ to $\tmmathbf{P}_f$, or equivalently, which lead to a final
momentum $\tmmathbf{P}_f$ due to a gain of momentum
$\tmmathbf{Q}=\tmmathbf{P}_f -\tmmathbf{P}_i$.

The loss term in (\ref{eq:poffdiag}) is given by the arithmetic mean of the
loss rates which appear in the corresponding classical equation. The latter
are related to the classical gain rate by
\begin{eqnarray}
  M_{\tmop{out}}^{\tmop{cl}} \left( \tmmathbf{P} \right) & = & \bigintlim
  \mathd \tmmathbf{Q} \, M_{\tmop{in}}^{\tmop{cl}} \left(
  \tmmathbf{P}+\tmmathbf{Q}; \tmmathbf{Q} \right) \, .  \label{eq:rateout}
\end{eqnarray}
This relationship, which ensures that the probability is conserved in the
classical case, follows directly from the structure of (\ref{eq:qlbe}), and it
guarantees the normalization of the statistical operator $\rho$.

If fact, by considering Eq.~(\ref{eq:poffdiag}) only on the diagonal,
\begin{eqnarray}
  \langle \tmmathbf{P}| \mathcal{L} \rho |\tmmathbf{P} \rangle & = & \int
  \mathd \tmmathbf{Q} \, M^{\tmop{cl}}_{\tmop{in}} \left( \tmmathbf{P};
  \tmmathbf{Q} \right) \langle \tmmathbf{P}-\tmmathbf{Q}| \rho
  |\tmmathbf{P}-\tmmathbf{Q} \rangle - M_{\tmop{out}}^{\tmop{cl}} \left(
  \tmmathbf{P} \right) \langle \tmmathbf{P}| \rho |\tmmathbf{P} \rangle, 
  \label{eq:pdiag}
\end{eqnarray}
one immediately recovers one of the possible expressions of the classical
linear Boltzmann equation, as we shall see in Sect.~\ref{sec:holevo}. It takes
the explicit form
\begin{eqnarray}
  \langle \tmmathbf{P}| \mathcal{L} \rho |\tmmathbf{P} \rangle & = &
  \frac{n_{\tmop{gas}} m}{m^2_{\ast}} \bigintlim \frac{\mathd \tmmathbf{Q}}{Q}
  \int_{\tmmathbf{Q}^{\bot}} \mathd \tmmathbf{k}_{\bot} \, \sigma \left(
  \tmop{rel} \left( \tmmathbf{k}_{\bot}, \tmmathbf{P}_{\bot \tmmathbf{Q}}
  \right) - \frac{\tmmathbf{Q}}{2}, \tmop{rel} \left( \tmmathbf{k}_{\bot},
  \tmmathbf{P}_{\bot \tmmathbf{Q}} \right) + \frac{\tmmathbf{Q}}{2} \right)
  \nonumber\\
  &  & \times \left[ \mu \left( \tmmathbf{k}_{\bot} \text{$\left. +
  \text{$\frac{m}{m_{\ast}} \frac{\tmmathbf{Q}}{2} + \frac{m}{M} \left(
  \tmmathbf{P}_{\| \tmmathbf{Q}} -\tmmathbf{Q} \right)$} \right)$} \langle
  \tmmathbf{P}-\tmmathbf{Q}| \rho |\tmmathbf{P}-\tmmathbf{Q} \rangle \right.
  \right. \nonumber\\
  &  & - \mu \left( \tmmathbf{k}_{\bot} \text{$\left. +
  \text{$\frac{m}{m_{\ast}} \frac{\tmmathbf{Q}}{2} + \frac{m}{M}
  \mathbf{\tmmathbf{P}_{\| \tmmathbf{Q}}}$} \right)$} \langle \tmmathbf{P}|
  \rho |\tmmathbf{P} \rangle \right] .  \label{eq:exp}
\end{eqnarray}

\subsection{Born approximation}

The weak coupling limit of the quantum linear Boltzmann equation is obtained
if one replaces the scattering amplitudes in Eq.~(\ref{eq:L}) by their Born
approximation. This simplifies the equation considerably, since the Born
amplitude associated to the interaction potential $V \left( \tmmathbf{x}
\right)$ depends only on the difference of the momenta according to
\begin{eqnarray}
  f_B \left( \tmmathbf{p}_f, \tmmathbf{p}_i \right) & = & - 4 \pi^2 \hbar
  m_{\ast} \langle \tmmathbf{p}_f |V \left( \mathsf{x} \right) |\tmmathbf{p}_i
  \rangle \nonumber\\
  & = & - \frac{m_{\ast}}{2 \pi \hbar^2} \int \mathd \tmmathbf{x} \, V \left(
  \tmmathbf{x} \right) \exp \left( - i \frac{\left( \tmmathbf{p}_f
  -\tmmathbf{p}_i \right) \cdot \tmmathbf{x}}{\hbar} \right) \bignone
  \nonumber\\
  & \equiv & f_B \left( \tmmathbf{p}_f -\tmmathbf{p}_i \right) . 
  \label{eq:fBorn}
\end{eqnarray}
This removes the operator-valuedness of the scattering amplitudes in
(\ref{eq:qlbe}) because the function (\ref{eq:L}) reduces to the form
\begin{eqnarray}
  L \left( \tmmathbf{p}, \tmmathbf{P}; \tmmathbf{Q} \right) & \rightarrow &
  \sqrt{\frac{n_{\tmop{gas}} m}{m_{\ast}^2 Q}} f_B \left( -\tmmathbf{Q}
  \right) \sqrt{\mu \left( \tmmathbf{p}_{\bot \tmmathbf{Q}} +
  \frac{m}{m_{\ast}} \frac{\tmmathbf{Q}}{2} + \frac{m}{M}
  \tmmathbf{P}_{\|\tmmathbf{Q}} \right)},  \label{eq:x1}
\end{eqnarray}
such that the $\mathd \tmmathbf{k}_{\bot}$-integration in Eq.~(\ref{eq:qlbe})
can be carried out if one assumes the Maxwell-Boltzmann distribution
(\ref{eq:muMB}). The resulting equation then reduces to the one proposed in
{\cite{Vacchini2001a}}. Introducing the operators
\begin{eqnarray}
  L_B \left( \mathsf{P} ; \tmmathbf{Q} \right) & = & \sqrt{\sqrt{\frac{\beta
  m}{2 \pi}} \frac{n_{\tmop{gas}} \sigma_B \left( \tmmathbf{Q}
  \right)}{m_{\ast}^2 Q}} \exp \left( - \beta \frac{\left( \left( 1 +
  \frac{m}{M} \right) Q^2 + 2 \frac{m}{M} \mathsf{P} \cdot \tmmathbf{Q}
  \right)^2}{16 mQ^2} \right),  \label{eq:Lb}
\end{eqnarray}
with $\sigma_B = \left| f_B \right|^2$, it takes the simpler form
\begin{eqnarray}
  \mathcal{L}_B \rho & = & \bigintlim \mathd \tmmathbf{Q} \left[
  \mathe^{i\tmmathbf{Q} \cdot \mathsf{X} / \hbar} L_B \left( \mathsf{P} ;
  \tmmathbf{Q} \right) \rho L_B^{\dag} \left( \mathsf{P} ; \tmmathbf{Q}
  \right) \mathe^{- i\tmmathbf{Q} \cdot \mathsf{X} / \hbar} - \frac{1}{2}
  \left\{ \rho, L_B^{\dag} \left( \mathsf{P} ; \tmmathbf{Q} \right) L_B \left(
  \mathsf{P} ; \tmmathbf{Q} \right) \right\} \right],  \label{eq:born}
\end{eqnarray}
where only the integration over the momentum transfers is left.

We note that Eq.~(\ref{eq:qlbe}) can be simplified also by other limiting
procedures. The minimally extended Caldeira-Leggett master equation is
obtained in the diffusive limit described in Sect.~\ref{sec:blim}, while the
master equation of pure collisional decoherence discussed in
Sect.~\ref{sec:decopos} is gained in the limit of a very massive tracer
particle, $m / M \rightarrow 0$.

\subsection{Survey and comparison of derivations in the
literature}\label{sec:survey}

The quantum linear Boltzmann equation has a long history, even though the
general form Eq.~(\ref{eq:qlbe}) was obtained only recently. The key
difficulty lies in the necessity to account for the collisional dynamics in a
microscopically realistic and non-perturbative way, leading to a dynamics
which is consistent with the semigroup structure of a proper Markovian
description and which displays the thermalization behavior known from the
classical description.

A Boltzmann-like master equation already appeared in the seminal work by Joos
and Zeh {\cite{Joos1985a}} which used a scattering theory description to treat
the decoherence effect of environmental collisions on the motion of a quantum
particle. Despite its similar name the master equation presented there is
quite different from Eq.~(\ref{eq:qlbe}) in that it linearizes the interaction
effect and neglects dissipation, thus leading to an infinite growth of the
kinetic energy of the test particle on long time scales. Indeed, the equation
is meant to describe only the short-time loss of coherence in the motional
state. However, due to the linearization involved the predicted localization
rate grows above all bounds for increasingly delocalized quantum states, which
may quickly lead to unrealistically large decoherence rates.

This result was improved by Gallis and Fleming {\cite{Gallis1990a}}, who
proposed a master equation using the full scattering matrix. This equation
predicts a saturation of the localization rate at large distances, as one
expects for local interactions with the environment, though at a numerically
incorrect value. Like the master equation by Joos and Zeh, this result does
not account for dissipation phenomena.

To our knowledge, the first proposal pointing to a quantum linear Boltzmann
equation which also describes energy relaxation and therefore grants the
existence of a stationary solution, was given by Di\`osi {\cite{Diosi1995a}}.
As discussed in Sect.~\ref{sec:altforms}, this result is not equivalent to
what we consider the correct quantum linear Boltzmann equation, although the
diagonal matrix elements in the momentum representation coincide, yielding the
expected classical equation.

The correct result including both decoherence and dissipative effects, though
restricted to a perturbative scattering cross-section evaluated in Born
approximation, was later obtained in
{\cite{Lanz1997a,Vacchini2000a,Vacchini2001a,Vacchini2001b}}. This work also
pointed out an important connection between the collision kernel in the
quantum linear Boltzmann equation and a two-point correlation function of the
medium known as dynamic structure factor, first introduced by van Hove for the
description of scattering of a test particle off a macroscopic sample. In the
master equation the dynamic structure factor, given by the Fourier transform
of the time dependent density autocorrelation function of the gas, appears
operator-valued and accounts for the existence of the canonical stationary
solution, as discussed in detail in Sect.~\ref{sec:dsf}. At the same time, the
connection between this quantum master equation and the general structure of
translation-covariant master equations obtained by Holevo {\cite{Holevo1998a}}
was first explored and clarified.

The final, non-perturbative version of the quantum linear Boltzmann equation,
which is discussed in the present article, was first derived in
{\cite{Hornberger2006b}}. It is noteworthy that this result is not obtained
from the weak coupling expression by naively replacing the Born approximation
of the scattering amplitude with the exact expression, since this procedure is
not unambiguous. Indeed, the substitution of $\sigma_B^{1 / 2}$ in
(\ref{eq:Lb}) by the exact scattering amplitude would yield an expression very
different from the correct equation as given by (\ref{eq:qlbe}) and
(\ref{eq:L}), which requires an additional two-dimensional integration.

Other recent attempts to derive master equations related to the effect of a
gas environment are due to Alicki {\cite{Alicki2002a}} and Dodd and Halliwell
{\cite{Dodd2003a}} (see also
{\cite{Clark2008a,Clark2009a,Clark2009b,Clark-xxx}} for recent mathematical
work building on the covariance under translations). These papers focus on the
collisional decoherence effect for the case of a very massive particle, where
dissipative effects are not accounted for. In this regard, a reexamination of
the derivation of the master equation by Gallis and Fleming was presented in
{\cite{Hornberger2003b}}. That article discusses several ways how to
consistently incorporate scattering theory into a Markovian description, which
served to ascertain the numerical value of the localization rate due to
collisional decoherence, thus confirming the prefactor implied by
{\cite{Diosi1995a,Vacchini2000a}}. The importance of an accurate description
of collisional decoherence was demonstrated by experimental measurements with
interfering fullerene molecules {\cite{Hornberger2003a,Hackermuller2003b}},
which were sensitive to the exact saturation value of the localization rate.
We note that the techniques used in {\cite{Hornberger2003b}} were also
instrumental for developing the methods required to combine asymptotic
scattering theory with a dynamic description of the state evolution, as
required in the derivation of the final form of the quantum linear Boltzmann
equation.

Retrospectively, one can distinguish two rather different approaches followed
in the various attempts at deriving the master equation discussed in the
present article. One possible line of argumentation starts from the
microscopic Hamiltonian for the joint system of test particle and gas, with an
interaction term specified by a potential depending on the relative distance
between gas particle and test particle. A natural mathematical framework in
this approach is the formalism of second quantization. In this
non-relativistic quantum field theoretical setting the interaction is
naturally expressed as a density-density coupling. In this context a second
order perturbation expansion leads to the Fourier transform of the density
autocorrelation function of the gas, that is to its dynamic structure factor
{\cite{Petruccione2005a}}. The result includes statistical corrections, which
are naturally accounted for in a field theoretical formalism. However, the
restriction to the second order term involves the Fourier transform of the
interaction potential, corresponding to the Born approximation for the
scattering cross-section. Going beyond the second order term, thus recovering
the full T-matrix including statistical corrections, becomes a very difficult
task, which has still not been fully accomplished. This standpoint has been
considered in {\cite{Lanz1996a,Altenmuller1997a,Vacchini2001a,Dodd2003a}}.

A second line of reasoning starts from the description of single collision
events, formulated by means of scattering theory, which has also recently been
the object of mathematical investigations
{\cite{Teta2004a,Durr2004a,Adami2004a,Cacciapuoti2005a,Adami2006a}}. While
this approach admits a non-perturbative description of the effect of a
collision, it cannot easily account for degeneracy effects, which may occur if
the gas particles are to be considered as being indistinguishable. The dynamic
effect of the gas is essentially described by weighting the effect of a single
collision, obtained by tracing over the scattered gas particle, with the rate
of collisions. This reasoning was pursued in a phenomenological sense in
{\cite{Stenholm1993a,Diosi1995a}}. However, since both the rate of collisions
and the effect of a single collision depend in general on the particle state,
it is not straightforward to derive an expression for the infinitesimal state
evolution which is linear in the density operator, as required in a Markovian
master equation. It is achieved in a line of reasoning termed the monitoring
approach {\cite{Hornberger2007b,Hornberger2007a}}, which is based on concepts
drawn from the theory of generalized and continuous measurements
{\cite{Holevo2001,Jacobs2006a,Barchielli2009}}. It introduces a positive
operator for the scattering rate and thus yields an expression for the dynamic
state evolution, involving the S-matrix and square roots of the rate operator,
which is manifestly of Lindblad form once the environmental trace is done.
This approach has been shown to yield nontrivial master equations that have
been established by other means {\cite{Hornberger2007b,Dumcke1985a}}. If
applied to the case of a tracer particle in a gas it leads in a stringent
calculation to the quantum linear Boltzmann equation, as discussed in some
detail in {\cite{Hornberger2008a}}.

Finally, we note that a large number of works can be found in the literature
which are related to the present subject. While it is impossible to do justice
to all authors, we mention a few recent articles related to the derivation of
a quantum kinetic equation in the spirit of the Boltzmann equation, both
linear
{\cite{Raffelt1993a,Tsonchev2000a,Kleckner2001a,Hellmich2004a,Pechen2004a,Pechen2005a,Barnett2005a,Adler2006a,Halliwell2007a,Dominguez-Clarimon2007a,Mintert200Xa}}
and nonlinear
{\cite{QKI,QKII,QK-PRLI,QKIII,QKIV,QK-PRLII,QKV,Erdos2004a,Benedetto2004a,Benedetto2006a,Benedetto2008a,Spohn2007a,Steinigeweg2007a,Kadiroglu2007a}}.

\section{A heuristic derivation}\label{sec:holevo}

We now present a heuristic motivation for the particular form of the quantum
linear Boltzmann equation as given by Eqs.~(\ref{eq:qlbe}) and (\ref{eq:L}).
It rests upon two basic requirements, its compatibility with (a) the classical
linear Boltzmann equation, and with (b) the structure of the general form of a
covariant completely positive master equation. Although our line of reasoning
can hardly be called a proper derivation, it is rather straightforward and
provides an intuitive reasoning why the particular form given above is
necessary.

\subsection{The classical linear Boltzmann equation}\label{sec:31}

It is helpful to first discuss the classical version of the linear Boltzmann
equation. In particular, we will rewrite it in a form that exhibits a natural
correspondence with the quantum version. As already explained in
Sect.~\ref{sec:intro}, the classical linear Boltzmann equation describes the
collisional dynamics of a test particle interacting with a homogeneous, ideal
background gas characterized by the momentum distribution $\mu \left(
\tmmathbf{p} \right)$. The nontrivial part of the equation is therefore the
collisional contribution to the time evolution of the distribution function $f
\left( \tmmathbf{P} \right)$, which we shall denote by
$\text{$\partial^{\tmop{coll}}_t f$} (\tmmathbf{P})$. We start with the form
\begin{eqnarray}
  \text{$\partial^{\tmop{coll}}_t f$} (\tmmathbf{P}) & = &
  \frac{n_{\tmop{gas}}}{m^2_{\ast}} \bigintlim \mathd \tmmathbf{P}' \int
  \mathd \tmmathbf{p}' \int \bignone \mathd \tmmathbf{p} \, \sigma \left(
  \tmop{rel} \left( \tmmathbf{p}, \tmmathbf{P} \right), \tmop{rel} \left(
  \tmmathbf{p}', \tmmathbf{P}' \right) \right)  \label{eq:clbeMB}\\
  &  & \times \delta \left( \frac{P'^2}{2 M} + \frac{p'^2}{2 m} -
  \frac{P^2}{2 M} - \frac{p^2}{2 m} \right) \delta^3 \left( \tmmathbf{P}'
  +\tmmathbf{p}' -\tmmathbf{P}-\tmmathbf{p} \right) \left[ \mu \left(
  \tmmathbf{p}' \right) f \left( \tmmathbf{P}' \right) - \mu \left(
  \tmmathbf{p} \right) f \left( \tmmathbf{P} \right) \right], \nonumber
\end{eqnarray}
where the fundamental role of energy and momentum conservation in each single
collision is put in the foreground. The differential scattering cross-section
$\sigma$, which is a function of the relative momenta before and after a
collision with a gas particle, is taken to satisfy $\sigma \left(
\tmmathbf{p}_f, \tmmathbf{p}_i \right) = \sigma \left( \tmmathbf{p}_i,
\tmmathbf{p}_f \right)$, reflecting the invariance of the collisional
interaction under an inversion of time and parity. We note that the collision
integral of the original, nonlinear Boltzmann is obtained from
(\ref{eq:clbeMB}) by replacing $\mu \left( \tmmathbf{p} \right)$ with $f
\left( \tmmathbf{p} \right)$, and by setting $M = m$.

As shown in Appendix~\ref{sec:a1}, by explicitly evaluating the two delta
functions in (\ref{eq:clbeMB}) one arrives at the expression
\begin{eqnarray}
  \text{$\partial^{\tmop{coll}}_t f$} (\tmmathbf{P}) & = &
  \frac{n_{\tmop{gas}} m}{m^2_{\ast}} \bigintlim \frac{\mathd \tmmathbf{Q}}{Q}
  \int_{\tmmathbf{Q}^{\bot}} \mathd \tmmathbf{k}_{\bot} \, \sigma \left(
  \tmop{rel} \left( \tmmathbf{k}_{\bot}, \tmmathbf{P}_{\perp \tmmathbf{Q}}
  \right) - \frac{\tmmathbf{Q}}{2}, \tmop{rel} \left( \tmmathbf{k}_{\bot},
  \tmmathbf{P}_{\perp \tmmathbf{Q}} \right) + \frac{\tmmathbf{Q}}{2} \right) 
  \label{eq:clbe}\\
  &  & \times \left[ \mu \left( \tmmathbf{k}_{\bot} \text{$\left. +
  \text{$\frac{m}{m_{\ast}} \frac{\tmmathbf{Q}}{2} + \frac{m}{M} \left(
  \tmmathbf{P}_{\| \tmmathbf{Q}} -\tmmathbf{Q} \right)$} \right)$} f \left(
  \tmmathbf{P}-\tmmathbf{Q} \right) - \mu \left( \tmmathbf{k}_{\bot} \text{$+
  \frac{m}{m_{\ast}} \frac{\tmmathbf{Q}}{2} + \frac{m}{M} \tmmathbf{P}_{\|
  \tmmathbf{Q}}$} \right) f \left( \tmmathbf{P} \right) \right] . \right.
  \nonumber
\end{eqnarray}
It has already the same form as the right-hand side of Eq.~(\ref{eq:exp}). By
exploiting the notation used in Eq.~(\ref{eq:ratein}) and
Eq.~(\ref{eq:rateout}) one can write (\ref{eq:clbe}) more compactly as
\begin{eqnarray}
  \text{$\partial^{\tmop{coll}}_t f$} (\tmmathbf{P}) & = & \int \mathd
  \tmmathbf{Q} \, M^{\tmop{cl}}_{\tmop{in}} \left( \tmmathbf{P}; \tmmathbf{Q}
  \right) f \left( \tmmathbf{P}-\tmmathbf{Q} \right) -
  M_{\tmop{out}}^{\tmop{cl}} \left( \tmmathbf{P} \right) f \left( \tmmathbf{P}
  \right),  \label{eq:inout}
\end{eqnarray}
that is, one arrives at Eq.~(\ref{eq:pdiag}) upon identifying the classical
momentum distribution function $f \left( \tmmathbf{P} \right)$ with $\langle
\tmmathbf{P}| \rho |\tmmathbf{P} \rangle$, the diagonal matrix elements of the
statistical operator in the momentum representation. Equation~(\ref{eq:inout})
makes it immediately apparent that the classical linear Boltzmann equation
takes exactly the form of a classical Markovian master equation.

\subsection{General form of a translation-covariant quantum master equation}

The classical result Eq.~(\ref{eq:clbe}) already provides a hint on the
quantum version of the linear Boltzmann equation, in that it tells how the
diagonal matrix elements in the momentum representation must look like.
However, this constraint does obviously not suffice to fix the quantum master
equation, which is required to describe not only the populations in a given
basis, but also the coherences corresponding to the off-diagonal elements.

As is well-known from the theory of open quantum systems {\cite{Breuer2007}},
the general structure of a Markovian master equation is fixed by the Lindblad
form, according to which the non-Hamiltonian part must have the structure
\begin{eqnarray}
  \text{} \text{$\mathcal{L} \rho$} & = & \sum_j \left[ \mathsf{L}_j \rho
  \mathsf{L}_j^{^{\dag}} - \frac{1}{2} \left\{ \mathsf{L}_j^{^{\dag}}
  \mathsf{L}_j, \rho \right\} \right] \bignone .  \label{eq:lindblad}
\end{eqnarray}
In the present case, however, a more powerful mathematical characterization of
the master equation is available, which builds on the translational covariance
of $\text{$\mathcal{L} \rho$}$, and thus provides further cues on the possible
expressions of the Lindblad operators $\mathsf{L}_j$. Early results on the
subject go back to one of the pioneering works on quantum dynamical semigroups
{\cite{Kossakowski1972a}}, and the problem was later reconsidered in
{\cite{Manita1991a,Botvich1991a}}. The final characterization was obtained by
Holevo in a series of papers
{\cite{Holevo1993a,Holevo1993b,Holevo1995a,Holevo1996a}}, by building on a
quantum generalization of the classical L\'evy-Khintchine formula.

The linear Boltzmann equation must be invariant under translations, both in
the classical and in the quantum version, because the background gas is taken
to be homogeneous, while the interactions are described by a two-body
potential depending on the relative distance between colliding particles. This
property is immediately apparent in the structure of the classical equation.
As discussed in more detail in Sect.~\ref{sec:ti}, it implies on the quantum
level that the master equation must be translation-covariant, that is, its
action must commute with the unitary representation of translations,
\begin{eqnarray}
  \mathcal{L} \left( \mathe^{- i \mathsf{\tmmathbf{A} \cdot P / \hbar}} \rho
  \mathe^{i \mathsf{\tmmathbf{A} \cdot P / \hbar}} \right) & = & \mathe^{- i
  \mathsf{\tmmathbf{A} \cdot P / \hbar}} \mathcal{L} \rho \mathe^{i
  \mathsf{\tmmathbf{A} \cdot P / \hbar}} .  \label{eq:cov}
\end{eqnarray}
It was shown by Holevo in the aforementioned work that the structure of a
master equation which complies with Eq.~(\ref{eq:cov}) is given by
\begin{eqnarray}
  \mathcal{L} \rho & = & \int \mathd \tmmathbf{Q} \bignone \sum_j \left[
  \mathe^{i\tmmathbf{Q} \cdot \mathsf{X} / \hbar} L_j \left( \mathsf{P} ;
  \tmmathbf{Q} \right) \rho L_j^{^{\dag}} \left( \mathsf{P} ; \tmmathbf{Q}
  \right) \mathe^{- i\tmmathbf{Q} \cdot \mathsf{X} / \hbar} - \frac{1}{2}
  \left\{ L_j^{^{\dag}} \left( \mathsf{P} ; \tmmathbf{Q} \right) L_j \left(
  \mathsf{P} ; \tmmathbf{Q} \right), \rho \right\} \right] \bignone, 
  \label{eq:holevo}
\end{eqnarray}
with arbitrary functions $L_j \left( \tmmathbf{P}; \tmmathbf{Q} \right)$,
while an additional Hamiltonian contribution may depend only on $\mathsf{P}$.
This assumes that the mapping $\mathcal{L}$ is bounded and that the underlying
statistical process is Poissonian. A more general expression including both a
Gaussian and a Poisson component has to be considered if one also allows for
unbounded mappings {\cite{Vacchini2001b,Petruccione2005a,Vacchini-xxx}}. An
example of such a Gaussian component is given by the master equation
Eq.~(\ref{eq:510}), obtained from the quantum linear Boltzmann equation in the
Brownian motion limit considered in Sect.~\ref{sec:qbm}.

Although the form of Eq.~(\ref{eq:holevo}) complies with the Lindblad result,
it provides much richer information on the structure of the Lindblad
operators. For instance, it states that the only possible dependence on the
position operator $\mathsf{X}$ can be in the unitary momentum boost operator
$\exp \left( i\tmmathbf{Q} \cdot \mathsf{X} / \hbar \right)$. In particular,
the diagonal matrix elements of Eq.~(\ref{eq:holevo}) in the momentum
representation are given by
\begin{eqnarray}
  \langle \tmmathbf{P}| \mathcal{L} \rho |\tmmathbf{P} \rangle & = & \int
  \mathd \tmmathbf{Q} \bignone \sum_j \left[ \left| L_j \left( \tmmathbf{Q},
  \tmmathbf{P}-\tmmathbf{Q} \right) \right|^2 \langle
  \tmmathbf{P}-\tmmathbf{Q}| \rho |\tmmathbf{P}-\tmmathbf{Q} \rangle - \left|
  L_j \left( \tmmathbf{Q}, \tmmathbf{P} \right) \right|^2 \langle
  \tmmathbf{P}| \rho |\tmmathbf{P} \rangle \right] .  \label{eq:hdiag}
\end{eqnarray}

\subsection{Combining translational invariance with the classical equation}

One can now ask whether a natural choice of the Lindblad operators for the
quantum linear Boltzmann equation is suggested already by combining the
requirements posed by its classical analog and by the characterization of
translation-covariant Markovian dynamics. Specifically, by comparing the
diagonal momentum matrix elements of the covariant master equation
(\ref{eq:hdiag}) with the classical expression Eq.~(\ref{eq:inout}) one finds
that the gain term $\sum_j \left| L_j \left( \tmmathbf{Q},
\tmmathbf{P}-\tmmathbf{Q} \right) \right|^2$ in Eq.~(\ref{eq:hdiag})
corresponds to that expression in Eq.~(\ref{eq:clbe}) which yields the
classical rate $M^{\tmop{cl}}_{\tmop{in}} \left( \tmmathbf{P}; \tmmathbf{Q}
\right)$ in Eq.~(\ref{eq:inout}).

It follows that the sum over the discrete set of Lindblad operators $L_j
\left( \mathsf{P} ; \tmmathbf{Q} \right)$ must be replaced by an integration
over a two-dimensional continuous index, in order to recover the five-fold
integration in Eq.~(\ref{eq:clbe}). Apart from the integral over the momentum
transfer $\tmmathbf{Q}$, this brings about an additional integration over the
plane of momenta perpendicular to $\tmmathbf{Q}$,
\begin{eqnarray}
  \sum_{_j} L_j \left( \mathsf{P} ; \tmmathbf{Q} \right) \rho L_j^{\dag}
  \left( \mathsf{P} ; \tmmathbf{Q} \right) \bignone & \rightarrow &
  \int_{\tmmathbf{Q}^{\bot}} \mathd \tmmathbf{k}_{\bot} \, L \left(
  \tmmathbf{k}_{\bot}, \mathsf{P} ; \tmmathbf{Q} \right) \rho L^{\dag} \left(
  \tmmathbf{k}_{\bot}, \mathsf{P} ; \tmmathbf{Q} \right)  \label{eq:proposal}
\end{eqnarray}
with $L \left( \tmmathbf{p}, \mathsf{P} ; \tmmathbf{Q} \right)$ the
accordingly continuous collection of Lindblad operators. It follows from
Eq.~(\ref{eq:clbe}) that they must satisfy
\begin{eqnarray}
  \left| L \left( \tmmathbf{p}, \mathsf{P} ; \tmmathbf{Q} \right) \right|^2 &
  = & \frac{n_{\tmop{gas}} m}{m^2_{\ast} Q} \sigma \left( \tmop{rel} \left(
  \mathbf{\tmmathbf{p}}_{\perp \tmmathbf{Q}}, \mathsf{P}_{\perp \tmmathbf{Q}}
  \right) - \frac{\tmmathbf{Q}}{2}, \tmop{rel} \left(
  \mathbf{\tmmathbf{p}}_{\perp \tmmathbf{Q}}, \mathsf{P}_{\perp \tmmathbf{Q}}
  \right) + \frac{\tmmathbf{Q}}{2} \right) \nonumber\\
  &  & \times \mu \left( \tmmathbf{p}_{\bot \tmmathbf{Q}} +
  \frac{m}{m_{\ast}} \frac{\tmmathbf{Q}}{2} + \frac{m}{M} 
  \mathsf{P}_{\|\tmmathbf{Q}} \right) .  \label{eq:ll}
\end{eqnarray}
The relation $\sigma \left( \tmmathbf{p}_f, \tmmathbf{p}_i \right) = \left| f
\left( \tmmathbf{p}_f, \tmmathbf{p}_i \right) \right|^2$ and the positivity of
$\mu$ now suggest a natural choice for the Lindblad operators,
\begin{eqnarray}
  L \left( \tmmathbf{p}, \mathsf{P} ; \tmmathbf{Q} \right) & = &
  \sqrt{\frac{n_{\tmop{gas}} m}{m_{\ast}^2 Q}} f \left( \tmop{rel} \left(
  \tmmathbf{p}_{\bot \tmmathbf{Q}}, \mathsf{P}_{\perp \tmmathbf{Q}} \right) -
  \frac{\tmmathbf{Q}}{2}, \tmop{rel} \left( \tmmathbf{p}_{\bot \tmmathbf{Q}},
  \mathsf{P}_{\perp \tmmathbf{Q}} \right) + \frac{\tmmathbf{Q}}{2} \right)
  \nonumber\\
  &  & \times \sqrt{\mu \left( \tmmathbf{p}_{\bot \tmmathbf{Q}} +
  \frac{m}{m_{\ast}} \frac{\tmmathbf{Q}}{2} + \frac{m}{M} 
  \mathsf{P}_{\|\tmmathbf{Q}} \right)} .  \label{eq:Lbis}
\end{eqnarray}
This expression, which involves a complex, operator-valued scattering
amplitude, was obtained here in a rather straightforward and intuitive way. It
is identical with the result~(\ref{eq:L}), derived stringently by means of
quantum scattering theory within the monitoring approach.

\subsection{Inequivalent expressions}\label{sec:altforms}

The previous section presented a rather suggestive line of reasoning in
support of the expression Eq.~(\ref{eq:L}). We emphasize, however, that these
arguments are not fully conclusive, since one can formulate quantum master
equations different from Eq.~(\ref{eq:qlbe}), which still comply with
Eq.~(\ref{eq:clbe}) and Eq.~(\ref{eq:holevo}). Incidentally, this is the case
for a proposal by Di\`osi {\cite{Diosi1995a}}, aiming at the generalization of
the classical linear Boltzmann equation. Clearly, the remaining freedom lies
in the fact that the diagonal matrix elements in the momentum representation
do not uniquely fix the mapping $\mathcal{L}$. The quantum linear Boltzmann
equation is an operator equation for the density matrix, after all, while its
classical version governs only a probability distribution.

In order to pinpoint the remaining freedom and the connection to Di\`osi's
result {\cite{Diosi1995a}} let us return to the classical expression
Eq.~(\ref{eq:clbe}). This equation is invariant under a change of integration
variables of the form
\begin{eqnarray}
  \tmmathbf{k}_{\bot} & \rightarrow & a (\tmmathbf{k}_{\bot}, \tmmathbf{P};
  \tmmathbf{Q})\tmmathbf{k}_{\bot} + b (\tmmathbf{k}_{\bot}, \tmmathbf{P};
  \tmmathbf{Q})\tmmathbf{P}_{\bot \tmmathbf{Q}},  \label{eq:trasla}
\end{eqnarray}
with $a$ and $b$ arbitrary scalar functions. This changes the dependence on
the test particle momentum $\tmmathbf{P}$ in the explicit expression of the
classical gain term, such that different master equations are obtained when
promoting $\tmmathbf{P}$ to an operator and reading off the Lindblad
operators, as done above in Eqs.~(\ref{eq:ll}) and (\ref{eq:Lbis}). These
various master equations coincide with respect to the diagonal in momentum
representation, because one can translate back the integration variables when
evaluating those matrix elements.

In particular, by choosing $a = 1$ and $b = m / M$ the classical linear
Boltzmann equation takes the form
\begin{eqnarray}
  \text{$\partial^{\tmop{coll}}_t f$} (\tmmathbf{P}) & = &
  \frac{n_{\tmop{gas}} m}{m^2_{\ast}} \bigintlim \frac{\mathd \tmmathbf{Q}}{Q}
  \int_{\tmmathbf{Q}^{\bot}} \mathd \tmmathbf{k}_{\bot} \, \sigma \left(
  \frac{m_{\ast}}{m} \tmmathbf{k}_{\bot} - \frac{\tmmathbf{Q}}{2},
  \frac{m_{\ast}}{m} \tmmathbf{k}_{\bot} + \frac{\tmmathbf{Q}}{2} \right)
  \nonumber\\
  &  & \times \left[ \mu \left( \tmmathbf{k}_{\bot} + \frac{m}{M}
  \tmmathbf{P}_{\bot \tmmathbf{Q}} + \text{$\frac{m}{m_{\ast}}
  \frac{\tmmathbf{Q}}{2} + \frac{m}{M} \left( \tmmathbf{P}_{\| \tmmathbf{Q}}
  -\tmmathbf{Q} \right)$} \right) f \left( \tmmathbf{P}-\tmmathbf{Q} \right)
  \right. \nonumber\\
  &  & \left. - \mu \left( \tmmathbf{k}_{\bot} + \frac{m}{M}
  \tmmathbf{P}_{\bot \tmmathbf{Q}} + \frac{m}{m_{\ast}} \frac{\tmmathbf{Q}}{2}
  + \frac{m}{M} \mathbf{\tmmathbf{P}}_{\|\tmmathbf{Q}} \right) f \left(
  \tmmathbf{P} \right) \right] .  \label{eq:diosic}
\end{eqnarray}
In this case, the constraint on the Lindblad operators corresponding to
Eq.~(\ref{eq:ll}) reads as
\begin{eqnarray}
  \left| L \left( \tmmathbf{p}, \mathsf{P} ; \tmmathbf{Q} \right) \right|^2 &
  = & \frac{n_{\tmop{gas}} m}{m^2_{\ast} Q} \sigma \left( \frac{m_{\ast}}{m}
  \mathbf{\tmmathbf{p}}_{\perp \tmmathbf{Q}} - \frac{\tmmathbf{Q}}{2},
  \frac{m_{\ast}}{m} \mathbf{\tmmathbf{p}}_{\perp \tmmathbf{Q}} +
  \frac{\tmmathbf{Q}}{2} \right) \nonumber\\
  &  & \times \mu \left( \tmmathbf{p} \mathbf{}_{\perp \tmmathbf{Q}} +
  \frac{m}{m_{\ast}} \frac{\tmmathbf{Q}}{2} + \frac{m}{M} \mathsf{P} \right) .
  \label{eq:llbis}
\end{eqnarray}
In the same spirit as above, this would lead to the identification
\begin{eqnarray}
  L_D \left( \tmmathbf{p}, \mathsf{P} ; \tmmathbf{Q} \right) & = &
  \sqrt{\frac{n_{\tmop{gas}} m}{m_{\ast}^2 Q}} f \left( \frac{m_{\ast}}{m}
  \mathbf{\tmmathbf{p}}_{\perp \tmmathbf{Q}} - \frac{\tmmathbf{Q}}{2},
  \frac{m_{\ast}}{m} \mathbf{\tmmathbf{p}}_{\perp \tmmathbf{Q}} +
  \frac{\tmmathbf{Q}}{2} \right) \nonumber\\
  &  & \times \sqrt{\mu \left( \tmmathbf{p}_{\bot \tmmathbf{Q}} +
  \frac{m}{m_{\ast}} \frac{\tmmathbf{Q}}{2} + \frac{m}{M} \mathsf{P} \right)},
  \label{eq:Ldiosi}
\end{eqnarray}
which corresponds to Di\`osi's result {\cite{Diosi1995a}} written in operator
form. Unlike the quantum linear Boltzmann equation~(\ref{eq:L}), the
scattering amplitude is here not operator-valued, since it is independent of
$\mathsf{P}$.

Other choices of $a$ and $b$ can lead to further equations of motion. They
all recover the classical linear Boltzmann equation, since their diagonal
momentum matrix elements coincide, granting in particular the existence of a
stationary solution. However, they differ in general as far as the description
of quantum effects is concerned. For instance, the diffusive limit considered
in Sect.~\ref{sec:qbm} will lead to different master equations of quantum
Brownian motion. In this respect, the assignment (\ref{eq:Lbis}), and
therefore the quantum linear Boltzmann equation, is distinguished by the fact
that it leads to the minimal correction of the Caldeira-Leggett master
equation {\cite{Breuer2007}}, as discussed in Sect.~\ref{sec:qbm}.

As another distinguishing feature of the assignment (\ref{eq:Lbis}), the
complex rate $M_{\tmop{in}} \left( \tmmathbf{P}, \tmmathbf{P}' ; \tmmathbf{Q}
\right)$ resulting from Eq.~(\ref{eq:Min3}) has a physically meaningful form.
It is an integration over {\tmem{all}} those pairs of scattering amplitudes
$f$ and $f^{\ast}$ which are defined by the value of $\tmmathbf{P}$ and of
$\tmmathbf{P}'$, respectively, together with the requirement of energy and
momentum conservation for a given choice of $\tmmathbf{Q}$ and the initial gas
particle momentum {\cite{Hornberger2007a}}. In an assignment like
(\ref{eq:Ldiosi}), in contrast, test particle momenta with different energies
$P^2 \neq \left( P' \right)^2$ would contribute with some common scattering
cross-section which does not correspond to the initial and final relative
momenta determined by $\tmmathbf{P}, \tmmathbf{P}'$, and $\tmmathbf{Q}$.

\section{Properties of the quantum linear Boltzmann
equation}\label{sec:relax}

We now address features of the quantum linear Boltzmann equation, which are
the quantum counterpart of properties also exhibited by the classical version.
We start by considering the behavior under symmetry transformations. Then the
relaxation dynamics predicted by the equation is addressed, and in particular
the approach to an equilibrium state. As in the classical case, the latter can
be understood independently of the details of scattering cross-section and gas
properties. This thermalization behavior is exemplified by means of numerical
results based on stochastic unravellings which exploit the particular form of
the master equation.

\subsection{Symmetry transformations}

\subsubsection{Spatial translations}\label{sec:ti}

A basic symmetry property of the quantum linear Boltzmann equation is its
invariance under spatial translations, which was already employed in
Sect.~\ref{sec:holevo}. One expects this to hold because the collisional
interaction is due to a two-body potential which depends on the relative
distance between test and gas particle. Together with the homogeneity of the
gas this implies that a transformation to a frame of reference translated by a
vector $\tmmathbf{A}$ should not affect the dynamics. This property is
expressed at the mathematical level by requiring that the mapping yielding the
infinitesimal time evolution commutes with the action of the unitary
representation of translations, as expressed by Eq.~(\ref{eq:cov}). It implies
in particular that if the statistical operator $\rho$ is a solution of the
master equation, the translated operator $\mathe^{- i \mathsf{\tmmathbf{A}
\cdot P / \hbar}} \rho \mathe^{i \mathsf{\tmmathbf{A} \cdot P / \hbar}}$ also
provides a solution.

To check this invariance of the quantum linear Boltzmann equation let us
consider the effect of a translation by a vector $\tmmathbf{A}$ on the
superoperator $\mathcal{M}$ appearing in Eq.~(\ref{eq:lvonn}). We here recall
its explicit expression
\begin{eqnarray}
  \mathcal{M} \rho & = & - \frac{i}{\hbar} \left[ \mathsf{H}_0 + H_{\text{n}}
  \left( \mathsf{P} \right), \rho \right] + \bigintlim \mathd \tmmathbf{Q}
  \int_{\tmmathbf{Q}^{\bot}} \mathd \tmmathbf{k}_{\bot} \left[
  \mathe^{i\tmmathbf{Q} \cdot \mathsf{X} / \hbar} L \left(
  \tmmathbf{k}_{\bot}, \mathsf{P} ; \tmmathbf{Q} \right) \rho L^{\dag} \left(
  \tmmathbf{k}_{\bot}, \mathsf{P} ; \tmmathbf{Q} \right) \mathe^{-
  i\tmmathbf{Q} \cdot \mathsf{X} / \hbar} \right. \nonumber\\
  &  & \left. - \frac{1}{2} \left\{ \rho, L^{\dag} \left(
  \tmmathbf{k}_{\bot}, \mathsf{P} ; \tmmathbf{Q} \right) L \left(
  \tmmathbf{k}_{\bot}, \mathsf{P} ; \tmmathbf{Q} \right) \right\} \right], 
  \label{eq:mdefbis}
\end{eqnarray}
with $L \left( \tmmathbf{p}, \mathsf{P} ; \tmmathbf{Q} \right)$ as in
Eq.~(\ref{eq:L}) and $H_{\text{n}} \left( \mathsf{P} \right)$ as in
Eq.~(\ref{eq:effectivebis}).

The mapping in the transformed frame of reference is then given by
\begin{eqnarray}
  \mathcal{M}_{\tmmathbf{A}}  \left[ \rho \right] & = & \mathe^{i
  \mathsf{\tmmathbf{A} \cdot P / \hbar}} \mathcal{M} \left[ \mathe^{- i
  \mathsf{\tmmathbf{A} \cdot P / \hbar}} \rho \mathe^{i \mathsf{\tmmathbf{A}
  \cdot P / \hbar}} \right] \mathe^{- i \mathsf{\tmmathbf{A} \cdot P /
  \hbar}},  \label{eq:Lcala}
\end{eqnarray}
and we have to verify that $\mathcal{\mathcal{M_{\tmmathbf{A}}} = M}$.

This is easily checked by separately considering the commutator term and the
incoherent part in (\ref{eq:mdefbis}). The Hamiltonian terms are only
functions of the momentum operator $\mathsf{P}$ and therefore invariant under
translations. In the incoherent part, the specific form of $L \left(
\tmmathbf{p}, \mathsf{P} ; \tmmathbf{Q} \right)$ is not important, since it
commutes in any case with $\exp \left( i \mathsf{\tmmathbf{A} \cdot P / \hbar}
\right)$, while the unitary operators $\exp \left( i\tmmathbf{Q} \cdot
\mathsf{X} / \hbar \right)$ and $\exp \left( - i\tmmathbf{Q} \cdot \mathsf{X}
/ \hbar \right)$ are modified by two opposite phases which cancel out. The
invariance clearly relies on the special way in which the position operator
$\mathsf{X}$ appears in Eq.~(\ref{eq:mdefbis}).

\subsubsection{Velocity transformations}

The collisional physics underlying the quantum linear Boltzmann equation not
only fixes its behavior under translations, but also under velocity
transformations. This is due to the fact that the interaction rate depends on
the relative velocity between the tracer particle and the gas particles, the
latter being drawn from the momentum distribution of the gas particles. A
transformation to a reference frame that moves with constant velocity
$\tmmathbf{V}$ should therefore be equivalent to performing the corresponding
shift in the momentum distribution of the gas particles.

Let us therefore consider the superoperator of the infinitesimal time
evolution $\mathcal{M}_{\tmmathbf{V}} $, valid in a frame of reference which
moves with velocity $\tmmathbf{V}$ compared to the original frame,
\begin{eqnarray}
  \mathcal{M}_{\tmmathbf{V}}  \left[ \rho \right] & = & \mathe^{iM\tmmathbf{V}
  \cdot \mathsf{X} / \hbar} \mathcal{M} \left[ \mathe^{- iM\tmmathbf{V} \cdot
  \mathsf{X} / \hbar} \rho \mathe^{iM\tmmathbf{V} \cdot \mathsf{X} / \hbar}
  \right] \mathe^{- iM\tmmathbf{V} \cdot \mathsf{X} / \hbar}, 
  \label{eq:Lcalv}
\end{eqnarray}
where $\mathcal{M}$ is given by Eq.~(\ref{eq:mdefbis}). According to the
microscopic considerations above, we expect that $\mathcal{M}_{\tmmathbf{V}}$
can be obtained as well by evaluating $\mathcal{M}$ with the transformed gas
momentum distribution $\mu \left( \tmmathbf{p} \right) \rightarrow
\mu^{\tmmathbf{V}} \left( \tmmathbf{p} \right) = \mu \left( \tmmathbf{p}-
m\tmmathbf{V} \right)$.

To check this symmetry, we first observe that the free kinetic term is left
untouched. The explicit form~(\ref{eq:effectivebis}) of the gas induced energy
shift also immediately grants that $H_{\text{n}} \left( \mathsf{P} -
M\tmmathbf{V} \right) = H^{\tmmathbf{V}}_{\text{n}} \left( \mathsf{P}
\right)$, where the superscript $\tmmathbf{V}$ indicates quantities evaluated
with the shifted momentum distribution $\mu^{\tmmathbf{V}}$.

Applying the same analysis to the remaining incoherent part, we note
that~(\ref{eq:Lcalv}) transforms the second term of Eq.~(\ref{eq:mdefbis})
into
\begin{eqnarray}
  \mathcal{L}_{\tmmathbf{V}} \left[ \rho \right] & = & \bigintlim \mathd
  \tmmathbf{Q} \int_{\tmmathbf{Q}^{\bot}} \mathd \tmmathbf{k}_{\bot} \left[
  \mathe^{i\tmmathbf{Q} \cdot \mathsf{X} / \hbar} L \left(
  \tmmathbf{k}_{\bot}, \mathsf{P} - M\tmmathbf{V}; \tmmathbf{Q} \right) \rho
  L^{^{\dag}} \left( \tmmathbf{k}_{\bot}, \mathsf{P} - M\tmmathbf{V};
  \tmmathbf{Q} \right) \mathe^{- i\tmmathbf{Q} \cdot \mathsf{X} / \hbar}
  \right. \nonumber\\
  &  & \left. - \frac{1}{2} \left\{ \rho, L^{^{\dag}} \left(
  \tmmathbf{k}_{\bot}, \mathsf{P} - M\tmmathbf{V}; \tmmathbf{Q} \right) L
  \left( \tmmathbf{k}_{\bot}, \mathsf{P} - M\tmmathbf{V}; \tmmathbf{Q} \right)
  \right\} \right] .  \label{eq:vv}
\end{eqnarray}
The definition of $L \left( \tmmathbf{p}, \tmmathbf{P}; \tmmathbf{Q} \right)$
in Eq.~(\ref{eq:L}) implies
\begin{eqnarray}
  L \left( \tmmathbf{p}, \tmmathbf{P}- M\tmmathbf{V}; \tmmathbf{Q} \right)  &
  = & \, L^{\tmmathbf{V}} \left( \tmmathbf{p}+ m\tmmathbf{V}_{\bot
  \tmmathbf{Q}}, \tmmathbf{P}; \tmmathbf{Q} \right) .  \label{eq:1}
\end{eqnarray}
By inserting this into Eq.~(\ref{eq:vv}) and performing a change of
integration variables $\tmmathbf{k}_{\bot} \rightarrow \tmmathbf{k}_{\bot}'
=\tmmathbf{k}_{\bot} - m\tmmathbf{V}_{\bot \tmmathbf{Q}}$ one also finds for
the incoherent part that the transformation (\ref{eq:Lcalv}) is tantamount to
the replacement $\mu \rightarrow \mu^{\tmmathbf{V}}$. In our notation this
proves that $\mathcal{M}_{\tmmathbf{V}} = \mathcal{M}^{\tmmathbf{V}}$, that
is, boosting the reference frame is equivalent to translating the gas
distribution in momentum space.

\subsubsection{Rotations}

On similar grounds one can consider the behavior of the quantum linear
Boltzmann equation under spatial rotations. Let us denote by $R$ the rotation
around a direction $\tmmathbf{n}$ by the angle $\alpha$. The mapping
$\mathcal{M}_R$ in the rotated frame is then related to the original mapping
$\mathcal{M}$ by
\begin{eqnarray}
  \mathcal{M}_R  \left[ \rho \right] & = & \mathe^{i \alpha \tmmathbf{n} \cdot
  \mathsf{J} / \hbar} \mathcal{M} \left[ \mathe^{- i \alpha \tmmathbf{n} \cdot
  \mathsf{J} / \hbar} \rho \mathe^{i \alpha \tmmathbf{n} \cdot \mathsf{J} /
  \hbar} \right] \mathe^{- i \alpha \tmmathbf{n} \cdot \mathsf{J} / \hbar}, 
  \label{eq:ruota}
\end{eqnarray}
where $\mathsf{J}$ is the angular momentum of the test particle, acting as the
generator of rotations.

Usually the collisional interaction between the particles depends only on the
relative orientation of incoming and outgoing momenta, $f \left(
\tmmathbf{p}_f, \tmmathbf{p}_i \right) = f \left( R\tmmathbf{p}_f,
R\tmmathbf{p}_i \right)$. One expects that the quantum linear Boltzmann
equation is then invariant under rotations provided the gas momentum
distribution is isotropic, i.e., $\mu \left( \tmmathbf{p} \right) = \mu \left(
R\tmmathbf{p} \right)$. Indeed, the identity $\mathcal{M}_R = \mathcal{M}$ is
then checked easily.

\subsection{Approach to equilibrium}\label{sec:eq}

\subsubsection{Detailed balance and the stationary solution}

One of the most important features of both the classical and the quantum
linear Boltzmann equation is the existence of a stationary solution of the
canonical form, e.g. proportional to the thermal momentum distribution
$\nu_{\tmop{EQ}} (\tmmathbf{P})$ given by
\begin{eqnarray}
  \text{$\nu_{\tmop{EQ}} \left( \tmmathbf{P} \right)$} & = & \frac{1}{\pi^{3 /
  2} P_{\beta}^3} \exp \left( - \frac{\tmmathbf{P}^2}{P_{\beta}^2} \right), 
  \label{eq:muMB2}
\end{eqnarray}
with $P_{\beta}^{} = \sqrt{2 M / \beta}$. It is reached asymptotically for any
initial state and depends only on the mass $M$ of the test particle and on the
temperature $T = 1 / \left( k_{\text{B}} \beta \right)$ of the gas, provided
the latter is described by the Maxwell-Boltzmann distribution (\ref{eq:muMB}).
We now show that the existence of this solution relies also in the quantum
case on the fact that the scattering rate $M_{\tmop{in}}^{\tmop{cl}}$ obeys
the so-called detailed balance condition. Here and in the following the
Hamiltonian part of the master equation does not contribute, so that we can
concentrate on the mapping $\mathcal{L}$ instead of $\mathcal{M}$.

As a first step, consider the quantum linear Boltzmann
equation~(\ref{eq:qlbe}) for a state $\nu ( \mathsf{P})$, which is only a
function of the momentum operator $\mathsf{P}$
\begin{eqnarray}
  \mathcal{L} \left[ \nu ( \mathsf{P}) \right] & = & \bigintlim \mathd
  \tmmathbf{Q} \left[ M^{\tmop{cl}}_{\tmop{in}} \left( \mathsf{P} ;
  \tmmathbf{Q} \right) \text{$\nu ( \mathsf{P} -\tmmathbf{Q})$} -
  M^{\tmop{cl}}_{\tmop{in}} \left( \mathsf{P} +\tmmathbf{Q}; \tmmathbf{Q}
  \right) \text{$\nu ( \mathsf{P})$} \right] .  \label{eq:x2}
\end{eqnarray}
One says that the transition rate $M^{\tmop{cl}} \left( \tmmathbf{P}_i
\rightarrow \tmmathbf{P}_f \right) \equiv M^{\tmop{cl}}_{\tmop{in}} \left(
\tmmathbf{P}_f ; \tmmathbf{P}_f -\tmmathbf{P}_i \right)$ obeys the detailed
balance condition provided
\begin{eqnarray}
  M^{\tmop{cl}} \left( \tmmathbf{P}_1 \rightarrow \tmmathbf{P}_2 \right) \exp
  \left( - \beta \frac{P_1^2}{2 M} \right) & = & M^{\tmop{cl}} \left(
  \tmmathbf{P}_2 \rightarrow \tmmathbf{P}_1 \right) \exp \left( - \beta
  \frac{P_2^2}{2 M} \right) .  \label{eq:dbcm}
\end{eqnarray}
For the gain rate Eq.~(\ref{eq:ratein}) this implies
\begin{eqnarray}
  M^{\tmop{cl}}_{\tmop{in}} \left( \tmmathbf{P}+\tmmathbf{Q}; \tmmathbf{Q}
  \right) & = & M^{\tmop{cl}}_{\tmop{in}} \left( \tmmathbf{P}; -\tmmathbf{Q}
  \right) \exp \left( - \beta E \left( \tmmathbf{Q}, \tmmathbf{P} \right)
  \right),  \label{eq:53}
\end{eqnarray}
with $E \left( \tmmathbf{Q}, \tmmathbf{P} \right)$ denoting the energy
transferred to the test particle in a collision where the momentum changes
from $\tmmathbf{P}$ to $\tmmathbf{P}+\tmmathbf{Q}$,
\begin{eqnarray}
  E \left( \tmmathbf{Q}, \tmmathbf{P} \right) & = & \frac{\left(
  \tmmathbf{P}+\tmmathbf{Q} \right)^2}{2 M} - \frac{P^2}{2 M} = \frac{Q^2}{2
  M} + \frac{\tmmathbf{Q} \cdot \tmmathbf{P}}{M} .  \label{eq:etransfer}
\end{eqnarray}
This quantity also relates the Maxwell-Boltzmann distribution at different
momenta,
\begin{eqnarray}
  \mathe^{- \beta E \left( \tmmathbf{Q}, \tmmathbf{P} \right)} \nu_{\tmop{EQ}}
  \left( \tmmathbf{P} \right) & = & \nu_{\tmop{EQ}} \left(
  \tmmathbf{P}+\tmmathbf{Q} \right) .  \label{eq:db2}
\end{eqnarray}
Based on the detailed balance condition~(\ref{eq:dbcm}) we therefore have
\begin{eqnarray}
  \mathcal{L} \nu_{\tmop{EQ}} ( \mathsf{P}) & = & \bigintlim \mathd
  \tmmathbf{Q} \left[ M^{\tmop{cl}}_{\tmop{in}} \left( \mathsf{P} ;
  \tmmathbf{Q} \right) \text{$\nu_{\tmop{EQ}} ( \mathsf{P} -\tmmathbf{Q})$} -
  M^{\tmop{cl}}_{\tmop{in}} \left( \mathsf{P} ; -\tmmathbf{Q} \right)
  \text{$\nu_{\tmop{EQ}} ( \mathsf{P} +\tmmathbf{Q})$} \right],  \label{eq:x3}
\end{eqnarray}
which is equal to zero because the integrand is manifestly odd in
$\tmmathbf{Q}$. If follows that $\nu_{\tmop{EQ}} ( \mathsf{P})$ is then indeed
a stationary solution.

The problem is therefore reduced to checking the validity of the detailed
balance condition (\ref{eq:dbcm}). This can be done considering the explicit
expression for the classical rate given by Eq.~(\ref{eq:diretta}) and taking
as the stationary distribution function $\mu$ for the gas particles the
Maxwell-Boltzmann distribution (\ref{eq:muMB}). Provided the scattering
cross-section obeys $\sigma \left( \tmmathbf{p}_f, \tmmathbf{p}_i \right) =
\sigma \left( \tmmathbf{p}_i, \tmmathbf{p}_f \right)$ one can then directly
confirm that (\ref{eq:53}) holds.

Finally, we emphasize that the existence of a stationary solution can be
proved in a more general and direct way by writing the quantum linear
Boltzmann equation in terms of the dynamic structure factor of the gas, as
will be discussed in Sect.~\ref{sec:dsf}.

\subsubsection{Entropy growth}

The celebrated H-theorem for the nonlinear Boltzmann equation
{\cite{Cercignani1975a}} states that the function
\begin{eqnarray}
  H \left( t \right) & = & \int \mathd \tmmathbf{P} \, \text{$f$}
  (\tmmathbf{P}) \log \bignone f \left( \tmmathbf{P} \right),  \label{eq:H}
\end{eqnarray}
satisfies $\partial_t H \left( t \right) \leqslant 0$. That is, the entropy $S
\left( t \right) = - k_{\text{B}} H \left( t \right)$ associated to the
momentum distribution function $f \left( \tmmathbf{P} \right)$ of the
self-interacting gas particles does not decrease under the evolution of the
nonlinear Boltzmann equation. In particular, $\partial_t H \left( t \right) =
0$ if and only if $f \left( \tmmathbf{P} \right) = f_{\tmop{EQ}} \left(
\tmmathbf{P} \right)$, with $f_{\tmop{EQ}} \left( \tmmathbf{P} \right)$ the
canonical stationary solution given by the Maxwell-Boltzmann distribution.
This tells us that $f_{\tmop{EQ}} \left( \tmmathbf{P} \right)$ is the only
stationary solution, reached for $t \rightarrow \infty$ independently of the
initial condition {\cite{Huang1987}}.

Unlike the proper Boltzmann equation, the classical linear Boltzmann equation
describes an open system, implying that the entropy associated to the
distribution function $f \left( \tmmathbf{P} \right)$ can increase or decrease
in general. In this case it is necessary to consider the relative entropy of
$f \left( \tmmathbf{P} \right)$, defined with respect to another reference
distribution function $g \left( \tmmathbf{P} \right)$,
\begin{eqnarray}
  H \left( f|g \right) & = & \int \mathd \tmmathbf{P} \, f \left( \tmmathbf{P}
  \right) \log \frac{f \left( \tmmathbf{P} \right)}{g \left( \tmmathbf{P}
  \right)} \bignone .  \label{eq:x5}
\end{eqnarray}
As a measure of how effectively one can distinguish the two distribution
functions, the relative entropy provides information on how close the two
distribution functions are. The quantity is non-negative, and equal to zero if
and only if the two distribution functions coincide. We prove in
Sect.~\ref{sec:a2} that any solution $f \left( \tmmathbf{P} \right)$ of the
classical linear Boltzmann equation satisfies
\begin{eqnarray}
  \frac{\mathd}{\mathd t} H \left( f|f_{\tmop{EQ}} \right) & \leqslant & 0, 
  \label{eq:521}
\end{eqnarray}
where $f_{\tmop{EQ}} \left( \tmmathbf{P} \right)$ is the canonical stationary
solution given by the Maxwell-Boltzmann distribution. Also in this case we
have $\partial_t H \left( f|f_{\tmop{EQ}} \right) = 0$ if and only if $f
\left( \tmmathbf{P} \right) = f_{\tmop{EQ}} \left( \tmmathbf{P} \right)$.
Eq.~(\ref{eq:521}) thus reflects the fact that independently of the initial
condition any solution approaches the stationary distribution function for $t
\rightarrow \infty$.

The situation is similar in the case of the quantum linear Boltzmann equation.
Here one considers the quantum correspondence of the relative entropy and
studies its behavior under the action of completely positive quantum dynamical
semigroups {\cite{Spohn1978b,Lindblad1983,Alicki2001}}.

One should emphasize that the following argumentation can be put on firmer
ground only by confining the whole system in a region of finite volume. In
this case the generator of the dynamics is given by a bounded mapping, and the
stationary solution of the canonical form is indeed a trace class operator. If
the test particle can move in an infinite volume, on the other hand, an
operator of this form is not of trace class since $\mathsf{P}$ has a
continuous spectrum. The particle is not bounded to a finite region, and it
gets completely delocalized with elapsing time.

The quantum analog of the relative entropy is given by the expression
\begin{eqnarray}
  S \left( \rho |w \right) & = & k_{\text{B}} \tmop{Tr} \rho \log \rho -
  k_{\text{B}} \tmop{Tr} \rho \log w,  \label{eq:relent}
\end{eqnarray}
where $\rho$ and $w$ are two statistical operators. According to Klein's
inequality this quantity is non-negative, and equal to zero if and only if the
two statistical operators are equal
{\cite{Wehrl1978,Lindblad1983,Alicki2001}}.

Taking the quantum relative entropy of the solution of the master equation
$\rho_t = \mathe^{t \mathcal{M}} \rho$ with respect to the stationary solution
$\rho_{\tmop{EQ}}$ one is led to consider the time derivative of $S \left(
\rho_t | \rho_{\tmop{EQ}} \right)$. We can now exploit the inequality
\begin{eqnarray}
  S \left( \mathcal{U} \rho | \mathcal{U} w \right) & \leqslant & S \left(
  \rho |w \right),  \label{eq:525}
\end{eqnarray}
valid for any completely positive mapping $\mathcal{U}$
{\cite{Wehrl1978,Lindblad1983,Alicki2001}}. It provides a quantum analog of
the H-theorem for dynamical evolutions described by quantum dynamical
semigroups. In particular, for $\mathcal{U} = \mathe^{t \mathcal{M}}$, with
$\mathcal{M}$ the generator of a Markovian master equation, and noting that
$\rho_{\tmop{EQ}}$ is a time translation invariant state, $\mathe^{t
\mathcal{M}} \rho_{\tmop{EQ}} = \rho_{\tmop{EQ}}$, it follows that
\begin{eqnarray}
  \frac{\mathd}{\mathd t} S \left( \rho_t | \rho_{\tmop{EQ}} \right) &
  \leqslant & 0.  \label{eq:527}
\end{eqnarray}
Eq.~(\ref{eq:527}) tells us that, even though the variation of the von Neumann
entropy associated to the solution $\rho_t = \mathe^{t \mathcal{M}} \rho$ does
not have a definite sign, reflecting the fact that the system is open, one
still has loss of information with elapsing time because different initial
states get less and less distinguishable.

In particular, the derivative of the relative entropy with respect to time
can be written as
\begin{eqnarray}
  \frac{\mathd}{\mathd t} S \left( \rho_t | \rho_{\tmop{EQ}} \right) & = &
  k_{\text{B}} \tmop{Tr} \left[ \mathcal{M} \rho_t \left( \log \rho_t - \log
  \rho_{\tmop{EQ}} \right) \right] .  \label{eq:529}
\end{eqnarray}
It is equal to zero if and only if $\rho_t$ equals the stationary solution,
which is unique as a consequence of the fact that only multiples of the
identity commute with the Lindblad operators $\exp \left( i\tmmathbf{Q} \cdot
\mathsf{X} / \hbar \right) L \left( \tmmathbf{p}, \mathsf{P} ; \tmmathbf{Q}
\right)$ appearing in the quantum linear Boltzmann equation~(\ref{eq:lvonn})
{\cite{Spohn1978b,Ingarden1997}}. This result is in complete correspondence
with the classical case.

Moreover, according to Eq.~(\ref{eq:527}) any solution of the master equation
gets increasingly close to the stationary distribution function. This implies
in particular that any initial state gets diagonal in momentum representation
for $t \rightarrow \infty$. That is, the momentum coherences described by the
off-diagonal elements decay, such that the quantum linear Boltzmann equation
goes over to the classical linear Boltzmann equation, which describes the
remaining populations.

\subsection{Relaxation dynamics}\label{sec:emrelax}{\tmsamp{}}

\subsubsection{Momentum and energy relaxation}

The previous section considered the evolution of the statistical operator,
which suggested to work in the Schr\"odinger picture. In contrast to that, we
will now use the Heisenberg picture when focussing on the dynamics of momentum
and kinetic energy, that is, on the two observables most relevant for the
description of a test particle in a gas.

The Heisenberg time evolution of observables requires the adjoint mapping
$\mathcal{M}^{\ast}$, with respect to the quantum linear Boltzmann equation.
It is defined through the duality relation
\begin{eqnarray}
  \tmop{Tr} \left( \mathsf{A} \mathcal{M} \rho \right) & = & \tmop{Tr} \left(
  \rho \mathcal{M}^{\ast} \mathsf{A} \right),  \label{eq:x6}
\end{eqnarray}
which connects the space of trace class operators, containing the statistical
operators $\rho$, with its dual, the space of bounded observables
$\mathsf{A}$. The dynamics of a Heisenberg observable $\mathsf{A}_t$ is thus
given by
\begin{eqnarray}
  \frac{\mathd}{\mathd t} \mathsf{A}_t & = & \frac{1}{i \hbar} \left[
  \mathsf{A}_t, \mathsf{H}_0 + H_n \left( \mathsf{P} \right) \right] +
  \mathcal{L}^{\ast} \mathsf{A},  \label{eq:x7}
\end{eqnarray}
with
\begin{eqnarray}
  \mathcal{L}^{\ast} \mathsf{A} & = & \bigintlim \mathd \tmmathbf{Q}
  \int_{\tmmathbf{Q}^{\bot}} \mathd \tmmathbf{k}_{\bot} \left[ L^{^{\dag}}
  \left( \tmmathbf{k}_{\bot}, \mathsf{P} ; \tmmathbf{Q} \right) \mathe^{-
  i\tmmathbf{Q} \cdot \mathsf{X} / \hbar} \mathsf{A} \mathe^{i\tmmathbf{Q}
  \cdot \mathsf{X} / \hbar} L \left( \tmmathbf{k}_{\bot}, \mathsf{P} ;
  \tmmathbf{Q} \right)  \right. \nonumber\\
  &  & \left. - \frac{1}{2} \left\{ L^{^{\dag}} \left( \tmmathbf{k}_{\bot},
  \mathsf{P} ; \tmmathbf{Q} \right) L \left( \tmmathbf{k}_{\bot}, \mathsf{P} ;
  \tmmathbf{Q} \right), \mathsf{A} \right\} \right] \bignone . 
  \label{eq:Hqlbe}
\end{eqnarray}
We will take the Heisenberg operator to coincide with the corresponding
Schr\"odinger observable at $t = 0$, i.e., $\mathsf{A}_0 = \mathsf{A}$.

An important class of observables are those which are only functions of the
momentum operator, $\mathsf{A}_t = A_t \left( \mathsf{P} \right)$. The
equation of motion in the Heisenberg picture then simplifies considerably,
\begin{eqnarray}
  \frac{\mathd}{\mathd t} \mathsf{A}_t & = & \int \bignone \mathd \tmmathbf{Q}
  \int_{\tmmathbf{Q}^{\bot}} \mathd \tmmathbf{k}_{\bot} \left| L \left(
  \tmmathbf{k}_{\bot}, \mathsf{P} ; \tmmathbf{Q} \right) \right|^2 \left[ 
  \text{$\mathe^{- i\tmmathbf{Q} \cdot \mathsf{X} / \hbar} \mathsf{A}_t
  \mathe^{i\tmmathbf{Q} \cdot \mathsf{X} / \hbar}$} - \mathsf{A}_t \right] . 
  \label{eq:x8}
\end{eqnarray}
Note that $\mathcal{L}$, and equivalently $\mathcal{L}^{\ast}$, is covariant
under translations in the sense of Eq.~(\ref{eq:cov}), which implies that the
algebra generated by the momentum operator is left invariant. Consequently,
$\mathsf{A}_0 = A_0 \left( \mathsf{P} \right)$ implies $\mathsf{A}_t = A_t (
\mathsf{P})$ for $t > 0$ and therefore $[ \mathsf{A}_t, \mathsf{P}] = 0$.
Recalling the definition of the quantum and the classical gain rates in
Eqs.~(\ref{eq:Min3}) and (\ref{eq:ratein}), we find that observables given by
a function of momentum obey
\begin{eqnarray}
  \frac{\mathd}{\mathd t} \text{$A_t \left( \mathsf{P} \right)$} & = &
  \bignone \int \bignone \mathd \tmmathbf{Q} \, M^{\tmop{cl}}_{\tmop{in}}
  \left( \mathsf{P} +\tmmathbf{Q}; \tmmathbf{Q} \right) \left[ A_t \left(
  \mathsf{P} +\tmmathbf{Q} \right) - A_t \left( \mathsf{P} \right) \right], 
  \label{eq:xx}
\end{eqnarray}
in strict analogy with the classical formulation.

Let us now focus on the time evolution of the expectation values of momentum
and kinetic energy, $A \left( \mathsf{P} \right) = \mathsf{P}$ and $A \left(
\mathsf{P} \right) = \mathsf{P}^2 / \left( 2 M \right)$. Using the momentum
representation one can evaluate Eq.~(\ref{eq:xx}) for these quantities by
working with real numbers rather than operators. In particular, for $A
(\tmmathbf{P}) =\tmmathbf{P}$ the quantity $\left[ A \left(
\tmmathbf{P}+\tmmathbf{Q} \right) - A \left( \tmmathbf{P} \right) \right]$
appearing at the right-hand side of Eq.~(\ref{eq:xx}) is simply the momentum
transfer $\tmmathbf{Q}$ in the single collision. For $A (\tmmathbf{P}) = P^2 /
\left( 2 M \right)$ it is the energy transfer $E \left( \tmmathbf{Q},
\tmmathbf{P} \right)$ as given by Eq.~(\ref{eq:etransfer}). The positivity of
$M^{\tmop{cl}}_{\tmop{in}}$ thus allows one straightforwardly to confirm from
Eq.~(\ref{eq:xx}) that the change of momentum is positive when the momentum
transfer is positive. Likewise, a positive energy transfer increases the
energy.

The dynamic equation~(\ref{eq:xx}) for the momentum and kinetic energy can be
evaluated explicitly if one assumes a constant scattering cross-section $|f
\left( \tmmathbf{p}_f, \tmmathbf{p}_i \right) |^2 = \sigma_{\tmop{tot}} / 4
\pi$ and a free gas of Maxwell-Boltzmann particles. Inserting the expression
of $M^{\tmop{cl}}_{\tmop{in}}$ into Eq.~(\ref{eq:xx}) it can be shown in this
case {\cite{Vacchini2007d}} that the time evolution of the momentum
expectation value is given by
\begin{eqnarray}
  \frac{\mathd}{\mathd t} \langle \mathsf{P} \rangle_{\rho_t} & = & -
  \frac{8}{3 \sqrt{\pi}} \frac{_{} m_{\ast}}{M} \Gamma_{\beta} \langle
  \mathsf{P}  \text{}_1 F_1 \left( - \frac{1}{2}, \frac{5}{2} ; - \left(
  \frac{\mathsf{P}}{Mv_{\beta}} \right)^2 \right) \rangle_{\rho_t} . 
  \label{eq:p}
\end{eqnarray}
Accordingly, the kinetic energy $\mathsf{E} = \mathsf{P}^2 / \left( 2 M
\right)$ evolves as
\begin{eqnarray}
  \frac{\mathd}{\mathd t} \langle \mathsf{E} \rangle_{\rho_t} & = & -
  \frac{16}{3 \sqrt{\pi}} \frac{_{} m_{\ast}}{M} \Gamma_{\beta} \langle
  \text{}_1 F_1 \left( - \frac{1}{2}, \frac{5}{2} ; - \beta \mathsf{E}
  \frac{m}{M} \right) \mathsf{E} - \frac{3}{2 \beta} \frac{_{} m_{\ast} }{m} 
  \text{}_1 F_1 \left( - \frac{3}{2}, \frac{3}{2} ; - \beta \mathsf{E}
  \frac{m}{M} \right) \rangle_{\rho_t} .  \label{eq:e}
\end{eqnarray}
In these equations we have denoted the most probable velocity of the gas
particles as $v_{\beta} = p_{\beta} / m = \sqrt{2 / \left( \beta m \right)}$.
Moreover,
\begin{eqnarray}
  \Gamma_{\beta} & = & n_{\tmop{gas}} v_{\beta} \sigma_{\tmop{tot}} 
  \label{eq:gamma0}
\end{eqnarray}
is the total scattering rate for a flux of particles with the most probable
velocity $v_{\beta}$, and the symbol $\text{}_1 F_1$ denotes confluent
hypergeometric functions. Their explicit expressions are given by
\begin{eqnarray}
  \text{}_1 F_1 \left( - \frac{1}{2}, \frac{5}{2} ; - x^2 \right) & = &
  \frac{3}{16} \frac{1}{x^2} \left\{ \left[ 1 + 2 x^2 \right] \mathe^{- x^2} -
  \left[ 1 - 4 x^2 - 4 x^4 \right] \frac{\sqrt{\pi}}{2} \frac{\text{erf}
  (x)}{x}  \right\}  \label{eq:g3}
\end{eqnarray}
and
\begin{eqnarray}
  \text{}_1 F_1 \left( - \frac{3}{2}, \frac{3}{2} ; - x^2 \right) & = &
  \frac{1}{8} \left\{  \left[ 5 + 2 x^2 \right] \mathe^{- x^2} + \left[ 3 + 12
  x^2 + 4 x^4 \right] \frac{\sqrt{\pi}}{2}  \frac{\text{erf} \left( x
  \right)}{x}  \right\},  \label{eq:g4}
\end{eqnarray}
which are both positive monotone increasing functions. From these expressions
it is immediately clear that in general there is no closed evolution equation
for either the first or the second moment of the momentum operator, $\langle
\mathsf{P} \rangle_{\rho_t}$ or $\langle \mathsf{P}^2 \rangle_{\rho_t}$, since
moments of arbitrary high order are involved in the equation, due to the
presence of the confluent hypergeometric functions. This becomes manifest in
strong deviations of the moments from Gaussian statistics, as can be observed
by studying the time dependence of high order cumulants of the momentum
distribution {\cite{Breuer2007c}}.

However, the equations do get closed in the limit of a very massive test
particle close to thermal equilibrium, which corresponds to the diffusive
limit considered in Sect.~\ref{sec:qbm}. In this case, the velocity $V = P /
M$ of the test particle is taken to be much smaller than the typical velocity
$v_{\beta}$ of the gas particles. Thus taking $V / v_{\beta} \ll 1$ and $m / M
\ll 1$ in Eq.~(\ref{eq:p}) and Eq.~(\ref{eq:e}) the confluent hypergeometric
functions can be replaced by unity, since $\text{}_1 F_1 \left( \alpha, \gamma
; 0 \right) = 1$, while the reduced mass $m_{\ast}$ is replaced by $m$. This
leads to
\begin{eqnarray}
  \frac{\mathd}{\mathd t} \langle \mathsf{P} \rangle_{\rho_t} & = & - \eta
  \langle \mathsf{P} \rangle_{\rho_t}  \label{eq:ap1}
\end{eqnarray}
which describes a friction proportional to velocity, leading to the expected
exponential relaxation to a vanishing mean momentum. Similarly, one finds
\begin{eqnarray}
  \frac{\mathd}{\mathd t} \langle \mathsf{E} \rangle_{\rho_t} & = & - 2 \eta
  \left( \langle \mathsf{E} \rangle_{\rho_t} - \frac{3}{2 \beta} \right) 
  \label{eq:ap2}
\end{eqnarray}
which shows that the mean kinetic energy relaxes exponentially to the
equipartition value ${3 / 2 k_{\text{B}} T}$. The friction coefficient $\eta$
in Eq.~(\ref{eq:ap1}) and (\ref{eq:ap2}) is given by
\begin{eqnarray}
  \eta & = &  \frac{8}{3 \sqrt{\pi}} \frac{m}{M} \Gamma_{\beta} . 
  \label{eq:friction}
\end{eqnarray}
\subsubsection{Simulation algorithm}\label{sec:mc}

The complexity of the quantum linear Boltzmann equation prevents its solutions
from being analytically tractable, even if one allows for special choices of
the scattering amplitude and the initial state. Nevertheless, one can
efficiently simulate the temporal behavior of various interesting quantities
by using quantum trajectory methods, which build on the Lindblad structure of
the master equation and its translation-covariance. These techniques are based
on stochastic differential equations for the state vector $| \psi \left( t
\right) \rangle$, whose solutions are a stochastic process in the Hilbert
space. A stochastic unravelling of the quantum master equation is achieved if
the ensemble average $\mathbbm{E} \left[ | \psi \left( t \right) \rangle
\langle \psi \left( t \right) | \right]$ yields a solution of the master
equation
{\cite{Diosi1986a,Gardiner1992a,Carmichael1993a,Molmer1993a,Breuer2007}}.

One way of motivating these unravelling techniques is to consider measurement
schemes, which lead to instantaneous changes of the wave vector conditioned on
the random outcomes. Together with the covariant structure of the master
equation described above, this consideration leads to a natural choice of the
unravelling. It is a piecewise deterministic process, described by a
deterministic time evolution, which gets interrupted by random jumps which
change the particle momentum. This unravelling corresponds to a measurement
scheme in which one monitors the momentum transferred between system and
environment.

For the sake of simplicity we will restrict our analysis to the Born
approximation of the equation, where the scattering cross-section depends only
on the momentum transfer. Although this is not a conceptual restriction, it
leads to major simplifications in the sampling of the jump events.

In Born approximation the quantum linear Boltzmann equation reads
\begin{eqnarray}
  \frac{\mathd}{\mathd t} \rho & = & - \frac{i}{\hbar} \left[ \mathsf{H}, \rho
  \right] + \bigintlim \mathd \tmmathbf{Q} \left[ \mathe^{i\tmmathbf{Q} \cdot
  \mathsf{X} / \hbar} L_B \left( \mathsf{P} ; \tmmathbf{Q} \right) \rho
  L_B^{\dag} \left( \mathsf{P} ; \tmmathbf{Q} \right) \mathe^{- i\tmmathbf{Q}
  \cdot \mathsf{X} / \hbar} - \frac{1}{2} \left\{ L_B^{\dag} \left( \mathsf{P}
  ; \tmmathbf{Q} \right) L_B \left( \mathsf{P} ; \tmmathbf{Q} \right), \rho
  \right\} \right], \nonumber\\
  &  &  \label{eq:c}
\end{eqnarray}
where we have used the more compact notation $\mathsf{H} = \mathsf{H}_0 +
H_{\text{n}} \left( \mathsf{P} \right)$ for the Hamiltonian term, and the
Lindblad operators are given by Eq.~(\ref{eq:Lb}).

In the Monte Carlo wave function method {\cite{Molmer1993a,Breuer2007}} one
then considers a stochastic differential equation for the stochastic wave
vector $\psi \left( t \right)$, which has the form
\begin{eqnarray}
  \mathd | \psi \left( t \right) \rangle & = & \left[ - \frac{i}{\hbar}
  \mathsf{H} - \frac{\mathsf{G}}{2} + \frac{1}{2} \bigintlim \mathd
  \tmmathbf{Q}\|L_B \left( \mathsf{P} ; \tmmathbf{Q} \right) | \psi \left( t
  \right) \rangle \|^2 \right] | \psi \left( t \right) \rangle \mathd t
  \nonumber\\
  &  & + \bigintlim \mathd \tmmathbf{Q} \left[ \frac{\mathe^{i\tmmathbf{Q}
  \cdot \mathsf{X} / \hbar} L_B \left( \mathsf{P} ; \tmmathbf{Q} \right) |
  \psi \left( t \right) \rangle}{\|L_B \left( \mathsf{P} ; \tmmathbf{Q}
  \right) | \psi \left( t \right) \rangle \|} - | \psi \left( t \right)
  \rangle \right] \tmop{dN}_{\tmmathbf{Q}} \left( t \right) .  \label{eq:d}
\end{eqnarray}
Here, the rate operator $\mathsf{G}$ is given by
\begin{eqnarray}
  \mathsf{G} & = & \bigintlim \mathd \tmmathbf{Q} \, L_B^{\dag} \left(
  \mathsf{P} ; \tmmathbf{Q} \right) L_B \left( \mathsf{P} ; \tmmathbf{Q}
  \right) .  \label{eq:lossrate}
\end{eqnarray}
The field of Poisson increments $\tmop{dN}_{\tmmathbf{Q}} \left( t \right)$
satisfies
\begin{eqnarray}
  \tmop{dN}_{\tmmathbf{Q}} \left( t \right) \tmop{dN}_{\tmmathbf{Q}'} \left( t
  \right) & = & \delta^3 \left( \tmmathbf{Q}-\tmmathbf{Q}' \right)
  \tmop{dN}_{\tmmathbf{Q}} \left( t \right)  \label{eq:x9}\\
  \mathbbm{E} \left[ \tmop{dN}_{\tmmathbf{Q}} \left( t \right) \right] & = &
  \|L_B \left( \mathsf{P} ; \tmmathbf{Q} \right) | \psi \left( t \right)
  \rangle \|^2 \mathd t, \nonumber
\end{eqnarray}
where the symbol $\mathbbm{E}$ denotes the ensemble mean over realizations of
the process. The solutions of~(\ref{eq:d}) starting from an initial state $|
\psi \left( 0 \right) \rangle$ provide an unravelling of the master
equation~(\ref{eq:c}) in the sense that
\begin{eqnarray}
  \rho_t & = & \mathbbm{E} \left[ | \psi \left( t \right) \rangle \langle \psi
  \left( t \right) | \right],  \label{eq:x10}
\end{eqnarray}
if $\rho_0 = | \psi \left( 0 \right) \rangle \langle \psi \left( 0 \right) |$.

In spite of the complicated form of the stochastic differential equation,
which involves a field of Poisson increments, the Monte Carlo wave function
method provides a rather simple simulation algorithm. The realizations of the
process are characterized by deterministic time evolutions interrupted by
jumps. If a jump into the state $| \psi \left( t \right) \rangle$ occurs at
time $t$, it will evolve until the time $t + \tau$ of the next jump according
to the deterministic equation
\begin{eqnarray}
  \text{$| \psi \left( t + \tau \right) \rangle$} & = & \frac{\mathe^{- i
  \mathsf{H} \tau / \hbar - \mathsf{G} \tau / 2} | \psi \left( t \right)
  \rangle}{\| \mathe^{- i \mathsf{H} \tau / \hbar - \mathsf{G} \tau / 2} |
  \psi \left( t \right) \rangle \|} .  \label{eq:det}
\end{eqnarray}
The waiting time $\tau$ between two jumps is determined by the cumulative
distribution function
\begin{eqnarray}
  F \left( \tau \right) & = & 1 -\| \mathe^{- i \mathsf{H} \tau / \hbar -
  \mathsf{G} \tau / 2} | \psi \left( t \right) \rangle \|^2 .  \label{eq:g}
\end{eqnarray}
A straightforward way to obtain the random variable $\tau$ numerically is to
draw a random number $\eta$ from a uniform distribution over the interval
$\left[ 0, 1 \right]$ and to invert the relation $\| \mathe^{- i \mathsf{H}
\tau / \hbar - \mathsf{G} \tau / 2} | \psi \left( t \right) \rangle \|^2 =
\eta$. The jump is then implemented by the replacement
\begin{eqnarray}
  | \psi \left( t + \tau \right) \rangle & \rightarrow &
  \frac{\mathe^{i\tmmathbf{Q} \cdot \mathsf{X} / \hbar} L_B \left( \mathsf{P}
  ; \tmmathbf{Q} \right) | \psi \left( t + \tau \right) \rangle}{\|L_B \left(
  \mathsf{P} ; \tmmathbf{Q} \right) | \psi \left( t + \tau \right) \rangle
  \|},  \label{eq:h}
\end{eqnarray}
where the unitary operator $\exp \left( i\tmmathbf{Q} \cdot \mathsf{X} / \hbar
\right)$ effects a momentum kick of amount $\tmmathbf{Q}$. These momentum
transfers must be drawn from the probability distribution
\begin{eqnarray}
  \mathcal{P} \left( \tmmathbf{Q} \right) & = & \frac{\|L_B \left( \mathsf{P}
  ; \tmmathbf{Q} \right) | \psi \left( t \right) \rangle \|^2}{\langle \psi
  \left( t \right) | \mathsf{G} | \psi \left( t \right) \rangle}, 
  \label{eq:i}
\end{eqnarray}
which is normalized by the total scattering rate out of the state $\psi \left(
t \right)$,
\begin{eqnarray}
  \langle \psi \left( t \right) | \mathsf{G} | \psi \left( t \right) \rangle &
  = & \bigintlim \mathd \tmmathbf{Q}\|L_B \left( \mathsf{P} ; \tmmathbf{Q}
  \right) | \psi \left( t \right) \rangle \|^2 .  \label{eq:l}
\end{eqnarray}
The random jumps in this unravelling are therefore naturally associated to the
collisions with the gas molecules experienced by the test particle. The
realizations of the process are thus given by a deterministic evolution
interrupted by random momentum kicks corresponding to collisions. The value of
the momentum transfers and the rate of collisions are fixed by the state
dependent quantities Eq.~(\ref{eq:i}) and Eq.~(\ref{eq:l}) respectively.

The described algorithm can be implemented in a particularly simple way if the
initial state is a discrete superposition of momentum eigenstates, $| \psi
\left( 0 \right) \rangle = \sum_{_{j = 1}}^N c_j |\tmmathbf{P}_j \rangle
\bignone$. Due to the translation-covariance (\ref{eq:Lcala}) of the master
equation superpositions of this kind are preserved during the stochastic time
evolution in the sense that the quantum trajectory remains a superposition of
$N$ momentum eigenstates at all times. The stochastic process is thus mapped
into the space of the different amplitudes $c_j$ and momenta $\tmmathbf{P}_j$.

\subsubsection{The loss term}{\hspace*{\fill}}\label{sec:loss}

When discussing the momentum and energy relaxation in Sect.~\ref{sec:emrelax}
it was not necessary to work out the loss term of the quantum linear Boltzmann
equation explicitly, since it cancels out in Eq.~(\ref{eq:xx}). However, this
expression is of independent interest; it appears in the Monte Carlo
simulation of the quantum linear Boltzmann equation through the rate
Eq.~(\ref{eq:lossrate}), and it also determines the loss of visibility due to
collisional decoherence in matter wave interference experiments, as we shall
discuss in Sect.~\ref{sec:decoh}. We therefore provide an explicit expression
assuming that the cross-section depends only on the relative velocity, which
includes the case of a constant cross-section discussed in
Sect.~\ref{sec:emrelax}.

The classical loss term in the quantum linear Boltzmann equation is given by
\begin{eqnarray}
  \mathcal{L_{\tmop{loss}}} \rho & = & - \frac{1}{2} \left\{
  M_{\tmop{out}}^{\tmop{cl}} ( \mathsf{P}), \rho \right\},  \label{eq:loss}
\end{eqnarray}
as evident from Eq.~(\ref{eq:poffdiag}). It is convenient to rewrite the
classical loss rate $M_{\tmop{out}}^{\tmop{cl}} \left( \tmmathbf{P} \right)$,
in terms of an effective scattering cross-section $\sigma_{\tmop{eff}} \left(
\tmmathbf{P} \right)$ defined as the area to be multiplied by the current
density $n_{\tmop{gas}} V$ of the test particle,
\begin{eqnarray}
  M_{\tmop{out}}^{\tmop{cl}} \left( \tmmathbf{P} \right) & = & n_{\tmop{gas}}
  \frac{P}{M} \sigma_{\tmop{eff}} (\tmmathbf{P}) .  \label{eq:macro}
\end{eqnarray}
The effective cross-section can be calculated exactly if the scattering
cross-section is independent of the scattering angle (s-wave scattering), and
if it depends on the energy only as a power of the relative velocity between
test particle and gas particles. The microscopic cross-section can then be
expressed as
\begin{eqnarray}
  \sigma \left( \tmop{rel} \left( \mathbf{\tmmathbf{p}}_{\perp \tmmathbf{Q}},
  \tmmathbf{P}_{\perp \tmmathbf{Q}} \right) - \frac{\tmmathbf{Q}}{2},
  \tmop{rel} \left( \mathbf{\tmmathbf{p}}_{\perp \tmmathbf{Q}},
  \tmmathbf{P}_{\perp \tmmathbf{Q}} \right) + \frac{\tmmathbf{Q}}{2} \right) 
  & = & c \left| \frac{\tmmathbf{p}}{m} - \frac{\tmmathbf{P}}{M} \right|^a = c
  \left| \tmmathbf{v}-\tmmathbf{V} \right|^a .  \label{eq:x11}
\end{eqnarray}
This yields the effective scattering cross-section {\cite{Vacchini2004a}}
\begin{eqnarray}
  \sigma_{\tmop{eff}} \left( \tmmathbf{P} \right) & = & 8 \sqrt{\pi} \Gamma
  \left( \frac{a}{2} + 2 \right) \frac{Mv_{\beta}^{a + 1} c}{P} \text{}_1 F_1 
  \left( - \left( \frac{a}{2} + \frac{1}{2} \right), \frac{3}{2} ; - \left(
  \frac{P}{Mv_{\beta}}  \right)^2 \right),  \label{eq:a}
\end{eqnarray}
where $\text{}_1 F_1$ denotes the confluent hypergeometric function and
$v_{\beta} = \sqrt{2 / \left( \beta m \right)}$ the most probable velocity in
the gas.

For $a = 0$ one recovers the case of a constant scattering cross-section
$\left| f \left( \tmmathbf{p}_f, \tmmathbf{p}_i \right) \right|^2 =
\sigma_{\tmop{tot}} / 4 \pi$, used below to perform a Monte Carlo simulation
of the quantum linear Boltzmann equation. The corresponding expression is
\begin{eqnarray}
  \sigma_{\tmop{eff}} \left( \tmmathbf{P} \right) & = & \sigma_{\tmop{tot}} 
  \frac{2}{\sqrt{\pi}} \frac{Mv_{\beta}}{P} \text{}_1 F_1 \left( -
  \frac{1}{2}, \frac{3}{2} ; - \left( \frac{P}{Mv_{\beta}} \right)^2 \right), 
  \label{eq:a=0}
\end{eqnarray}
where
\begin{eqnarray}
  \text{}_1 F_1 \left( - \frac{1}{2}, \frac{3}{2} ; - x^2 \right) & = &
  \left\{  \frac{1}{2} \mathe^{- x^2} + \left[ 1 + 2 x^2 \right]
  \frac{\sqrt{\pi}}{4}  \frac{\text{erf} \left( x \right)}{x}  \right\} . 
  \label{eq:g5}
\end{eqnarray}
For small velocities of the test particle $V / v_{\beta} \ll 1$ the expression
Eq.~(\ref{eq:a=0}) simplifies to
\begin{eqnarray}
  \sigma_{\tmop{eff}} \left( \tmmathbf{P} \right) & \longrightarrowlim^{V \ll
  v_{\beta}} & \sigma_{\tmop{tot}}  \frac{2}{\sqrt{\pi}} \frac{Mv_{\beta}}{P}
  = \sigma_{\tmop{tot}} \frac{\bar{v}}{V},  \label{eq:x12}
\end{eqnarray}
with $\bar{v} = \sqrt{8 / \pi m \beta}$ the average thermal velocity. This
corresponds to the famous $1 / V$ dependence of the total scattering
cross-section for small velocities, expected when the scattering probability
is independent of the velocity of the tracer particle, as verified in neutron
scattering {\cite{Williams1966}}.

\subsubsection{Simulating the momentum and energy dynamics }\label{sec:Usim}

We now discuss a numerical study of the relaxation properties of the quantum
linear Boltzmann equation {\cite{Breuer2007c}}, which is based on the
unravelling given by Eq.~(\ref{eq:d}) and the algorithm presented above. The
use of dimensionless variables, naturally suggested by the typical physical
lengths of the problem, turns out to be very convenient. Specifically, we
introduce the variables\begin{figure}[tb]
  \resizebox{120mm}{!}{\epsfig{file=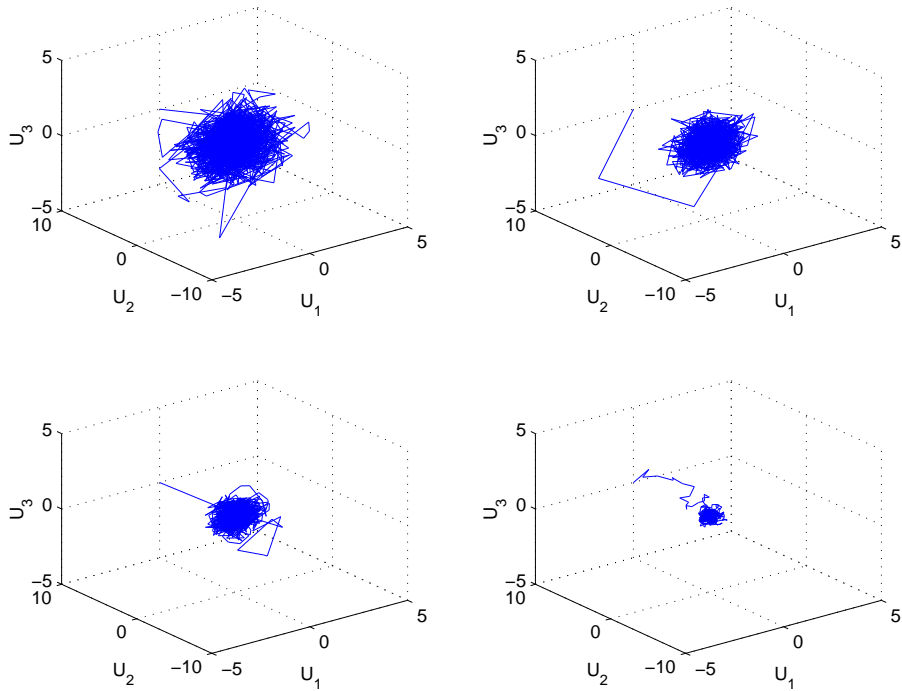}}
  \caption{(Color online) Single quantum trajectory realizations of the scaled
  momentum $\tmmathbf{U} \left( t \right)$, as described in Sect.
  \ref{sec:Usim}. The panels differ in the ratio $m / M$ between the gas and
  particle mass, which decreases from top left to right bottom (top row: $m /
  M = 2$ and $m / M = 1$, bottom row: $m / M = 0.5$ and $m / M = 0.1$). One
  observes that the number of collisions required to get into the thermal
  cloud increases as $m / M$ gets smaller, while the width of the cloud
  decreases. \ \ \ \ \ \ \ \ \label{fig:realizations} }
\end{figure}
\begin{eqnarray}
  \tmmathbf{U}= \frac{\tmmathbf{P}}{Mv_{\beta}} =
  \frac{\tmmathbf{V}}{v_{\beta}}, &  &  \label{eq:U}
\end{eqnarray}
and
\begin{eqnarray}
  \tmmathbf{K}= \frac{\tmmathbf{Q}}{m_{\ast} v_{\beta}} . &  &  \label{eq:K}
\end{eqnarray}
They provide the momentum of the test particle, scaled with respect to the
most probable velocity $v_{\beta}$ in the gas, as well as the scaled momentum
transfer in a single collision.

For the sake of simplicity we also focus on the case of a constant
cross-section. In terms of the new variables the Lindblad operators then read
as
\begin{eqnarray}
  \tilde{L} _B \left( \mathsf{U} ; \tmmathbf{K} \right) & = &
  \sqrt{\frac{\Gamma_{\beta} }{\sqrt{\pi} K}} \exp \left( - \frac{1}{2} \left(
  \frac{K}{2} + \frac{\mathsf{U} \cdot \tmmathbf{K}}{K} \right)^2 \right) . 
  \label{eq:Lku}
\end{eqnarray}
We assume that the initial state is a momentum eigenstate $|\tmmathbf{U}
\rangle$, labeled by the dimensionless momentum (\ref{eq:U}) of the test
particle. The algorithm considered in Sect.~\ref{sec:mc} then leads to a pure
jump process in $\tmmathbf{U}$, whose realizations can be obtained by means of
the standard algorithm used for the stochastic simulation of classical
Markovian master equations {\cite{Gillespie1992}}. This is due to the fact
that the deterministic evolution given by Eq.~(\ref{eq:det}) only affects the
normalization in this case, since both $\mathsf{H}$ and $\mathsf{G}$ are only
functions of the momentum operator.

According to Eq.~(\ref{eq:g}) and Eq.~(\ref{eq:l}) the waiting time
distribution for a jump is characterized by the cumulative distribution
function $F \left( \tau \right) = 1 - \exp \left[ - \tilde{\Gamma} \left( U
\right) \tau \right]$ in our case, where $\tilde{\Gamma} \left( U \right) =
\tilde{M}_{\tmop{out}}^{\tmop{cl}} \left( \tmmathbf{U} \right)$ is the
dimensionless version of the loss rate. According to Eq.~(\ref{eq:macro}) and
Eq.~(\ref{eq:a=0}) it reads
\begin{eqnarray}
  \tilde{\Gamma} \left( U \right) & = & \Gamma_{\beta}  \frac{2}{\sqrt{\pi}} 
  \text{}_1 F_1 \left( - \frac{1}{2}, \frac{3}{2} ; - U^2 \right), 
  \label{eq:x14}
\end{eqnarray}
with $\Gamma_{\beta}$ as in Eq.~(\ref{eq:gamma0}) and $\text{}_1 F_1$ as in
Eq.~(\ref{eq:g5}).

Each jump is characterized by a momentum transfer drawn from the probability
density~(\ref{eq:i}). In the present case, this distribution depends only on
the modulus $K$ of the momentum transfer and on the cosine $\xi =\tmmathbf{U}
\cdot \tmmathbf{K}/ \left( UK \right)$ of the angle between $\tmmathbf{U}$ and
$\tmmathbf{K}$,
\begin{eqnarray}
  \mathcal{P} \left( K, \xi \right) & = & \frac{\Gamma_{\beta}}{2 \sqrt{\pi} 
  \tilde{\Gamma} \left( U \right)} K \exp \left[ - \left( \frac{K}{2} + U \xi
  \right)^2 \right],  \label{eq:kl}
\end{eqnarray}
while components of the momentum transfer perpendicular to $\tmmathbf{U}$ are
uniformly distributed. For each jump one has to draw random numbers $\left( K,
\xi \right)$ according to the probability density~(\ref{eq:kl}) and perform
the shift $\tmmathbf{U} \rightarrow \tmmathbf{U}+\tmmathbf{K}m_{\ast} / M$
with $\tmmathbf{K}$ given by
\begin{eqnarray}
  \tmmathbf{K} & = & K \xi \frac{\tmmathbf{U}}{U} + K \sqrt{1 - \xi^2}
  \frac{\tmmathbf{U} \times \tmmathbf{e}}{\left| \tmmathbf{U} \times
  \tmmathbf{e} \right|},  \label{eq:x15}
\end{eqnarray}
with $\tmmathbf{e}$ a random unit vector.

Single realizations of the process for different values of the mass ratio are
shown in Fig. \ref{fig:realizations} as three-dimensional plots. One observes
that the ratio between the masses of the gas and test particle affects the
number of collisions which are necessary in order to drive the test particle
momentum close to its equilibrium value. At smaller mass ratios $m / M$, an
increasingly large number of collisions is required to thermalize the test
particle. The fluctuations around the equilibrium value, on the other hand,
decrease in this limit.\begin{figure}[tb]
  \resizebox{120mm}{!}{\epsfig{file=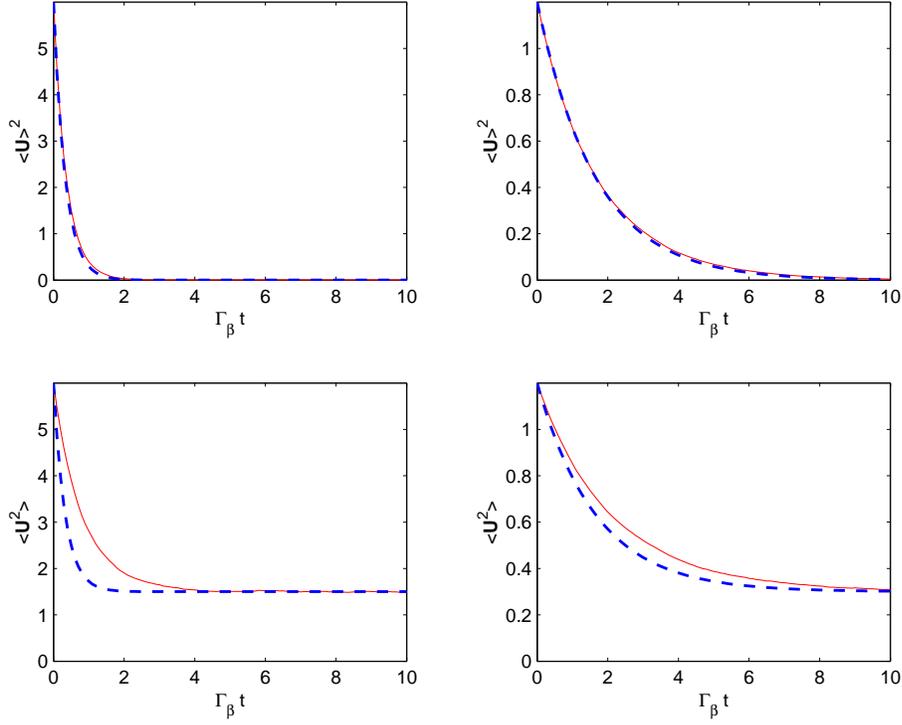}}
  \caption{(Color online) Averages of squared mean momentum (top) and mean
  squared momentum (bottom) for the mass ratios $m / M = 1$ (left) and $m / M
  = 0.2$ (right). The solid line gives the results obtained from the Monte
  Carlo method, while the dashed lines correspond to the analytic predictions
  in Eq. (\ref{eq:ana1}) and Eq. (\ref{eq:ana2}), respectively. As one
  expects, the analytic approximation deteriorates if one increases the mass
  ratio (left panels). In this regime the decay rates for the (squared)
  momentum and the kinetic energy also no longer coincide.
  \label{fig:ensembles} }
\end{figure}

By averaging over a sufficiently large number of realizations one can study
the relaxation of the momentum and of the kinetic energy of the test particle.
According to the scaling introduced in Eq.~(\ref{eq:U}) the expected
equilibrium value $3 / \left( 2 \beta \right)$ of the kinetic energy
corresponds to $\langle \tmmathbf{U}^2 \rangle_{\tmop{eq}} = 3 m / \left( 2 M
\right)$, while the momentum should relax to $\langle \tmmathbf{U}
\rangle_{\tmop{eq}} = 0$. This behavior is confirmed well by the numerical
simulations reported in Fig. \ref{fig:ensembles}, where the squared mean
momentum $\langle \tmmathbf{U} \rangle^2_{\tmop{eq}}$ and the mean squared
momentum $\langle \tmmathbf{U}^2 \rangle_{\tmop{eq}}$ are plotted as a
function of time in units of the inverse of the total scattering rate
$\Gamma_{\beta}$.

The numerical results in Fig. \ref{fig:ensembles} are compared to the
analytic predictions Eq.~(\ref{eq:ap1}) and Eq.~(\ref{eq:ap2}) for the mean
momentum and mean kinetic energy respectively, which are valid for $m / M \ll
1$. In terms of the dimensionless variable $\tmmathbf{U}$ the solutions of
Eq.~(\ref{eq:ap1}) and Eq.~(\ref{eq:ap2}) read as
\begin{eqnarray}
  \langle \tmmathbf{U} \rangle^2_{\rho_t} & = & \langle \tmmathbf{U}
  \rangle^2_{\rho_0} \mathe^{- 2 \eta t}  \label{eq:ana1}
\end{eqnarray}
and
\begin{eqnarray}
  \langle \tmmathbf{U}^2 \rangle_{\rho_t} & = & \langle \tmmathbf{U}^2
  \rangle_{\tmop{eq}} + \left[ \langle \tmmathbf{U}^2 \rangle_{\rho_0} -
  \langle \tmmathbf{U}^2 \rangle_{\tmop{eq}} \right] \mathe^{- 2 \eta t}, 
  \label{eq:ana2}
\end{eqnarray}
with $\eta$ given by Eq.~(\ref{eq:friction}). As one observes from Fig.
\ref{fig:ensembles} the analytic approximation is indeed very good for small
mass ratios $m / M \ll 1$, while deviations appear for higher values of the
ratio. These changes show up in particular through the appearance of different
decay rates for the momentum and the kinetic energy. For large $m / M$ one can
also observe strong deviations from Gaussian statistics, e.g., by studying the
behavior of cumulants of order higher than two {\cite{Breuer2007c}}.

\section{The limit of quantum Brownian motion}\label{sec:qbm}

One of the most important limits of the quantum linear Boltzmann equation is
the situation where the test particle has a much greater mass than the gas
particles. In the classical case, this limit turns the linear Boltzmann
equation into the Kramers equation for the description of Brownian motion. As
discussed in the following, the situation is quite similar in the quantum
case, where one obtains a quantum Fokker-Planck equation describing quantum
Brownian motion.

The term {\tmem{quantum Brownian motion}} is generally used for master
equations intended to phenomenologically describe the quantum analog of
classical Brownian motion
{\cite{Grabert1988a,Ingold2002a,Hanggi2005a,Weiss2008}}, without accounting
for the microscopic details of the underlying dynamics. In contrast to such
models, which usually assume a linear coupling to a harmonic bath, we focus in
the following on the specific situation of a massive test particle suffering
collisions with a surrounding Maxwell-Boltzmann gas. The microscopic
expressions of the friction and diffusion coefficients are worked out as a
function of the cross-section, and we put into evidence the new quantum terms
appearing in the master equation corresponding to the Kramers equation.

\subsection{The classical Kramers equation}

In order to discuss the Brownian motion limit of the quantum linear Boltzmann
equation it is useful to first recall the relationship between the classical
linear Boltzmann equation and Kramers' Fokker-Planck equation. It was Rayleigh
who first investigated the limiting case of a very heavy test particle in the
collision integral of the linear Boltzmann equation
{\cite{Rayleigh1891a,Green1951a}}. The resulting equation in this so-called
Rayleigh limit, or Brownian limit, is the well-known Kramers equation. For the
homogeneous case and in the absence of an external force it reads as
\begin{eqnarray}
  \text{$\partial_t f$} (\tmmathbf{P}) & = & \eta \sum^3_{i = 1} \left[
  \frac{\partial}{\partial P_i} \left( P_i f \left( \tmmathbf{P} \right)
  \right) + \frac{M}{\beta} \frac{\partial^2}{\partial P^2_i} f \left(
  \tmmathbf{P} \right) \right] .  \label{eq:fp}
\end{eqnarray}
The friction coefficient $\eta$ is given by
{\cite{Uhlenbeck1970a,Ferrari1987a}}
\begin{eqnarray}
  \eta & = & \frac{16}{3}  \sqrt{\pi}  \frac{m}{M} n_{\tmop{gas}}
  \sqrt{\frac{2}{\beta m}}  \int^{\infty}_0 \mathd u \, u^5 \mathe^{- u^2}
  \int^{\pi}_0 \mathd \bignone \vartheta \sin \vartheta \bignone  \left( 1 -
  \cos \vartheta \right) \sigma (\vartheta ; up_{\beta}),  \label{eq:etac}
\end{eqnarray}
with $\sigma (\vartheta ; up_{\beta})$ the differential scattering
cross-section as a function of the scattering angle and of the momentum of the
gas particle $p = up_{\beta}$.

The derivation of the microscopic expression of the friction coefficient is
not easy to find in the literature. We therefore briefly recall the derivation
of (\ref{eq:fp}) and (\ref{eq:etac}) in Sect.~\ref{sec:a3} for the reader's
convenience and for the sake of comparison with the quantum calculation. The
crucial assumption required to derive the Fokker-Planck Eq.~(\ref{eq:fp}) from
the classical linear Boltzmann equation~(\ref{eq:clbeMB}) is that the test
particle is much heavier than the gas particles, $m / M \ll 1$, and that its
momentum is close to the typical thermal value $P_{\beta} = \sqrt{2 M /
\beta}$. The collisions with the gas particles are then characterized by small
momentum and energy transfers. The typical value of the momentum transfer $Q$
is of the order of the momentum of the colliding gas particles
$\text{$p_{\beta} = \sqrt{2 m / \beta}$}$, and therefore much smaller than the
momentum $P_{\beta}$ of the test particle, while the energy transfer is on the
order of $E \left( \tmmathbf{Q}, \tmmathbf{P} \right) \approx 1 / \beta
\sqrt{m / M}$, which is smaller than the thermal energy $1 / \beta$. As
discussed in Sect.~\ref{sec:a3}, under these conditions one can expand $f
\left( \tmmathbf{P}' \right)$ in Eq.~(\ref{eq:clbeMB}) around $\tmmathbf{P}$
for small momentum transfers $\tmmathbf{Q}=\tmmathbf{P}' -\tmmathbf{P}$, and
arrive at Eq.~(\ref{eq:fp}) by sorting the orders in an $m / M$ expansion.

\subsection{The quantum Brownian limit }\label{sec:blim}

\subsubsection{Conditions of validity}

The situation is more complicated in the quantum case, since one is dealing
with operators whose values can be estimated in a meaningful way only by
considering suitable matrix elements. We will show that the quantum master
equation which corresponds to the classical Fokker-Planck equation is obtained
by expanding the scattering kernel up to second order in the operators
$\mathsf{X}$ and $\mathsf{P}$. It is important to stress that the result is
{\tmem{not}} equivalent to a direct application of the correspondence
principle on the classical result, which would lead to a master equation which
does not generally preserve the positivity of the statistical operator
{\cite{Ambegaokar1991a,Pechukas1991a}}.

The expansion can be performed provided the test particle is very massive and
not far from thermal equilibrium, that is to say, close to diagonal in
momentum representation. More precisely, the off-diagonal elements $\langle
\tmmathbf{P}| \rho |\tmmathbf{P}' \rangle$ may differ significantly from zero
only for $\Delta P \assign \left| \tmmathbf{P}-\tmmathbf{P}' \right| \lesssim
\sqrt{M / \beta}$. In the position representation, the validity of the
expansion requires that the motional state exhibits coherences only over a
length scale of the order of the thermal wavelength of the test particle, such
that $\langle \tmmathbf{X}| \rho |\tmmathbf{X}' \rangle$ is appreciably
different from zero only within a range given by the thermal de Broglie
wavelength $\lambda_{\tmop{th}} = \sqrt{2 \pi \hbar^2 \beta / M}$, i.e., for
$\Delta X \assign \left| \tmmathbf{X}-\tmmathbf{X}' \right| \lesssim
\lambda_{\tmop{th}}$. The small mass ratio $m / M \ll 1$ implies in particular
that when the test particle gets close to thermal equilibrium its coherence
length of the order of its thermal de Broglie wavelength is much smaller than
the coherence length of the environment given by the thermal wavelength of the
gas.

\subsubsection{Diffusive limit of the quantum linear Boltzmann equation }

In order to study the Brownian motion limit of the quantum linear Boltzmann
equation, we start with the incoherent part given by Eq.~(\ref{eq:qlbe}).
Using the explicit expression of the Lindblad operators Eq.~(\ref{eq:L}) with
the distribution function (\ref{eq:muMB}) of a Maxwell-Boltzmann gas, one
obtains the exact expression
\begin{eqnarray}
  \mathcal{L} \rho & = & n_{\tmop{gas}}  \frac{m}{m^2_{\ast}} \bigintlim
  \frac{\mathd \tmmathbf{Q}}{Q} \int_{\tmmathbf{Q}^{\bot}} \mathd
  \tmmathbf{k}_{\bot} \, \mu_{\beta} \left( \tmmathbf{k}_{\bot} \text{$+
  \frac{m}{m_{\ast}} \frac{\tmmathbf{Q}}{2}$} \right)  \left[ f \left(
  \tmop{rel} \left( \tmmathbf{k}_{\bot}, \mathsf{P}_{\bot \tmmathbf{Q}}
  \right) - \frac{\tmmathbf{Q}}{2}, \tmop{rel} \left( \tmmathbf{k}_{\bot},
  \mathsf{P}_{\bot \tmmathbf{Q}} \right) + \frac{\tmmathbf{Q}}{2} \right)
  \right. \nonumber\\
  &  & \times \mathe^{i\tmmathbf{Q} \cdot \mathsf{X} / \hbar} \exp \left( -
  \beta \frac{m \mathsf{P}^2_{\|\tmmathbf{Q}}}{4 M^2} \mathsf{} - \beta
  \frac{m\tmmathbf{Q} \cdot \mathsf{P}}{4 Mm_{\ast}} \right) {\color{black} }
  \rho \exp \left( - \beta \frac{m \mathsf{P}^2_{\|\tmmathbf{Q}}}{4 M^2}
  \mathsf{} - \beta \frac{m\tmmathbf{Q} \cdot \mathsf{P}}{4 Mm_{\ast}} \right)
  \mathe^{- i\tmmathbf{Q} \cdot \mathsf{X} / \hbar} \nonumber\\
  &  & \times f^{\dag} \left( \tmop{rel} \left( \tmmathbf{k}_{\bot},
  \mathsf{P}_{\bot \tmmathbf{Q}} \right) - \frac{\tmmathbf{Q}}{2}, \tmop{rel}
  \left( \tmmathbf{k}_{\bot}, \mathsf{P}_{\bot \tmmathbf{Q}} \right) +
  \frac{\tmmathbf{Q}}{2} \right) \nonumber\\
  &  & \left. - \frac{1}{2} \left\{ \exp \left( - \beta \frac{m
  \mathsf{P}^2_{\|\tmmathbf{Q}}}{4 M^2} \mathsf{} - \beta \frac{m\tmmathbf{Q}
  \cdot \mathsf{P}}{4 Mm_{\ast}} \right) \left| f \left( \tmop{rel} \left(
  \tmmathbf{k}_{\bot}, \mathsf{P}_{\bot \tmmathbf{Q}} \right) -
  \frac{\tmmathbf{Q}}{2}, \tmop{rel} \left( \tmmathbf{k}_{\bot},
  \mathsf{P}_{\bot \tmmathbf{Q}} \right) + \frac{\tmmathbf{Q}}{2} \right)
  \right|^2, \rho \right\} \right] .  \label{eq:a1}
\end{eqnarray}
In the expansion up to second order in $\mathsf{P}$ the contributions from the
terms $\exp \left( - \beta m \mathsf{P^2_{\|\tmmathbf{Q}}} / \left( 4 M^2
\right) \right)$ cancel out, such that they can be replaced by unity. We
further expand the various quantities under the assumption that the change in
momentum of the Brownian particle is small compared to the scales involved in
its motional state. More specifically, one has to assume that for typical
values of the momentum transfer $Q$ the relevant matrix elements of the
statistical operator vanish unless
\begin{eqnarray}
  \frac{Q}{\hbar} \Delta X & \ll & 1 
\end{eqnarray}
and
\begin{eqnarray}
  \frac{\beta Q}{M} \Delta P & \ll & 1. 
\end{eqnarray}
These conditions are both satisfied if the Brownian state is close to thermal
and $M \gg m$, such that $m_{\ast} \approx m$, since the momentum transfer $Q$
is then typically of the order of the momentum $p_{\beta} = \sqrt{2 m /
\beta}$ of the colliding gas particles. This implies in particular that
\begin{eqnarray}
  \frac{Q}{\hbar} \approx \sqrt{\frac{m}{2 \pi \hbar^2 \beta}} =
  \frac{1}{\lambda_{\tmop{th}}^{\tmop{gas}}} \ll \frac{1}{\lambda_{\tmop{th}}}
  \lesssim \frac{1}{\Delta X} &  & 
\end{eqnarray}
and
\begin{eqnarray}
  \frac{\beta Q}{M} \approx \sqrt{\frac{m}{M}} \sqrt{\frac{\beta}{M}} \ll
  \sqrt{\frac{\beta}{M}} \lesssim \frac{1}{\Delta P} . &  & 
\end{eqnarray}
Apart from the exponentials in Eq.~(\ref{eq:a1}) one also has to expand the
scattering amplitude, which appears as an operator-valued expression. However,
for the assumed small mass ratio the argument of the scattering amplitude
becomes
\begin{eqnarray}
  \tmop{rel} \left( \tmmathbf{k}_{\bot}, \tmmathbf{P}_{\bot \tmmathbf{Q}}
  \right) - \frac{\tmmathbf{Q}}{2} & \approx & \tmmathbf{k}_{\bot} -
  \frac{m}{M} \tmmathbf{P}_{\bot \tmmathbf{Q}} - \frac{\tmmathbf{Q}}{2}, 
  \label{eq:x16}
\end{eqnarray}
such that the leading term in an expansion in $m / M$ does not depend any
longer on $\tmmathbf{P}_{\bot \tmmathbf{Q}}$. This removes an operator
dependence arising from the scattering amplitude. By further expanding the
exponentials involving $\tmmathbf{Q} \cdot \mathsf{P}$ in Eq.~(\ref{eq:a1})
one is then left with
\begin{eqnarray}
  \mathcal{L} \rho & = & - \frac{n_{\tmop{gas}}}{m} \frac{1}{2} \bigintlim
  \frac{\mathd \tmmathbf{Q}}{Q} \sum^3_{j, k = 1} Q_j Q_k
  \int_{\tmmathbf{Q}^{\bot}} \mathd \tmmathbf{k}_{\bot} \, \mu_{\beta} \left(
  \tmmathbf{k}_{\bot} \text{$+ \frac{\tmmathbf{Q}}{2}$} \right)  \left| f
  \left( \tmmathbf{k}_{\bot} - \frac{\tmmathbf{Q}}{2}, \tmmathbf{k}_{\bot} +
  \frac{\tmmathbf{Q}}{2} \right) \right|^2  \label{eq:e1}\\
  &  & \times \left[ \frac{1}{\hbar^2} \left[ \mathsf{X}_j, \left[
  \mathsf{X}_k, \rho \right] \right] + \left( \frac{\beta}{4 M} \right)^2
  \left[ \mathsf{P}_j, \left[ \mathsf{P}_k, \rho \right] \right] +
  \frac{i}{\hbar} \frac{\beta}{2 M} \left[ \mathsf{X}_j, \left\{ \mathsf{P}_k,
  \rho \right\} \right] \right] . \nonumber
\end{eqnarray}
We here omitted the term linear in the momentum transfer, since it vanishes
upon integration due to the isotropy of the integrand. By exploiting also the
invariance of the scattering cross-section under parity transformations the
quantity $Q_j Q_k$ can be replaced by $\delta_{jk} Q^2 / 3$. We have thus put
into evidence the operator structure of the limiting expression of the quantum
linear Boltzmann equation for the case of quantum Brownian motion,
\begin{eqnarray}
  \mathcal{L} \rho & = & - \eta \sum^3_{j = 1} \left[ \frac{M}{\hbar^2 \beta}
  \left[ \mathsf{X}_j, \left[ \mathsf{X}_j, \rho \right] \right] +
  \frac{\beta}{16 M} \left[ \mathsf{P}_j, \left[ \mathsf{P}_j, \rho \right]
  \right] + \frac{i}{2 \hbar} \left[ \mathsf{X}_j, \left\{ \mathsf{P}_j, \rho
  \right\} \right] \right] .  \label{eq:510}
\end{eqnarray}
This equation is still covariant under translations
{\cite{Vacchini2001b,Petruccione2005a,Vacchini-xxx}}, since the limiting
procedure does not spoil the symmetry properties.

\subsubsection{Microscopic expression for the friction constant}

We now evaluate the friction coefficient $\eta$ in Eq.~(\ref{eq:510}), which
is defined by
\begin{eqnarray}
  \eta & = & \frac{\beta}{6} \frac{n_{\tmop{gas}}}{Mm} \bigintlim \mathd
  \tmmathbf{Q} \, Q \int_{\tmmathbf{Q}^{\bot}} \mathd \tmmathbf{k}_{\bot} \,
  \mu_{\beta} \left( \tmmathbf{k}_{\bot} \text{$+ \frac{\tmmathbf{Q}}{2}$}
  \right)  \left| f \left( \tmmathbf{k}_{\bot} - \frac{\tmmathbf{Q}}{2},
  \tmmathbf{k}_{\bot} + \frac{\tmmathbf{Q}}{2} \right) \right|^2 . 
  \label{eq:e2}
\end{eqnarray}
The exact expression of $\eta$ will turn out to be the same as the classical
friction coefficient (\ref{eq:etac}). Comparing the procedure followed here
with the calculations in Appendix~\ref{sec:a3} it turns out that the
microscopic derivation of the friction coefficient is more straightforward in
the quantum case. Upon using $\delta \left( \tmmathbf{p} \cdot \tmmathbf{Q}
\right) = \delta \left( p_{\|} \right) / Q$ and introducing the new variables
$\tmmathbf{p}_i =\tmmathbf{p} \text{$+ \tmmathbf{Q}/ 2$}$ and $\tmmathbf{p}_f
=\tmmathbf{p}-\tmmathbf{Q}/ 2$ the expression for $\eta$ becomes
\begin{eqnarray}
  \eta & = & \frac{\beta}{6} \frac{n_{\tmop{gas}}}{Mm} \int \mathd
  \tmmathbf{p}_i \int \mathd \tmmathbf{p}_f \, \delta \left( \frac{p^2_i -
  p^2_f}{2} \right) \left| \tmmathbf{p}_i -\tmmathbf{p}_f \right|^2
  \mu_{\beta} \left( \tmmathbf{p}_i \right) \left| f \left( \tmmathbf{p}_f,
  \tmmathbf{p}_i \right) \right|^2 .  \label{eq:x17}
\end{eqnarray}
As a final step, we introduce the angle $\vartheta$ between $\tmmathbf{p}_i$
and $\tmmathbf{p}_f$ as an integration variable, and recall that the
scattering cross-section is assumed to depend only on $p_i$ and $\vartheta$,
thus finally obtaining for a Maxwell-Boltzmann gas distribution
\begin{eqnarray}
  \eta & = & \frac{16}{3}  \sqrt{\pi}  \frac{m}{M} n_{\tmop{gas}} 
  \sqrt{\frac{2}{m \beta}} \int^{\infty}_0 \mathd u \, u^5 \mathe^{- u^2}
  \int^{\pi}_0 \mathd \bignone \vartheta \sin \vartheta \bignone  \left( 1 -
  \cos \vartheta \right) \left| f \left( \vartheta ; up_{\beta} \right)
  \right|^2,  \label{eq:e3}
\end{eqnarray}
with $u$ as in (\ref{eq:etac}). This expression coincides with the classical
result {\cite{Uhlenbeck1970a}}, recalled in Eq.~(\ref{eq:etac}). For the case
of a constant scattering cross-section it reads as
\begin{eqnarray}
  \eta & = & \frac{8}{3 \sqrt{\pi}}  \frac{m}{M} \Gamma_{\beta}, 
\end{eqnarray}
which also agrees with the result (\ref{eq:friction}) obtained in
Sect.~\ref{sec:emrelax}.

\subsection{The master equation of quantum Brownian motion}

Putting the master equation for quantum Brownian motion Eq.~(\ref{eq:510}) in
the standard form, it finally reads
\begin{eqnarray}
  \mathcal{L} \rho & = & \frac{1}{i \hbar} \frac{\eta}{2} \sum^3_{j = 1}
  \left[ \mathsf{X}_j, \left\{ \mathsf{P}_j, \rho \right\} \right] -
  \frac{D_{pp}}{\hbar^2} \sum^3_{j = 1} \left[ \mathsf{X}_j, \left[
  \mathsf{X}_j, \rho \right] \right] - \frac{D_{xx}}{\hbar^2} \sum^3_{j = 1}
  \left[ \mathsf{P}_j, \left[ \mathsf{P}_j, \rho \right] \right] . 
  \label{eq:me}
\end{eqnarray}
The friction coefficient $\eta$ is given by Eq.~(\ref{eq:e3}), while $D_{pp}$
has the meaning of a diffusion coefficient,
\begin{eqnarray}
  D_{pp} & = & \eta \frac{M}{\beta},  \label{eq:dpp}
\end{eqnarray}
and $D_{xx}$ is related to $D_{pp}$ and to $\eta$ by
\begin{eqnarray}
  D_{xx} & = & \left( \frac{\beta \hbar}{4 M} \right)^2 D_{pp} = \frac{\beta
  \hbar^2}{16 M} \eta .  \label{eq:dxx}
\end{eqnarray}
This equation admits an exact solution, though the explicit expression is a
bit cumbersome {\cite{Bassi2005b}}. We shall consider it in
Sect.~\ref{sec:decohqbm}, when studying how this master equation describes
decoherence effects.

We note that the Caldeira-Leggett master equation, describing the
high-temperature regime of the Caldeira-Leggett model
{\cite{Caldeira1983a,Caldeira1983b,Leggett1987a,Weiss2008}}, is related to
Eq.~(\ref{eq:me}) by neglecting the third term in (\ref{eq:me}), i.e., by
setting $D_{xx} = 0$. This equation, which cannot be brought into Lindblad
form, is also obtained from the classical Kramers equation if one takes the
naive operator correspondence $\partial / \partial P_j \rightarrow
\mathsf{X}_j / \left( i \hbar \right)$.

\subsubsection{The diffusion coefficients and complete positivity}

The diffusion coefficient $D_{pp}$ can also be related to the second moment of
the thermal momentum $\langle P^2 \rangle_{\beta} = 3 M / \beta$ according to
\begin{eqnarray}
  D_{pp} & = & \frac{\eta}{3} \langle P^2 \rangle_{\beta},  \label{eq:pth}
\end{eqnarray}
while $D_{xx}$, which is sometimes denoted as ``quantum diffusion
coefficient'' or ``momentum localization coefficient'', can be written by
means of the thermal de Broglie wavelength $\lambda_{\tmop{th}} = \sqrt{2 \pi
\hbar^2 \beta / M}$ of the test particle as
\begin{eqnarray}
  D_{xx} & = & \frac{\eta^{}}{32 \pi} \lambda_{\tmop{th}}^2 .  \label{eq:xth}
\end{eqnarray}
As discussed in
{\cite{Barchielli1983b,Sandulescu1987a,Isar1999a,Vacchini2002b,Isar1994a,Breuer2007}}
a necessary and sufficient condition for master equations of the
form~(\ref{eq:me}) to yield a completely positive dynamics is
\begin{eqnarray}
  D_{xx} D_{pp} & \geqslant & \frac{\eta^2 \hbar^2}{16},  \label{eq:cons}
\end{eqnarray}
since only then they can be cast into Lindblad form. This is an important
statement, because in many approaches the coefficients $\eta$, $D_{xx}$, and
$D_{pp}$ are treated as free, phenomenological parameters. In contrast to
that, the derivation from the quantum linear Boltzmann equation yielded
microscopically formulated definitions and relations of the coefficients
(\ref{eq:me})-(\ref{eq:dxx}). It is therefore reassuring that $D_{xx}$ and
$D_{pp}$ from Eq.~(\ref{eq:dxx}) and Eq.~(\ref{eq:dpp}) satisfy
Eq.~(\ref{eq:cons}) with an equality sign.

For a given choice of $\eta$ and $D_{pp} = \eta M / \beta$ the coefficient
$D_{xx}$ obtained in Eq.~(\ref{eq:dxx}) therefore takes the minimal value
allowed within the framework of a Lindblad master equation. This implies that
the equation of quantum Brownian motion obtained from the quantum linear
Boltzmann equation is as close to the classical Kramers equation as it can
possibly be, within the bounds set by the requirement of complete positivity.

\subsubsection{The quantum diffusion term}

It was discussed by several authors
{\cite{Barchielli1983b,Sandulescu1987a,Isar1999a,Vacchini2002b,Jacobs2009a,Breuer2007}}
that the lack of complete positivity of the Caldeira Leggett master equation
can be corrected by adding a ``position diffusion term'' of the form $\left[
\mathsf{P}_j, \left[ \mathsf{P}_j, \rho \right] \right]$. In view of the fact
that the Brownian limit of the quantum linear Boltzmann equation yields the
minimal correction, it is worth noting that this is not achieved in other
microscopic approaches.

Di\`osi's proposal {\cite{Diosi1995a}}, for instance, leads to an expression
of the coefficient $D_{xx}$, whose relation to $D_{pp}$ is not given
exclusively by thermodynamic quantities. Rather, it involves a complicated
average over the scattering cross-section. More refined derivations within the
Caldeira Leggett model, on the other hand, which seek for correction terms
beyond the high temperature approximation {\cite{Diosi1993a,Diosi1993b}}, lead
to a coefficient $D_{xx}$ which exceeds the present, minimal value by a factor
of 4/3.

As already mentioned, the term $\sum^3_{j = 1} \left[ \mathsf{P}_j, \left[
\mathsf{P}_j, \rho \right] \right]$ in Eq.~(\ref{eq:me}) has no counterpart in
the classical Fokker-Planck equation~(\ref{eq:fp}). It describes a position
diffusion process, which is expected to appear together with momentum
diffusion, as implied by the inequality Eq.~(\ref{eq:cons}). That this is a
pure quantum phenomenon is confirmed by the fact that the coefficient $D_{xx}$
depends explicitly on $\hbar$; it vanishes in the semiclassical limit $\hbar
\rightarrow 0$.

Even though the occurrence of this quantum effect leading to momentum
localization was predicted in many theoretical papers, it is still unclear how
to observe the effect experimentally. The main difficulty lies in the fact
that the position localization term, with its direct classical analog,
dominates under typical experimental conditions, due to its strong temperature
dependence, while the quantum diffusion term vanishes in the limit of high
temperatures. Effects of the latter might show up by either inducing
decoherence with respect to the momentum basis, or by looking for corrections
to the classical diffusion coefficient. The last possibility was considered
for the strong friction limit of the quantum master equation
{\cite{Vacchini2002a}}. Compared to the classical case, the Einstein diffusion
coefficient appears multiplied by a factor $1 + \left( \eta \beta \hbar
\right)^2 / 16$, involving the ratio between two characteristic times $\beta
\hbar$ for the bath of gas particles and $1 / \eta$ for the system. The
separation of time scales typical for a Markov description of the dynamics
implies that this correction should be extremely small under typical
conditions.

\section{Matter wave optics in gaseous samples}\label{sec:optics}

So far, we mostly dealt with features of the quantum linear Boltzmann equation
that are also shared by its classical counterpart. In particular, its symmetry
and relaxation properties, as well as the existence of the stationary
solution, are already described by the behavior of the momentum populations,
i.e., of the diagonal matrix elements of Eq.~(\ref{eq:poffdiag}). In the
following two sections, we will now consider properties of the quantum linear
Boltzmann equation that become manifest in interference phenomena, where it is
decisive that the motion of the test particle may be in a coherently
delocalized, wavelike state.

An important quantum effect that one expects to be described by
Eq.~(\ref{eq:qlbe}) is the dynamical transition of an initially delocalized
superposition state into an incoherent, particlelike mixture. Such decoherence
phenomena are described in Sect.~\ref{sec:decoh}. But before that, we focus on
the dynamics described by the coherent part of the quantum linear Boltzmann
equation and its loss term. They play an important role in experiments
involving spatially separate beams.

The claim that massive particles can exhibit wave-like behavior is one of the
central predictions of quantum mechanics; it is confirmed experimentally with
neutrons {\cite{Sears1989a,Werner2000a}}, atoms
{\cite{Adams1994a,Miffre2006a,Cronin2009a}}, and even complex molecules
{\cite{Arndt2009a}}. In such experiments, the de Broglie wave length
$\lambda_{\tmop{dB}} = h / \left( MV \right)$ of the matter waves is usually
much shorter than the scale of change of the potential produced by the
gratings and collimation slits, which act as optical elements. It is then
often permissible to neglect diffraction phenomena by describing the matter
wave in terms of well defined beams, which are coherently split and rejoined
by means of interference gratings. The effect of an additional external
potential on those beams can then be accounted for, in complete analogy to the
optical case, by introducing an index of refraction.

In a number of Mach-Zehnder-type interference experiments a truly macroscopic
separation of the interfering beams was achieved. This makes it possible to
perform experiments where the interfering particles interact with an external
system in just one arm of the interferometer. The modification of the
interference fringe pattern then provides sensitive information of the
strength of interaction. Its theoretical description requires the use of
quantum kinetic equations if the interaction is due to a dilute background
gas, as realized in experiments with sodium and lithium atoms
{\cite{Schmiedmayer1995a,Vigue1995a,Jacquey2007a}}, or due to a homogeneous
sample of thermalized condensed matter, as in the experiments performed with
the perfect crystal neutron interferometer {\cite{Rauch1974a,Werner2000a}}.

\subsection{The gas induced energy shift}

The presence of a background gas manifests itself not only in the incoherent
part $\mathcal{L}$ of the quantum linear Boltzmann equation, but also in a
modified Hamiltonian $\mathsf{H} = \mathsf{P}^2 / \left( 2 M \right) +
H_{\text{n}} \left( \mathsf{P} \right)$ appearing in its coherent part. The
additional term $H_{\text{n}} \left( \mathsf{P} \right)$, which has been
neglected so far, describes the modification of the free matter wave
dispersion relation due to the coherent interaction with the background gas.
It becomes observable in the abovementioned Mach-Zehnder interferometers as a
gas-induced phase shift.

According to Eq.~(\ref{eq:effectivebis}), derived in {\cite{Hornberger2008a}},
the energy shift $H_{\text{n}} \left( \mathsf{P} \right)$ is related to a
``forward scattering process'' which does not change the momenta of the gas
and the test particle. The forward scattering amplitude $f_0 \left(
\tmmathbf{p} \right) \equiv f \left( \tmmathbf{p}, \tmmathbf{p} \right)$
appears in a thermal average over the momentum distribution of the gas,
\begin{eqnarray}
  \text{$\langle f_0 \left( \tmmathbf{P} \right) \rangle _{\mu}$} & = & \int
  \bignone \mathd \tmmathbf{p} \, \mu \left( \tmmathbf{p} \right) f \left(
  \tmop{rel} \left( \tmmathbf{p}, \tmmathbf{P} \right), \tmop{rel} \left(
  \tmmathbf{p}, \tmmathbf{P} \right) \right),  \label{eq:fbeta}
\end{eqnarray}
allowing us to write more compactly
\begin{eqnarray}
  H_{\text{n}} \left( \mathsf{P} \right) & = & - 2 \pi \hbar^2
  \frac{n_{\tmop{gas}}}{m_{\ast}} \tmop{Re} \text{$\langle f_0 \left(
  \mathsf{P} \right) \rangle _{\mu}$} .  \label{eq:for}
\end{eqnarray}
To observe the effect of Eq.~(\ref{eq:for}) an experimental situation is
required where the interfering matter wave beams are strongly collimated and
directed into narrow apertures. In this case any proper, momentum changing
collision due to a dilute background gas will unavoidably lead to a loss of
the particle. This implies that all particles arriving in the detector
experienced at most the effect of ``forward scattering''. It can be observed
as a phase shift in the interference pattern, as done in the experiments
{\cite{Schmiedmayer1995a,Vigue1995a,Jacquey2007a}}.

If the particles do not get lost, their motional state is modified at most
coherently in the described experimental situation. The purity of the matter
waves impinging at the detector therefore remains unchanged by the presence of
the background gas, rendering a description by means of a wave equation
possible. The effect of particle loss can be accounted for by an attenuation
of the amplitude, as argued below in Sect.~\ref{sec:iof}. Incidentally, the
loss term introduced in Eq.~(\ref{eq:pdiag}) and discussed in
Sec.~\ref{sec:loss}, can be related to the {\tmem{imaginary}} part of
Eq.~(\ref{eq:fbeta}). This fact naturally leads to the introduction of an
optical potential.

\subsection{The optical potential}

It is helpful to express the damping of the wave amplitude due to collisional
losses in term of the forward scattering amplitude. Let us start from
Eq.~(\ref{eq:clbeMB}) for the loss term in the classical linear Boltzmann
equation. Switching to center-of-mass variables with respect to
$\tmmathbf{p}'$ and $\tmmathbf{P}'$, denoting $\tmmathbf{k}= \tmop{rel} \left(
\tmmathbf{p}', \tmmathbf{P}' \right)$ and $\tmmathbf{K}=\tmmathbf{p}'
+\tmmathbf{P}'$ one obtains
\begin{eqnarray}
  M^{\tmop{cl}}_{\tmop{out}} \left( \mathsf{\tmmathbf{P}} \right) & = &
  \frac{n_{\tmop{gas}}}{m_{\ast}} \int \mathd k \, k \int \bignone \mathd
  \tmmathbf{p} \, \mu \left( \tmmathbf{p} \right) \delta \left( k - |
  \tmop{rel} \left( \tmmathbf{p}, \tmmathbf{P} \right) | \right) \int \mathd
  \Omega_{\tmmathbf{k}} \, \sigma \left( \tmmathbf{k}, \tmop{rel} \left(
  \tmmathbf{p}, \tmmathbf{P} \right) \right) . 
\end{eqnarray}
We also introduce the total scattering cross-section
\begin{eqnarray}
  \sigma_{\tmop{tot}} \left( \tmmathbf{p} \right) & = & \int \mathd
  \Omega_{\tmmathbf{n}} \, \sigma \left( p\tmmathbf{n}, \tmmathbf{p} \right), 
  \label{eq:totcross}
\end{eqnarray}
where $\tmmathbf{n}$ is a unit vector with $\mathd \Omega_{\tmmathbf{n}}$ the
associated solid angle element. By exploiting the optical theorem
{\cite{Taylor1972a}},
\begin{eqnarray}
  \sigma_{\tmop{tot}} \left( \tmmathbf{p} \right) \text{} & = & \frac{4 \pi
  \hbar}{p} \tmop{Im} \left[ f \left( \tmmathbf{p}, \tmmathbf{p} \right)
  \right], 
\end{eqnarray}
and by using the thermal average defined in Eq.~(\ref{eq:fbeta}), one finds
that the loss term takes the simple form
\begin{eqnarray}
  M^{\tmop{cl}}_{\tmop{out}} \left( \mathsf{\tmmathbf{P}} \right) & = &
  \frac{n_{\tmop{gas}}}{m_{\ast}} 4 \pi \hbar \tmop{Im} \text{$\langle f_0
  \left( \tmmathbf{P} \right) \rangle _{\mu}$} .  \label{eq:imf}
\end{eqnarray}
Recalling Eq.~(\ref{eq:poffdiag}) and Eq.~(\ref{eq:for}) the quantum linear
Boltzmann equation can therefore be written as

\begin{eqnarray}
  \frac{\mathd}{\mathd t} \rho & = & - \frac{i}{\hbar} \left[ \mathsf{H}_0 - 2
  \pi \hbar^2 \frac{n_{\tmop{gas}}}{m_{\ast}} \tmop{Re} \text{$\langle f_0
  \left( \mathsf{P} \right) \rangle _{\mu}$}, \rho \right] - \frac{1}{2}
  \left\{ \frac{n_{\tmop{gas}}}{m_{\ast}} 4 \pi \hbar \tmop{Im} \text{$\langle
  f_0 \left( \mathsf{P} \right) \rangle _{\mu}$}, \rho \right\} 
  \label{eq:preottica}\\
  &  & + \bigintlim \mathd \tmmathbf{Q} \int_{\tmmathbf{Q}^{\bot}} \mathd
  \tmmathbf{k}_{\bot} \, \mathe^{i\tmmathbf{Q} \cdot \mathsf{X} / \hbar} L
  \left( \tmmathbf{k}_{\bot}, \mathsf{P} ; \tmmathbf{Q} \right) \rho L^{\dag}
  \left( \tmmathbf{k}_{\bot}, \mathsf{P} ; \tmmathbf{Q} \right) \mathe^{-
  i\tmmathbf{Q} \cdot \mathsf{X} / \hbar} . \nonumber
\end{eqnarray}
This suggests to introduce a non-hermitian operator, often denoted as optical
potential, of the form
\begin{eqnarray}
  V_{\tmop{opt}} \left( \mathsf{P} \right) & = & - 2 \pi \hbar^2
  \frac{n_{\tmop{gas}}}{m_{\ast}} \langle f_0 \left( \mathsf{P} \right)
  \rangle_{\mu} .  \label{eq:opt}
\end{eqnarray}
One thus arrives at a compact and suggestive expression for the quantum linear
Boltzmann equation
\begin{eqnarray}
  \frac{\mathd}{\mathd t} \rho & = & \frac{1}{i \hbar} \left[ \mathsf{H}_0,
  \rho \right] + \frac{1}{i \hbar} \left( V_{\tmop{opt}} \left( \mathsf{P}
  \right) \rho - \rho V_{\tmop{opt}}^{\dag} \left( \mathsf{P} \right) \right)
  \nonumber\\
  &  & + \bigintlim \mathd \tmmathbf{Q} \int_{\tmmathbf{Q}^{\bot}} \mathd
  \tmmathbf{k}_{\bot} \, \mathe^{i\tmmathbf{Q} \cdot \mathsf{X} / \hbar} L
  \left( \tmmathbf{k}_{\bot}, \mathsf{P} ; \tmmathbf{Q} \right) \rho L^{\dag}
  \left( \tmmathbf{k}_{\bot}, \mathsf{P} ; \tmmathbf{Q} \right) \mathe^{-
  i\tmmathbf{Q} \cdot \mathsf{X} / \hbar} .  \label{eq:ottica}
\end{eqnarray}
The first three terms at the right-hand side describe a coherent dynamics.
They preserve the pure states, but not their normalization, due to the fact
that the optical potential is not self-adjoint. The completely positive
mapping expressed by the last term describes an incoherent contribution. It
transforms pure states into mixtures and accounts for the preservation of the
trace.

\subsection{Index of refraction}\label{sec:iof}

As described above, typical matter wave interference experiments are
characterized by a well-collimated, stationary beam of particles, with a
momentum centered around a reference value $\tmmathbf{P}= \hbar \tmmathbf{K}$
in the longitudinal direction. One can then exploit the formal analogy between
the Schr\"odinger equation and the Helmholtz wave equation
{\cite{Sears1989a,Adams1994a,Arndt2009a}} in order to introduce an index of
refraction for matter waves.

For the case of a constant, complex potential,
\begin{eqnarray}
  V_{\tmop{opt}} & = & - 2 \pi \hbar^2 \frac{n_{\tmop{gas}}}{m_{\ast}} \langle
  f_0 \left( \hbar \tmmathbf{K} \right) \rangle_{\mu}, 
\end{eqnarray}
the stationary Schr\"odinger equation takes the form
\begin{eqnarray}
  \nabla^2 \psi \left( \tmmathbf{r} \right) + \frac{2 M}{\hbar^2} \left[ E -
  V_{\tmop{opt}} \mathsf{} \right] \psi \left( \tmmathbf{r} \right) & = & 0. 
\end{eqnarray}
By analogy with the Helmholtz equation one can introduce a wave number
associated to the propagation of the scalar field $\psi \left( \tmmathbf{r}
\right)$ with energy $E$,
\begin{eqnarray}
  K' & = & \sqrt{\frac{2 M}{\hbar^2} \left[ E - V_{\tmop{opt}} \mathsf{}
  \right]} . 
\end{eqnarray}
The effect of the optical potential on the dynamics can thus be equivalently
described by an effective index of refraction
\begin{eqnarray}
  n \hspace{0.6em} = \hspace{0.6em} \frac{K'}{K} \hspace{0.6em} =
  \hspace{0.6em} \sqrt{1 - \frac{V_{\tmop{opt}} \mathsf{}}{E}}, \hspace{0.6em}
  &  &  \label{eq:index}
\end{eqnarray}
where $K = \sqrt{2 ME} / \hbar$ is the wave number associated to the de
Broglie wavelength of the tracer particle. Since the optical potential can be
non-Hermitian in general, the index of refraction is in general complex.

In the situations described above, where the last, incoherent term of
Eq.~(\ref{eq:ottica}) can be neglected, the matter wave dynamics can thus be
effectively described by a wave equation with the optical potential
Eq.~(\ref{eq:opt}). The beam state may then be equivalently expressed as a
wave propagation in a medium characterized by the refractive
index~(\ref{eq:index}), in complete analogy to the optical case.

For dilute gases the complex index of refraction is very close to unity, such
that
\begin{eqnarray}
  n & \approx & 1 + 2 \pi \frac{n_{\tmop{gas}}}{K^2}  \frac{M}{m_{\ast}}
  \text{$\langle f_0 \left( \hbar \tmmathbf{K} \right) \rangle_{\mu}$} . 
  \label{eq:nindex}
\end{eqnarray}
Its real part
\begin{eqnarray}
  n_1 & = & 1 + 2 \pi \frac{n_{\tmop{gas}}}{K^2}  \frac{M}{m_{\ast}} \tmop{Re}
  \langle f_0 \left( \hbar \tmmathbf{K} \right) \rangle_{\mu},  \label{eq:n1}
\end{eqnarray}
describes the phase shift due to the background gas, while its imaginary part
\begin{eqnarray}
  n_2 & = & 2 \pi \frac{n_{\tmop{gas}}}{K^2}  \frac{M}{m_{\ast}} \tmop{Im}
  \langle f_0 \left( \hbar \tmmathbf{K} \right) \rangle_{\mu}  \label{eq:n2}\\
  & = & \frac{M_{\tmop{cl}}^{\tmop{out}} \left( \hbar \tmmathbf{K} \right)}{2
  \hbar K^2 / M}, \nonumber
\end{eqnarray}
determined by the loss term of the classical linear Boltzmann equation,
accounts for the attenuation of the coherent beam.

We stress that this analysis in terms of a refractive index relies on
neglecting the last term in Eq.~(\ref{eq:ottica}). This is only valid under
very particular experimental conditions. These might be realized in a
Mach-Zender interferometer provided just one of the spatially separated paths
interacts with a background gas and particles whose momentum has changed due
to this interaction are blocked by the interferometer apertures. In this case
only the part of the incoming beam scattered in the forward direction
contributes to the signal collected at the outlet of the interferometer. Note
in particular that the preservation of the trace implies that the loss and
gain term do have the same weight, such that one is not allowed to neglect the
last term of Eq.~(\ref{eq:ottica}) because of its smallness with respect to
the anticommutator term.

At variance with the typical intuition developed for electromagnetic waves the
imaginary part of the index of refraction, which describes the exponential
decay of the beam intensity, does not involve true absorption. The tracer
particles are obviously not absorbed by the background gas, but the coherent
collimated part of the beam entering the interferometer and responsible for
the final interference pattern gets reduced. This is the reason why the
reduction of intensity in the signal does not lead to a corresponding
reduction of the visibility as in the case of collisional decoherence
{\cite{Hornberger2003a}}. Over a distance $L$ travelled with velocity $V = P /
M$ the reduction factor is given by the exponential $\exp \left( -
M_{\tmop{cl}}^{\tmop{out}} \left( \tmmathbf{P} \right) L / V \right)$.
Comparing this quantity with the damped intensity $\exp \left( - 2 n_2 KL
\right)$ of an electromagnetic wave with wavenumber $K$ in a medium
characterized by an index of refraction with imaginary part $n_2$ gives indeed
back Eq.~(\ref{eq:n2}).

Specializing to a rotationally invariant scattering amplitude, $f \left(
\tmmathbf{p}_f, \tmmathbf{p}_i \right) = f \left( \vartheta ; E_{\tmop{rel}}
\right)$, with $\vartheta$ the angle between incoming and outgoing relative
momentum, and $E_{\tmop{rel}} = p_i^2 / 2 m_{\ast}$ the kinetic energy in the
center of mass, one can explicitly evaluate the thermal average introduced in
Eq.~(\ref{eq:fbeta})
\begin{eqnarray}
  \text{$\langle f_0 \left( \tmmathbf{P} \right) \rangle_{\mu}$} & = &
  \frac{2}{\sqrt{\pi}} \int_0^{\infty} \frac{\mathd v}{v_{\beta}}  \frac{v}{V}
  \sinh \left( \frac{2 vV}{v_{\beta}^2} \right) \exp \left( - \frac{v^2 +
  V^2}{v_{\beta}^2} \right) f \left( 0 ; \frac{m_{\ast} v^2}{2} \right) . 
  \label{eq:champenois}
\end{eqnarray}
Here $V = P / M$ is the velocity of the interfering particle and a
Maxwell-Boltzmann distribution (\ref{eq:muMB}) is assumed with $v_{\beta} =
p_{\beta} / m$.

This expression, derived here from the quantum linear Boltzmann equation
{\cite{Hornberger2008a}}, agrees with the formula in
{\cite{Champenois1999a,Champenois2008a}}, which was recently obtained along a
very different line of thought. While it has been used in the most recent
interferometric experiments with lithium atoms {\cite{Jacquey2007a}}, it
deviates from previous theoretical results
{\cite{Forrey1996a,Kharchenko2001a}} used for the analysis of the earlier
experiments {\cite{Schmiedmayer1995a,Vigue1995a,Roberts2002a}}. The difference
lies in the thermal averaging procedure employed, and the averages in
Eqs.~(\ref{eq:fbeta}), (\ref{eq:champenois}), which follow naturally from the
linear Boltzmann equation, should be considered correct, as also pointed out
in {\cite{Champenois2008a}}. We note that still different methods of deriving
a gas induced index of refraction were recently presented in
{\cite{Dominguez-Clarimon2007a,Sanders2009a}}.

It was already emphasized that the notion of a gas induced index of refraction
can only be applied to a restricted set of experimental situations. One may
describe phase shift and particle loss phenomena with this concept, but it
does not apply to proper decoherence phenomena, where the purity of the
motional state in the beams gets dynamically reduced due to the interaction
with a background gas. As described in the following section, we have to
return to a density matrix description in order to account for such effects.

\section{Collisional decoherence of matter waves}\label{sec:decoh}

The concept of environmental decoherence
{\cite{Zurek1991a,Joos2003,Zurek2003a,Schlosshauer2007,Hornberger2009a,Breuer2007}}
is often drawn upon to explain how the emergence of classical behavior comes
about if physical systems get larger and more complex. Though formulated in
the same framework of open quantum systems also used to describe friction and
thermalization, decoherence should be distinguished from such dissipative
phenomena since it usually takes place on much shorter time scales and even in
the absence of relaxation processes with a classical analog.

The main point is to acknowledge the crucial role played by the quantum
correlations arising between the system and environmental degrees of freedom.
The combined state of system and environment is no longer separable after the
interaction, resulting in a loss of purity in the reduced quantum state, which
is obtained by disregarding the environmental state by means of a partial
trace. In a complementary view, which highlights the quantum aspects of the
correlation, one may view the interaction as constituting an information
transfer from the system to the environment.

In our case of a quantum test particle, collisions with a background gas have
the effect of unsharp position measurements because the scattered state of a
gas particle carries away some information on the position of the tracer which
would be available, in principle, to an outside observer. One therefore
expects that if the test particle is initially in a superposition of two
positions, as in an interferometer, the coherence of this state will be
reduced the more the better a typical scattering event can distinguish between
the two positions. Once the coherence expressed by the off-diagonal elements
in the position representation has vanished, the motional state will be
indistinguishable from a classical mixture.

In the following we discuss how the dynamics of such collisional decoherence
phenomena are described by the quantum linear Boltzmann equation. We start
with the case of {\tmem{pure collisional decoherence}}, which is present even
in the limit of an infinitely massive particle, where no dissipation takes
place. We also provide the quantitative connection to recent interference
experiments {\cite{Arndt2005a}}, where decoherence due to a background gas was
observed as a loss of visibility in the quantum interference fringes. After
that, we discuss the related topics of the loss of coherence in case of
momentum superpositions and in the Brownian motion limit.

\subsection{The limit of a massive tracer particle}\label{sec:decopos}

The incoherent part of the quantum linear Boltzmann equation~(\ref{eq:qlbe})
simplifies if the test particle is much more massive than the gas particles.
The function $L$ from Eq.~(\ref{eq:L}), which provides the details of the
collisional interaction, is then replaced by
\begin{eqnarray}
  L \left( \tmmathbf{p}, \tmmathbf{P}; \tmmathbf{Q} \right) & \overset{M \gg
  m}{\rightarrow} & \sqrt{\frac{n_{\tmop{gas}}}{mQ}} f \left(
  \tmmathbf{p}_{\bot \tmmathbf{Q}} - \frac{\tmmathbf{Q}}{2},
  \tmmathbf{p}_{\bot \tmmathbf{Q}} + \frac{\tmmathbf{Q}}{2} \right) \sqrt{\mu
  \left( \tmmathbf{p}_{\bot \tmmathbf{Q}} \text{$+ \frac{\tmmathbf{Q}}{2}$}
  \right)} .  \label{eq:lm}
\end{eqnarray}
Hence, in this limit the Lindblad operators no longer depend on the test
particle momentum $\mathsf{\mathsf{P}}$. One can now perform the $\mathd
\tmmathbf{k}_{\bot}$-integration in Eq.~(\ref{eq:qlbe}), such that, recalling
Eq.~(\ref{eq:Min3}) and Eq.~(\ref{eq:ratein}), one gets the much simpler
expression
\begin{eqnarray}
  \mathcal{L} \rho & = & - \int \mathd \tmmathbf{Q} \,
  M_{\tmop{in}}^{\tmop{cl}} \left( \tmmathbf{Q}; \tmmathbf{Q} \right)  \left[
  \rho - \mathe^{i\tmmathbf{Q} \cdot \mathsf{X} / \hbar} \rho \mathe^{-
  i\tmmathbf{Q} \cdot \mathsf{X} / \hbar} \right] .  \label{eq:100}
\end{eqnarray}

Here the classical rate $M_{\tmop{in}}^{\tmop{cl}} \left( \tmmathbf{Q};
\tmmathbf{Q} \right)$ must be evaluated with $m_{\ast} \rightarrow m$. It
gives the rate for a particle with vanishing momentum $\tmmathbf{P}= 0$ to
experience a momentum change of $\tmmathbf{Q}$.

It is now helpful to introduce the probability distribution of the possible
momentum transfers according to
\begin{eqnarray}
  \mathcal{P} \left( \tmmathbf{Q} \right) & = &
  \frac{M_{\tmop{in}}^{\tmop{cl}} \left( \tmmathbf{Q}; \tmmathbf{Q}
  \right)}{\int \mathd \bignone \tmmathbf{Q}' \, M_{\tmop{in}}^{\tmop{cl}}
  \left( \tmmathbf{Q}' ; \tmmathbf{Q}' \right)} .  \label{eq:102}
\end{eqnarray}
The normalization, i.e., the total rate of collisions is given by the
classical loss rate known from Eq.~(\ref{eq:rateout}),
\begin{eqnarray}
  \Gamma_{\tmop{tot}} & \equiv & M_{\tmop{out}}^{\tmop{cl}} \left( 0 \right) =
  \int \mathd \bignone \tmmathbf{Q}' \, M_{\tmop{in}}^{\tmop{cl}} \left(
  \tmmathbf{Q}' ; \tmmathbf{Q}' \right) .  \label{eq:norma}
\end{eqnarray}
It is determined by the total scattering cross-section and therefore always
finite.

We can thus write Eq.~(\ref{eq:100}) in the compact form {\cite{Alicki2002a}}

\begin{eqnarray}
  \mathcal{L \rho} & = & - \Gamma_{\tmop{tot}} \int \mathd \tmmathbf{Q} \,
  \mathcal{P} \left( \tmmathbf{Q} \right) \left( \rho - \mathe^{i\tmmathbf{Q}
  \cdot \mathsf{X} / \hbar} \rho \mathe^{- i\tmmathbf{Q} \cdot \mathsf{X} /
  \hbar} \right) .  \label{eq:101}
\end{eqnarray}
It puts into evidence that the incoherent dynamics is simply given by momentum
kicks
\begin{eqnarray}
  \rho & \rightarrow & \mathe^{i\tmmathbf{Q} \cdot \mathsf{X} / \hbar} \rho
  \mathe^{- i\tmmathbf{Q} \cdot \mathsf{X} / \hbar},  \label{eq:z1}
\end{eqnarray}
characterized by a random momentum transfer $\tmmathbf{Q}$, which is
distributed according to Eq.~(\ref{eq:102}). As discussed below,
Eq.~(\ref{eq:101}) induces an exponential decay of the off-diagonal elements
in position representation, i.e., a ``localization'' of a spatial
superposition into a statistical mixture.

A useful explicit expression is obtained from Eq. (\ref{eq:100}) by
introducing the variables $\tmmathbf{p}_i \equiv p\tmmathbf{n}_i =\tmmathbf{p}
\text{$+ \tmmathbf{Q}/ 2$}$ and $\tmmathbf{p}_f \equiv p\tmmathbf{n}_f
=\tmmathbf{p} \text{$- \tmmathbf{Q}/ 2$}$,
\begin{eqnarray}
  \mathcal{L} \rho & = & - n_{\tmop{gas}} \int_0^{\infty} \mathd p \,
  \bar{\nu} \left( p \right)  \frac{p}{m}  \int \frac{\mathd
  \Omega_{\tmmathbf{n}_i}}{4 \pi} \bignone \int \mathd \Omega_{\tmmathbf{n}_f}
  \bignone \left| f \left( p\tmmathbf{n}_f, p\tmmathbf{n}_i \right) \right|^2
  \left[ \rho - \mathe^{i (\tmmathbf{p}_i -\tmmathbf{p}_f) \mathsf{\cdot X} /
  \hbar} \rho \mathe^{- i (\tmmathbf{p}_i -\tmmathbf{p}_f) \mathsf{\cdot X} /
  \hbar} \right],  \label{eq:103}
\end{eqnarray}
where $\bar{\nu} \left( p \right) = 4 \pi p^2 \mu_{\beta} \left( p\tmmathbf{n}
\right)$ is the Maxwell distribution of the momentum magnitudes in the gas.

Equation (\ref{eq:103}) corresponds to the result of Gallis and Fleming
{\cite{Gallis1990a}}, once amended for an incorrect factor of $2 \pi$. This
little flaw, which is however experimentally relevant, was not due to a simple
calculational error, but goes back to an incorrect treatment of a ``square of
$\delta$-functions'' in {\cite{Joos1985a}}. Such a strictly ill-defined
expression arises if asymptotic scattering theory is naively incorporated into
a dynamic description. One can fix the factor either by circumventing the
S-matrix in a weak-coupling treatment
{\cite{Vacchini2000a,Vacchini2005b,Hornberger2003b,Dodd2003a,Adler2006a}} or
by formally modifying the S-matrix such that asymptotically outgoing states
are not transformed {\cite{Hornberger2003b,Hornberger2007a,Hornberger2008a}}.

The equation of pure collisional decoherence discussed so far was obtained by
the naive limit $M \rightarrow \infty$ of the quantum linear Boltzmann
equation. As a result, the incoherent part Eq. (\ref{eq:qlbe}) depends only on
the position operator $\mathsf{X}$ of the test particle. Also in the coherent
part of Eq. (\ref{eq:lvonn}) the dependence on $\mathsf{P}$ gets lost in this
limit, since the kinetic energy $\mathsf{P}^2 / \left( 2 M \right)$ vanishes,
while the gas induced energy shift $H_{\text{n}} \left( \mathsf{P} \right)$
turns into an (unobservable) constant. It follows that the dynamics of the
momentum distribution of the test particle state remains frozen under Eq.
(\ref{eq:103}).

The full quantum linear Boltzmann equation provides of course a much more
refined description of the dynamics. However, for a massive test particle and
at short time scales, where the effects of an energy exchange between particle
and gas are negligible, an improved version of Eq. (\ref{eq:100}) may already
be sufficient in many situations. A natural improvement is suggested by the
fact that the loss term (\ref{eq:norma}) is evaluated for a test particle at
rest.

Provided that the test particle is very massive, such that its momentum varies
slowly on the short time scale of decoherence, it may be allowed to replace
the momentum operator in Eq.~(\ref{eq:qlbe}) by its initial expectation value
$\tmmathbf{P}_0 = \langle \mathsf{P} \rangle_{\rho_0}$. This leads to an
equation very similar to Eq. (\ref{eq:100}),
\begin{eqnarray}
  \mathcal{L \rho} & = & - \int \mathd \tmmathbf{Q} \,
  M_{\tmop{in}}^{\tmop{cl}} \left( \tmmathbf{P}_0 +\tmmathbf{Q}; \tmmathbf{Q}
  \right) \left[ \rho - \mathe^{i\tmmathbf{Q} \cdot \mathsf{X} / \hbar} \rho
  \mathe^{- i\tmmathbf{Q} \cdot \mathsf{X} / \hbar} \right] .  \label{eq:104}
\end{eqnarray}
Introducing the normalized probability density
\begin{eqnarray}
  \mathcal{P}_{\tmmathbf{P}_0} \left( \tmmathbf{Q} \right) & = &
  \frac{M_{\tmop{in}}^{\tmop{cl}} \left( \tmmathbf{P}_0 +\tmmathbf{Q};
  \tmmathbf{Q} \right)}{\int \mathd \tmmathbf{Q}' \, M_{\tmop{in}}^{\tmop{cl}}
  \left( \tmmathbf{P}_0 +\tmmathbf{Q}' ; \tmmathbf{Q}' \right)}, 
  \label{eq:z2}
\end{eqnarray}
one arrives at
\begin{eqnarray}
  \mathcal{L \rho} & = & - \Gamma_{\tmop{tot}} \left( \tmmathbf{P}_0 \right)
  \int \mathd \tmmathbf{Q} \, \mathcal{P}_{\tmmathbf{P}_0} \left( \tmmathbf{Q}
  \right) \left[ \rho - \mathe^{i\tmmathbf{Q} \cdot \mathsf{X} / \hbar} \rho
  \mathe^{- i\tmmathbf{Q} \cdot \mathsf{X} / \hbar} \right],  \label{eq:105}
\end{eqnarray}
where the rate $\Gamma_{\tmop{tot}} \left( \tmmathbf{P}_0 \right) =
M_{\tmop{out}}^{\tmop{cl}} \left( \tmmathbf{P}_0 \right)$ gives the classical
rate of collisions experienced by a test particle with momentum
$\tmmathbf{P}_0$.

The difference with respect to Eq.~(\ref{eq:101}) lies in the residual,
parametric dependence on the momentum of the test particle $\tmmathbf{P}_0$,
leading to a more realistic estimate for the total scattering rate
$\Gamma_{\tmop{tot}} \left( \tmmathbf{P}_0 \right)$ and thus of collisional
decoherence. This is of relevance for the quantitative description of
experiments on gas-induced decoherence
{\cite{Hornberger2003a,Hackermuller2003b,Hornberger2004a,Vacchini2004a}}.

It is clear that Eq.~(\ref{eq:105}) cannot account for the long time dynamics
of momentum and kinetic energy, since the momentum of the test particle here
only appears as a classical label. As far as the short time dynamics is
concerned, we note that the corresponding random momentum transfers
(\ref{eq:z1}) are here incorporated in full, without the assumption $\left| Q
\Delta X / \hbar \right| \ll 1$, used in Sect.~\ref{sec:qbm} when considering
the limit of quantum Brownian motion.

\subsection{Decoherence in position }\label{sec:posdeco}

\subsubsection{The exact solution of pure decoherence}\label{sec:colldeco}

In position representation Eq.~(\ref{eq:101}) takes the simple form
\begin{eqnarray}
  \text{$\langle \tmmathbf{X}| \mathcal{L} \rho |\tmmathbf{X}' \rangle$} & = &
  - \Gamma_{\tmop{tot}} \left[ 1 - \Phi_{\mathcal{P}} \left(
  \tmmathbf{X}-\tmmathbf{X}' \right) \right] \text{$\langle \tmmathbf{X}| \rho
  |\tmmathbf{X}' \rangle$},  \label{eq:cf11}
\end{eqnarray}
where $\Phi_{\mathcal{P}} \left( \tmmathbf{X} \right)$ denotes the Fourier
transform of the probability density (\ref{eq:102}) of momentum transfers,
i.e., its characteristic function {\cite{Feller1971}}
\begin{eqnarray}
  \Phi_{\mathcal{P}} \left( \tmmathbf{S} \right) & = & \int \mathd
  \tmmathbf{Q} \, \mathcal{P} \left( \tmmathbf{Q} \right) \mathe^{i
  \mathsf{\tmmathbf{Q} \cdot \tmmathbf{S}/ \hbar}} .  \label{eq:cf0}
\end{eqnarray}
The solution of the master equation $\mathd \rho / \mathd t = \mathcal{L}
\rho$ then shows an exponential decay of the position off-diagonal elements,
\begin{eqnarray}
  \text{$\langle \tmmathbf{X}| \mathcal{} \rho_t |\tmmathbf{X}' \rangle$} & =
  & \mathe^{- \Gamma_{\tmop{tot}} \left[ 1 - \Phi_{\mathcal{P}} \left(
  \tmmathbf{X}-\tmmathbf{X}' \right) \right] t} \langle \tmmathbf{X}| \rho_0
  |\tmmathbf{X}' \rangle .  \label{eq:cf1}
\end{eqnarray}
The position density is not affected, $\langle \tmmathbf{X}| \mathcal{} \rho_t
|\tmmathbf{X} \rangle = \langle \tmmathbf{X}| \mathcal{} \rho_0 |\tmmathbf{X}
\rangle$, since $\Phi_{\mathcal{P}} \left( 0 \right) = 1$.

The function $\Phi_{\mathcal{P}} \left( \tmmathbf{X}-\tmmathbf{X}' \right)$
describes the effect of a single collision event on the statistical operator
in the position representation, $\langle \tmmathbf{X}| \rho |\tmmathbf{X}'
\rangle \rightarrow \Phi_{\mathcal{P}} \left( \tmmathbf{X}-\tmmathbf{X}'
\right) \langle \tmmathbf{X}| \rho |\tmmathbf{X}' \rangle$. It is sometimes
called the ``decoherence function'' since it amounts to the overlap of two
scattered gas particle states displaced by $\tmmathbf{X}-\tmmathbf{X}'$, which
quantifies to what extent the gas can distinguish between the two positions.

It follows from Eq. (\ref{eq:102}) that in terms of the scattering amplitude
it is explicitly given by
\begin{eqnarray}
  \Phi_{\mathcal{P}} \left( \tmmathbf{S} \right) & = &
  \frac{n_{\tmop{gas}}}{\Gamma_{\tmop{tot}}} \int \bignone \mathd p \,
  \bar{\nu} \left( p \right) \frac{p}{m}  \int \frac{\mathd
  \Omega_{\tmmathbf{n}_i}}{4 \pi} \bignone \int \mathd \Omega_{\tmmathbf{n}_f}
  \bignone \left| f \left( p\tmmathbf{n}_f, p\tmmathbf{n}_i \right) \right|^2
  \mathe^{ip (\tmmathbf{n}_i -\tmmathbf{n}_f) \cdot \tmmathbf{S}/ \hbar} . 
  \label{eq:z3}
\end{eqnarray}
In the case of isotropic scattering, the decoherence effect depends only on
the distance $S = \left| \tmmathbf{X}-\tmmathbf{X}' \right|$ of the positions
considered. It then takes the form
\begin{eqnarray}
  \Phi_{\mathcal{P}} \left( S \right) & = & 2 \pi
  \frac{n_{\tmop{gas}}}{\Gamma_{\tmop{tot}}} \int \bignone \mathd p \,
  \bar{\nu} \left( p \right) \frac{p}{m}  \int_{- 1}^1 \mathd \left( \cos
  \theta \right) \bignone \left| f \left( \cos \theta ; \frac{p^2}{2 m}
  \right) \right|^2 \tmop{sinc} \left( 2 p \sin \left( \frac{\theta}{2}
  \right) \frac{S}{\hbar} \right),  \label{eq:715}
\end{eqnarray}
with $\theta$ the scattering angle and $\tmop{sinc} \left( x \right) = \sin
\left( x \right) / x$. Due to the infinite test mass the momentum transfer is
given by the expression $Q = 2 p \sin \left( \theta / 2 \right)$, which
appears in the argument of the sinc-function. The effect of this function in
Eq. (\ref{eq:715}) is thus to suppress position coherences the more the better
the scattered state with its resolving capacity $\hbar / Q$ can distinguish
between two possible scattering positions with distance $S$.

The second ingredient to Eq.~(\ref{eq:cf1}) is the total rate of collisions
$\Gamma_{\tmop{tot}}$ defined in Eq.~(\ref{eq:norma}). The effect and the rate
of collisions are thus fixed by $\Phi_{\mathcal{P}}$ and $\Gamma_{\tmop{tot}}$
separately, which can be of great importance in physical applications. We note
in particular that the function
\begin{eqnarray}
  \Psi \left( \tmmathbf{S}, t \right) & = & \mathe^{- \Gamma_{\tmop{tot}}
  \left[ 1 - \Phi_{\mathcal{P}} \left( \tmmathbf{S} \right) \right] t} 
  \label{eq:fcarat}
\end{eqnarray}
appearing in Eq.~(\ref{eq:cf1}) is itself a characteristic function, namely
that of a compound Poisson process {\cite{Lukacs1966a,Feller1971}}. Such a
process differs from a simple Poisson process in that the jump events, here
the collisions, do not have a fixed size, while the waiting time distribution
between the events is Poissonian. Rather, the jump size is itself a random
variable, here the momentum transfer in a single collision, which is
distributed according to $\mathcal{P} \left( \tmmathbf{Q} \right)$.

This physical picture behind the collisional dynamics is best seen by
expanding the exponential in Eq.~(\ref{eq:cf1}), which corresponds to a jump
expansion of the solution of the master equation
\begin{eqnarray}
  \text{$\langle \tmmathbf{X}| \mathcal{} \rho_t |\tmmathbf{X}' \rangle$} & =
  & \sum^{\infty}_{n = 0} \bignone p_n \left( t \right) \Phi^n_{\mathcal{P}}
  \left( \tmmathbf{X}-\tmmathbf{X}' \right) \langle \tmmathbf{X}| \rho_0
  |\tmmathbf{X}' \rangle .  \label{eq:cf2}
\end{eqnarray}
Here $\bignone p_n \left( t \right) = \mathe^{- \Gamma_{\tmop{tot}} t} \left(
\Gamma_{\tmop{tot}} t \right)^n / n!$ is the probability to have $n$
collisions during the time $t$, given by the Poisson distribution with
parameter $\Gamma_{\tmop{tot}}$. The associated effect is the multiplication
of the initial statistical operator in the position representation with the
$n$-th power of the decoherence function $\Phi_{\mathcal{P}}$ for the effect
of a single collision.

Equation~(\ref{eq:cf1}) is a particular case of a more general class of
solutions of master equations for the description of decoherence, applying in
the presence of translational invariance. In the general case, the
characteristic function $\Psi \left( \tmmathbf{X}, t \right)$ corresponding to
Eq.~(\ref{eq:fcarat}) is the characteristic function of a L\'evy process, of
which compound Poisson processes are a particular example
{\cite{Breuer2007,Feller1971}}. This result can be obtained considering the
general structure of translation-covariant Markovian master equations obtained
by Holevo, and considering its limiting expression relevant for the
description of decoherence. This provides a common theoretical framework for
the description of quite different decoherence experiments
{\cite{Vacchini2005a}}. The solution of such master equations reads
\begin{eqnarray}
  \langle \tmmathbf{X}| \mathcal{} \rho_t |\tmmathbf{X}' \rangle & = & \Psi
  \left( \tmmathbf{X}-\tmmathbf{X}', t \right) \langle \tmmathbf{X}|
  \mathcal{} \rho_0 |\tmmathbf{X}' \rangle,  \label{eq:z4}
\end{eqnarray}
and due to the general properties of characteristic functions it naturally
describes decoherence effects. In fact, it is a defining property of
characteristic functions that $\Psi \left( 0, t \right) = 1$, while $| \Psi
\left( \tmmathbf{X}-\tmmathbf{X}', t \right) | \leqslant 1$, such that
diagonal matrix elements are not affected, while off-diagonal matrix elements
are generally suppressed. In particular, for the characteristic function of a
L\'evy process one has $\lim_{t \rightarrow \infty} \Psi \left(
\tmmathbf{X}-\tmmathbf{X}', t \right) = 0$ for $\tmmathbf{X} \neq
\tmmathbf{X}'$ {\cite{Lukacs1966a}}, such that all coherences are completely
lost with elapsing time. For fixed interaction time $t$, on the other hand, a
variety of position dependencies are possible, from an exponential loss of
coherence to revivals ending up with a constant decoherence rate. All such
behaviors are comprised in the general expression of the characteristic
function of a L\'evy process, provided by the L\'evy-Khintchine formula, one
of the beautiful results of probability theory {\cite{Feller1971}}.

\subsubsection{Experimental tests}

One of the most important experimental verifications of the quantum linear
Boltzmann equation comes from its application to the quantitative explanation
of collisional decoherence observed in near-field interferometry experiments
with large molecules {\cite{Hornberger2003a,Hackermuller2003b}}. In such
experiments a velocity selected molecular beam from a thermal source passes an
interferometer made up of material gratings. After flooding the vacuum chamber
with various background gases one records the reduction of visibility of the
interference fringes as a function of the gas pressure. This decrease of the
interference contrast is directly related to the loss of coherence in the
position off-diagonal elements in the state of motion, which was discussed in
the previous section.

An important requirement for the experimental observation of collisional
decoherence is that the interfering molecule is not removed from the
interferometer by the interaction with the background gas. As described in
Sect. \ref{sec:optics}, such a loss process would lead to a reduction of the
count rate, but it would not diminish the interference contrast of the
detected molecules. This is the reason why relatively heavy C$_{70}$ fullerene
molecules were required to observe the effect in
{\cite{Hornberger2003a,Hackermuller2003b}}, while the Mach-Zehnder-type atom
interferometers realized so far cannot observe collisional decoherence with
room temperature gases. We note that the loss of visibility in a sodium atom
interferometer reported in {\cite{Uys2005a}} is therefore due to a different
mechanism, namely the gas-induced noise at the preparation, i.e., the slit
collimation and at the detection stage of the interferometer beam. The atomic
motion between preparation and detection is there coherent, in contrast, such
that this gas-induced effect is not described by the theory of Sect.
\ref{sec:colldeco}.

Due to the large mass of the interfering particles and the short time of
flight in {\cite{Hornberger2003a,Hackermuller2003b}}, one is allowed to
neglect relaxation effects in the quantum description of the interfering test
particle. The limit of Sect. \ref{sec:decopos} thus applies, and one expects
in particular that the dynamics is correctly described by Eq.~(\ref{eq:105}),
which accounts for the finite velocity of the molecular beam with respect to
the thermal gas. In the position representation the equation reads as
\begin{eqnarray}
  \text{$\langle \tmmathbf{X}| \mathcal{L} \rho |\tmmathbf{X}' \rangle$} & = &
  - \Gamma_{\tmop{tot}} \left( \tmmathbf{P}_0 \right) \left[ 1 -
  \Phi_{\mathcal{P}_{\tmmathbf{P}_0}} \left( \tmmathbf{X}-\tmmathbf{X}'
  \right) \right] \text{$\langle \tmmathbf{X}| \rho |\tmmathbf{X}' \rangle$}, 
  \label{eq:z5}
\end{eqnarray}
where $\tmmathbf{P}_0$ is now the mean momentum of the incoming fullerenes. It
determines the rate $\Gamma_{\tmop{tot}} \left( \tmmathbf{P}_0 \right) =
M_{\tmop{out}}^{\tmop{cl}} \left( \tmmathbf{P}_0 \right)$, as well as the
decoherence function
\begin{eqnarray}
  \Phi_{\mathcal{P}_{\tmmathbf{P}_0}} \left( \tmmathbf{X} \right) & = & \int
  \mathd \tmmathbf{Q} \, M_{\tmop{in}}^{\tmop{cl}} \left( \tmmathbf{P}_0
  +\tmmathbf{Q}; \tmmathbf{Q} \right) \mathe^{i \mathsf{\tmmathbf{Q} \cdot
  \tmmathbf{X}/ \hbar}} . \nonumber
\end{eqnarray}
The classical rate $M_{\tmop{in}}^{\tmop{cl}}$, defined in
Eq.~(\ref{eq:diretta}), includes the differential scattering cross-section for
collisions between fullerene and gas molecules.

The interaction between the gas molecules and the fullerenes is well
described in the experiment by the isotropic part of the London dispersion
force. It corresponds to a potential of the form $U \left( r \right) = - C_6 /
r^6$, with an interaction strength $C_6$, which can be calculated for
different gases {\cite{Ruiz1997a,Hornberger2003a}}. The associated total
scattering cross-section can be obtained in a semiclassical calculation
{\cite{Maitland1981a}}; it depends on the relative velocity as
\begin{eqnarray}
  \sigma_{\tmop{tot}} \left( \tmmathbf{v}-\tmmathbf{V} \right) =
  \frac{\pi^2}{\sin \left( \pi / 5 \right) \Gamma \left( 2 / 5 \right)} 
  \left( \frac{3 \pi C_6}{8 \hbar} \right)^{2 / 5} \left|
  \tmmathbf{v}-\tmmathbf{V} \right|^{- \frac{2}{5}} . &  &  \label{eq:c6}
\end{eqnarray}

A quantitative description of the general loss of coherence in a near field
interferometer due to master equations of the form (\ref{eq:105}) was
presented in {\cite{Hornberger2004a}}. However, the situation is somewhat
simpler in the experiment described in
{\cite{Hornberger2003a,Hackermuller2003b}} since the background gases are at
room temperature. This implies that for practically all positions in the
interferometer the collisions are characterized by large momentum transfers
$Q$ in the sense that $Q \gg \hbar / S$, if $S$ denotes the separation of the
interference paths. As follows from Eq. (\ref{eq:715}), already a single
collision then suffices to completely reduce the off-diagonal elements of the
position representation, $\Phi_{\mathcal{P}} \left( S \neq 0 \right) \simeq
0$, while $\Phi_{\mathcal{P}} \left( 0 \right) = 1$. Only those molecules then
contribute to the contrast of the interference pattern which experienced no
collisions, while the others contribute only to the unstructured background.

In this situation the loss of visibility is determined only by the collision
rate $\Gamma_{\tmop{tot}} \left( \tmmathbf{P}_0 \right)$ in Eq. (\ref{eq:z5}),
determined by the loss term $\Gamma_{\tmop{tot}} \left( \tmmathbf{P}_0 \right)
= M_{\tmop{out}}^{\tmop{cl}} \left( \tmmathbf{P}_0 \right)$ in the linear
Boltzmann equation {\cite{Vacchini2004a,Dominguez-Clarimon2007a}}, according
to the formula {\cite{Hornberger2003a}}
\begin{eqnarray}
  \mathcal{V} & = & \mathcal{V}_0 \mathe^{- \Gamma_{\tmop{tot}} \left(
  \tmmathbf{P}_0 \right) t \mathsf{}} .  \label{eq:v}
\end{eqnarray}
Here $\mathcal{V}_0$ is the visibility in the absence of the gas and $t$ is
the time of flight of the fullerenes crossing the interferometer.

\begin{figure}[tb]
  \begin{center}
    \resizebox{10cm}{!}{\epsfig{file=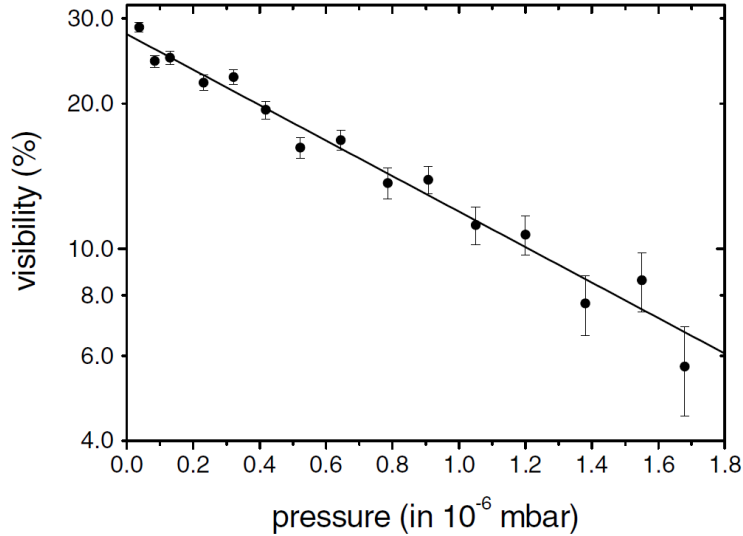}}
  \end{center}
  \caption{Experimentally observed visibility of the interference fringes of
  C$_{70}$ fullerenes in an argon gas at room temperature, as a function of
  the gas pressure (dots). The solid line gives the prediction of decoherence
  theory. It is based on Eq.~(\ref{eq:z11}), but also accounts for gas-induced
  modifications of the particular gravitational velocity selection scheme
  employed in the experiment. (Figure taken from {\cite{Hackermuller2003b}}) \
  \ \label{fig:experiment}}
\end{figure}

It is therefore sufficient to evaluate the collision frequency
$\Gamma_{\tmop{tot}} \left( \tmmathbf{P}_0 \right)$, which is best described
by an effective scattering cross-section
\begin{eqnarray}
  \Gamma_{\tmop{tot}} \left( \tmmathbf{P}_0 \right) & = & n_{\tmop{gas}}
  \frac{P_0}{M} \sigma_{\tmop{eff}} \left( P_0 \right) .  \label{eq:z7}
\end{eqnarray}
The expression $\sigma_{\tmop{eff}} \left( P_0 \right)$ for a total
cross-section of the form Eq.~(\ref{eq:c6}) has already been presented in
Sect.~\ref{sec:loss}. According to Eq.~(\ref{eq:a}), evaluated for $a = - 2 /
5$, we have
\begin{eqnarray}
  \Gamma_{\tmop{tot}} \left( \tmmathbf{P}_0 \right) & = & n_{\tmop{gas}}
  \frac{4 \pi \Gamma \left( 9 / 10 \right)}{5 \sin \left( \pi / 5 \right)} 
  \left( \frac{3 \pi C_6}{2 \hbar} \right)^{2 / 5} v^{\frac{3}{5}}_{\beta} 
  \text{}_1 F_1  \left( - \frac{3}{10}, \frac{3}{2} ; - \left(
  \frac{P_0}{Mv_{\beta}}  \right)^2 \right) .  \label{eq:z8}
\end{eqnarray}
In the experiment, the beam velocity is small compared to the thermal
velocities of the gas, $P_0 / \left( Mv_{\beta} \right) \ll 1$. The expression
can then be approximated as
{\cite{Hornberger2003a,Hackermuller2003b,Vacchini2004a}}
\begin{eqnarray}
  \Gamma_{\tmop{tot}} \left( \tmmathbf{P}_0 \right) & = & n_{\tmop{gas}}
  \frac{4 \pi \Gamma \left( 9 / 10 \right)}{5 \sin \left( \pi / 5 \right)} 
  \left( \frac{3 \pi C_6}{2 \hbar} \right)^{2 / 5} v^{\frac{3}{5}}_{\beta}
  \left[ 1 + \frac{1}{5} \left( \frac{P_0}{Mv_{\beta}}  \right)^2 +
  \mathcal{O} \left( \left( \frac{P_0}{Mv_{\beta}}  \right)^4 \right) \right]
  .  \label{eq:z9}
\end{eqnarray}
One can introduce a critical gas pressure for the onset of decoherence in
terms of the effective scattering cross-section introduced in
Eq.~(\ref{eq:z7}),
\begin{eqnarray}
  p_0 & = & \frac{Mk_{\text{B}} T}{P_0 \sigma_{\tmop{eff}} \left( P_0 \right)
  t}  \label{eq:z10}
\end{eqnarray}
One therefore expects an exponential loss of visibility of the interference
fringes with growing gas pressure according to
\begin{eqnarray}
  \mathcal{V} & = & \mathcal{V}_0 \mathe^{- p / p_0 \mathsf{}} . 
  \label{eq:z11}
\end{eqnarray}
This behavior was observed in the experiments
{\cite{Hornberger2003a,Hackermuller2003b}}, in quantitative agreement with
theory, see Fig.~\ref{fig:experiment}. \ 

Finally, we mention that the reduction of interference contrast due to
momentum exchange effects was the subject of several other experimental
{\cite{Cheng1999a,Pfau1994a,Chapman1995a,Kokorowski2001a,Hackermuller2004a,Hornberger2009b}}
and theoretical
{\cite{Hornberger2004a,Facchi2004a,Vacchini2005a,Hornberger2005a,Hornberger2006a,Nimmrichter2008b}}
investigations. While the physical mechanism is here the scattering,
absorption, or emission of photons, the dynamic effect on the motional state
of the interfering particle can still be described with a master equation of
the form~(\ref{eq:101}), (\ref{eq:cf11}). All that needs to be done is to
adjust the rate of events $\Gamma_{\tmop{tot}}$ according to the physical
mechanism considered, as well as the decoherence function $\Phi_{\mathcal{P}}$
for the effect of a single event {\cite{Vacchini2005a}}. Studies of other
environmental decoherence effects in interferometers were presented in
{\cite{Viale2003a,Lombardo2005a,Villar2007a,Qureshi2008a}}. These are based on
master equations characteristic of linear coupling models, which are of the
form of the Brownian motion master equation discussed in Sect.
\ref{sec:decohqbm}.

\subsection{Decoherence in momentum}

As described in the previous section, it is relatively simple to evaluate and
to experimentally test the decay of coherence in superpositions of position
eigenstates. The analogous case of superpositions of momentum eigenstates is
more involved. On the theoretical side, the momentum operator of the test
particle can no longer be considered a classical label in the limiting
expression of the quantum linear Boltzmann equation, but one deals with a
master equation for the operators $\mathsf{X}$ and $\mathsf{P}$ of the test
particle. From the experimental point of view, it is difficult to conceive
experiments testing decoherence in momentum space. The source must generate a
state showing longitudinal momentum coherences, which requires a
non-stationary beam preparation, as well as an interferometric setup
permitting to monitor the presence of such coherences
{\cite{Rubenstein1999a,Rubenstein1999b}}. This is complicated by the fact that
the decoherence in position discussed in Sect.~\ref{sec:posdeco} is always
present and may quickly dominate. It affects superposition states of different
momenta, because the free evolution of such states unavoidably leads to a
superposition of spatially separated wave packets.

In order to study the decoherence dynamics of momentum superpositions we
exploit the quantum trajectory algorithm presented in Sect.~\ref{sec:mc},
taking as initial condition a coherent superposition of momentum eigenstates.
We start with an initial state of the form
\begin{eqnarray}
  | \psi \left( 0 \right) \rangle & = & \sum_{i = 1}^N \alpha_i \left( 0
  \right) |\tmmathbf{U}_i \left( 0 \right) \rangle, \bignone  \label{eq:z12}
\end{eqnarray}
where the complex amplitudes $\alpha_i$ satisfy the normalization condition
$\sum_{i = 1}^N \left| \alpha_i \left( 0 \right) \right|^2 = 1$. As discussed
in Sect.~\ref{sec:mc}, the state at a later time $t$ will still be in a
superposition of $N$ momentum eigenstates, but with different amplitudes
$\alpha_i \left( t \right)$ and scaled momenta $\tmmathbf{U}_i \left( t
\right)$
\begin{eqnarray}
  | \psi \left( t \right) \rangle & = & \sum_{i = 1}^N \alpha_i \left( t
  \right) |\tmmathbf{U}_i \left( t \right) \rangle .  \label{eq:t}
\end{eqnarray}
Given a jump into the state Eq.~(\ref{eq:t}) occurring at time $t$, the
deterministic time evolution described by Eq.~(\ref{eq:det}) only modifies the
amplitudes,
\begin{eqnarray}
  \alpha_i \left( t + \tau \right) & = & \frac{\mathe^{- iE \left( U_i \left(
  t \right) \right) \tau / \hbar} \mathe^{- \tilde{\Gamma} \left( U_i \left( t
  \right) \right) \tau / 2}}{\sqrt{\sum_{j = 1}^N \left| \alpha_j \left( t
  \right) \right|^2 \mathe^{- \tilde{\Gamma} \left( U_j \left( t \right)
  \right) \tau}}} \alpha_i \left( t \right),  \label{eq:z13}
\end{eqnarray}
with $E \left( U \right) = Mv_{\beta}^2 U^2 / 2$ the kinetic energy. The
cumulative distribution function $F \left( \tau \right)$ of Eq.~(\ref{eq:g})
for the waiting time until the next jump is now given by a sum of exponential
functions,
\begin{eqnarray}
  F \left( \tau \right) & = & 1 - \sum_{j = 1}^N \left| \alpha_j \left( t
  \right) \right|^2 \mathe^{- \tilde{\Gamma} \left( U_j \left( t \right)
  \right) \tau} .  \label{eq:z14}
\end{eqnarray}
The jump modifies amplitudes and momenta as
\begin{eqnarray}
  \tmmathbf{U}_i \left( t \right) & \rightarrow & \tmmathbf{U}_i \left( t
  \right) + \frac{m_{\ast}}{M} \tmmathbf{K}  \label{eq:z15}\\
  \alpha_i \left( t + \tau \right) & \rightarrow & f_i \alpha_i \left( t +
  \tau \right) \nonumber
\end{eqnarray}
with
\begin{eqnarray}
  f_i & = & \frac{\mathe^{- \left( K / 2 +\tmmathbf{K} \cdot \tmmathbf{U}_i
  \left( t \right) \right)^2 / 2}}{\sqrt{\sum_{j = 1}^N \left| \alpha_j \left(
  t \right) \right|^2 \mathe^{- \left( K / 2 +\tmmathbf{K} \cdot
  \tmmathbf{U}_j \left( t \right) \right)^2}}} .  \label{eq:ff}
\end{eqnarray}
While all momentum eigenstates in the superposition are shifted by the same
amount, their different amplitudes are modified in different ways. The
momentum transfer $\tmmathbf{K}$ characterizing the jump is distributed
according to Eq.~(\ref{eq:i}) with the state $| \psi \left( t \right)
\rangle$given by Eq.~(\ref{eq:t}). The probability density Eq.~(\ref{eq:kl})
is therefore replaced by the mixture
\begin{eqnarray}
  \widetilde{\mathcal{P}} \left( K, \xi \right) & = & \sum_{i = 1}^N \lambda_i
  \mathcal{} \mathcal{P} \left( K, \xi_i \right),  \label{eq:mix}
\end{eqnarray}
where the weights $\lambda_i \mathcal{}$ are given by
\begin{eqnarray}
  \lambda_i \mathcal{} & = & \frac{\left| \alpha_i \left( t \right) \right|^2
  \mathe^{- \tilde{\Gamma} \left( U_i \left( t \right) \right) \tau} 
  \tilde{\Gamma} \left( U_i \left( t \right) \right)}{\sum_{j = 1}^N \left|
  \alpha_j \left( t \right) \right|^2 \mathe^{- \tilde{\Gamma} \left( U_j
  \left( t \right) \right) \tau}  \tilde{\Gamma} \left( U_j \left( t \right)
  \right)} . \nonumber
\end{eqnarray}
The probability density $\mathcal{P} \left( K, \xi \right)$ is defined in
Eq.~(\ref{eq:kl}) and $\xi_i =\tmmathbf{U}_i \cdot \tmmathbf{K}/ \left( U_i K
\right)$ is the cosine of the angle between $\tmmathbf{K}$ and $\tmmathbf{U}_i
\left( t \right)$.

\begin{figure}[tb]
  \begin{center}
    \resizebox{90mm}{!}{\epsfig{file=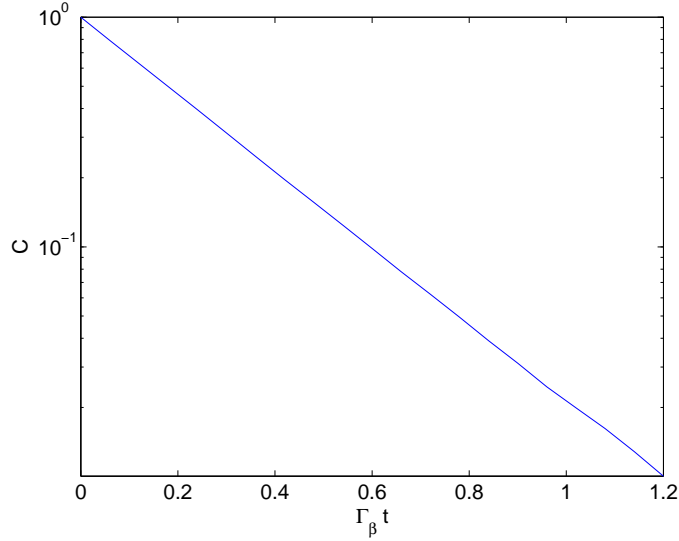}}
  \end{center}
  \caption{Semilogarithmic plot of the coherence $C \left( t \right)$ of a
  superposition of momentum eigenstates, as defined in Eqs.~(\ref{eq:q1}),
  (\ref{eq:ct}). One observes a clear exponential decay, in analogy to the
  behavior of position superpositions. The plot corresponds to a constant
  cross-section with $U_0 = 4$ and a mass ratio of $m / M = 1$; the time is
  given in units of the inverse total scattering rate (\ref{eq:gamma0}). \ \ \
  \ \ \ \ \label{fig:decoherence}}
\end{figure}

We now specialize to a superposition of two momentum eigenstates with equal
amplitudes $\alpha_1 \left( 0 \right) = \alpha_2 \left( 0 \right) = 1 /
\sqrt{2}$,
\begin{eqnarray}
  | \psi \left( 0 \right) \rangle & = & \frac{1}{\sqrt{2}} \left(
  |\tmmathbf{U}_1 \left( 0 \right) \rangle + |\tmmathbf{U}_2 \left( 0 \right)
  \rangle \right),  \label{eq:q1}
\end{eqnarray}
and evaluate the temporal behavior of the matrix elements in position
representation. The off-diagonal elements provide a measure for the coherence
in the state of motion
\begin{eqnarray}
  C \left( t \right) & = & \mathbbm{E} \left[ \left| \frac{\alpha_1 \left( t
  \right) \alpha_2^{\ast} \left( t \right)}{\alpha_1 \left( 0 \right)
  \alpha_2^{\ast} \left( 0 \right)} \right| \right] .  \label{eq:ct}
\end{eqnarray}
This quantity is plotted in Fig. \ref{fig:decoherence} for the choice
$\tmmathbf{U}_1 \left( 0 \right) = -\tmmathbf{U}_2 \left( 0 \right)
=\tmmathbf{U}_0$, such that the distance in momentum space is given by $\left|
\tmmathbf{U}_1 \left( 0 \right) -\tmmathbf{U}_2 \left( 0 \right) \right| = 2
U_0$. One observes an exponential decay in time in analogy to the behavior
(\ref{eq:cf1}) of the position coherences.

Based on this observation we now assume that the decay has the generic form
\begin{eqnarray}
  C \left( t \right) & = & \mathe^{- \Lambda \left( U_0 \right) t} . 
  \label{eq:q2}
\end{eqnarray}
We proceed to evaluate how the decay constant $\Lambda$ depends on the value
$U_0$ characterizing the initial momentum separation and to compare the
analytic result with the simulations.

\begin{figure}[tb]
  \begin{center}
    \resizebox{100mm}{!}{\epsfig{file=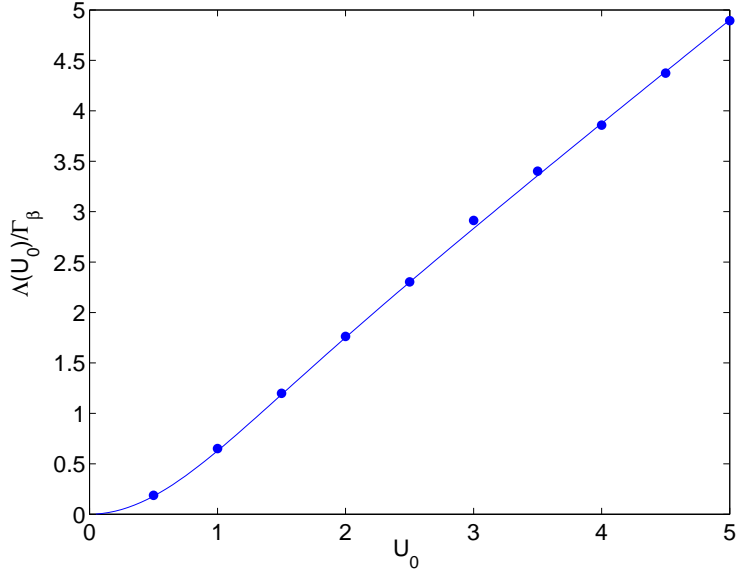}}
  \end{center}
  \caption{ Plot of the decoherence rate $\Lambda$ in units of the collision
  rate $\Gamma_{\beta}$ as a function of the initial scaled momentum $U_0$.
  The points represent least square fits of the simulated data for the mass $m
  / M = 1$, while the line represents the analytic estimate given by Eq.
  (\ref{eq:esatto}). \ \ \ \ \ \ \label{fig:decoherence-rate}}
\end{figure}

The quantum trajectory method presented in Sect.~\ref{sec:mc} is equivalent
to solving the master equation. One can thus assess the decay constant by
noting that for short times $C \left( t \right)$ either does not change, with
a probability $1 - \tilde{\Gamma} \left( U_0 \right) t$, or it becomes $f_1
f_2$ if a jump takes place. Neglecting higher order jump processes one thus
has
\begin{eqnarray}
  C \left( t \right) \approx 1 - \Lambda \left( U_0 \right) t \approx 1 -
  \tilde{\Gamma} \left( U_0 \right) t + \tilde{\Gamma} \left( U_0 \right) t
  \text{$\langle f_1 f_2$} \rangle, &  &  \label{eq:q3}
\end{eqnarray}
where $\langle \cdots \rangle$ denotes the average over the different possible
momentum transfers. Exploiting the expression Eq.~(\ref{eq:ff}) and averaging
with respect to Eq.~(\ref{eq:mix}) one comes to
\begin{eqnarray}
  \Lambda \left( U_0 \right) & \approx & \tilde{\Gamma} \left( U_0 \right)
  \left( 1 - \text{$\langle f_1 f_2$} \rangle \right) .  \label{eq:ratel}
\end{eqnarray}
For the case of a constant scattering cross-section, it takes the explicit
form
\begin{eqnarray}
  \Lambda \left( U_0 \right) & = & \tilde{\Gamma} \left( U_0 \right) -
  \Gamma_{\beta} \frac{\tmop{erf} \left( U_0 \right)}{U_0},  \label{eq:esatto}
\end{eqnarray}
with $\Gamma_{\beta}$ defined in Eq.~(\ref{eq:gamma0}). As shown in Fig.
\ref{fig:decoherence-rate}, this prediction is in very good agreement with the
values of $\Lambda \left( U_0 \right)$ extracted from the numerical
simulations for various initial values of the momentum $U_0$. Similar results
can be obtained for different scattering cross-sections, the degree of
accuracy of the analytic formula depending on the considered scattering
cross-section.

A direct comparison of relaxation and decoherence rate in momentum space can
be made for the case $m / M \ll 1$ and assuming that the momentum separation
is large compared to the thermal momentum of the gas $U_0 \gg 1$. The
relaxation rate is then given by $\eta$ as in Eq.~(\ref{eq:friction}), while
the decoherence rate Eq.~(\ref{eq:esatto}) can be approximated for $U_0 \gg 1$
by $\Lambda \left( U_0 \right) \approx \Gamma_{\beta} U_0$. The ratio of
relaxation and decoherence rate
\begin{eqnarray}
  \frac{\Lambda}{\eta} \approx \frac{3 \sqrt{\pi}}{8} \frac{M}{m} U_0 \gg 1 & 
  &  \label{eq:comparerd}
\end{eqnarray}
indicates that momentum decoherence takes place on a much shorter time scale
than relaxation effects.

The study of how decoherence takes place in momentum space is relevant also
for the description of the transition from the quantum linear Boltzmann
equation to the classical linear Boltzmann equation, already discussed at the
end of Sect.~\ref{sec:eq}. As observed in Sect.~\ref{sec:mom}, the diagonal
matrix elements of the quantum linear Boltzmann equation coincide with the
expression of the classical linear Boltzmann equation, provided one considers
the quantum scattering cross-section. The dynamical suppression of the
off-diagonal matrix elements in the momentum representation as a consequence
of decoherence thus amounts to the onset of classical dynamics as far as the
momentum observable is concerned.

\subsection{Decoherence in the quantum Brownian motion
limit}\label{sec:decohqbm}

Having discussed how the full quantum linear Boltzmann equation can be used to
describe decoherence both in position and momentum space, let us now address
its description in the Brownian motion limit of the equation. The resulting
master equation, which corresponds to approximating the environment by a bath
of linearly coupled harmonic oscillators, is most often used to discuss
decoherence effects
{\cite{Zurek1991a,Joos2003,Zurek2003a,Schlosshauer2007,Hornberger2009a,Breuer2007}}.
In the simplest case, it leads to an exponential decay rate of the
off-diagonal matrix elements in the position representation, which depends
like a Gaussian on the position separation. This picture corresponds to the
common lore, which however obviously cannot cover all interesting physical
situations, neither theoretically
{\cite{Anglin1997a,Kusnezov1999a,Lutz2002a,Schomerus2007a}}, nor
experimentally
{\cite{Cheng1999a,Pfau1994a,Chapman1995a,Kokorowski2001a,Hornberger2003a,Hackermuller2004a}}.

We start with the master equation obtained by neglecting the friction term in
the equation (\ref{eq:me}) of quantum Brownian motion.
\begin{eqnarray}
  \mathcal{L} \rho & = & - \frac{i}{\hbar} \left[ \mathsf{H_0}, \rho \right] -
  \frac{D_{pp}}{\hbar^2} \sum^3_{j = 1} \left[ \mathsf{X}_j, \left[
  \mathsf{X}_j, \rho \right] \right] - \frac{D_{xx}}{\hbar^2} \sum^3_{j = 1}
  \left[ \mathsf{P}_j, \left[ \mathsf{P}_j, \rho \right] \right] . 
  \label{eq:ld}
\end{eqnarray}
Since the friction term in Eq.~(\ref{eq:me}) accounts for energy dissipation,
one expects that Eq.~(\ref{eq:ld}) still describes the short time decoherence
phenomena associated with quantum Brownian motion.

\subsubsection{Position representation}

In the position representation, the general solution of Eq.~(\ref{eq:ld}) can
be most easily expressed in terms of the solution of the free Schr\"odinger
equation {\cite{Bassi2005b,Bellomo2007a,Fehti-xxx}}. Denoting the freely
evolved state at time $t$ as $\rho^S_t$, we have
\begin{eqnarray}
  \langle \tmmathbf{X}| \rho_t |\tmmathbf{X}' \rangle & = & \exp \left( -
  \frac{D_{pp}}{\hbar^2} \left( \tmmathbf{X}-\tmmathbf{X}' \right)^2 t \left[
  1 - \frac{\left( D_{pp} / 4 M^2 \right) t^2}{\left[ D_{xx} + \left( D_{pp} /
  3 M^2 \right) t^2 \right]} \right] \right) \nonumber\\
  &  & \times \left( \frac{1}{4 \pi \left[ D_{xx} + \left( D_{pp} / 3 M^2
  \right) t^2 \right] t} \right)^{3 / 2} \int \mathd \tmmathbf{Y} \exp \left(
  \frac{i}{\hbar} \frac{D_{pp}}{2 M} \frac{\tmmathbf{Y} \cdot \left(
  \tmmathbf{X}-\tmmathbf{X}' \right) t}{\left[ D_{xx} + \left( D_{pp} / 3 M^2
  \right) t^2 \right]} \right) \nonumber\\
  &  & \times \exp \left( - \frac{Y^2}{4 \left[ D_{xx} + \left( D_{pp} / 3
  M^2 \right) t^2 \right] t} \right) \langle \tmmathbf{X}-\tmmathbf{Y}|
  \rho^S_t |\tmmathbf{X}' -\tmmathbf{Y} \rangle .  \label{eq:x}
\end{eqnarray}
It is easy to recover from Eq.~(\ref{eq:x}) the standard result of elementary
decoherence models, according to which the loss of coherence in position space
is governed by a Gaussian function in the distance $\left|
\tmmathbf{X}-\tmmathbf{X}' \right|$ {\cite{Wigner1983a,Joos1985a}}. In the
limit of a very massive particle and short times, the spreading due to the
Hamiltonian term can be neglected, $\left( D_{pp} / 3 M^2 \right) t^2 \simeq
0$. One is then left with a convolution of the initial statistical operator in
position representation with a Gaussian of width $2 D_{xx} t$. On short time
scales, and for high temperatures, the Gaussian is strongly peaked. One is
then left with the expected expression
\begin{eqnarray}
  \langle \tmmathbf{X}| \rho_t |\tmmathbf{X}' \rangle & = & \exp \left( -
  \frac{D_{pp}}{\hbar^2} \left( \tmmathbf{X}-\tmmathbf{X}' \right)^2 t \right)
  \langle \tmmathbf{X}| \rho_0 |\tmmathbf{X}' \rangle .  \label{eq:q4}
\end{eqnarray}

\subsubsection{Momentum representation}

In the momentum representation, it is more convenient to directly express the
solution with reference to the initial state $\rho_0$
\begin{eqnarray}
  \langle \tmmathbf{P}| \rho_t |\tmmathbf{P}' \rangle & = & \exp \left( -
  \frac{D_{xx}}{\hbar^2} \left( \tmmathbf{P}-\tmmathbf{P}' \right)^2 t \right)
  \exp \left( - \frac{1}{12} \frac{D_{pp}}{\hbar^2} \left(
  \frac{\tmmathbf{P}-\tmmathbf{P}'}{M} t \right)^2 t \right) \exp \left( -
  \frac{i}{\hbar} \frac{P^2 - P'^2}{2 M} t \right) \nonumber\\
  &  & \times \left( \frac{1}{4 \pi D_{pp} t} \right)^{3 / 2} \int \mathd
  \tmmathbf{Q} \exp \left( - \frac{Q^2}{4 D_{pp} t} \right) \langle
  \tmmathbf{P}-\tmmathbf{Q}| \rho_0 |\tmmathbf{P}' -\tmmathbf{Q} \rangle . 
  \label{eq:p0}
\end{eqnarray}
Looking at the position matrix elements Eq.~(\ref{eq:x}), one notes that the
quantum position diffusion term with coefficient $D_{xx}$ mainly corrects the
detailed behavior in the exponential suppression of position coherences. In
the momentum representation, one has two distinct decoherence mechanisms
instead.

The first factor at the right-hand side of Eq.~(\ref{eq:p0}) is just due to
the $\sum^3_{j = 1} \left[ \mathsf{P}_j, \left[ \mathsf{P}_j, \rho \right]
\right]$ term in Eq.~(\ref{eq:ld}), in perfect analogy with the $\sum^3_{j =
1} \left[ \mathsf{X}_j, \left[ \mathsf{X}_j, \rho \right] \right]$ term
causing decoherence in position. The remaining part comes from the joint
effect of free evolution and the term causing decoherence in position. As
already mentioned, initial superpositions of different momenta lead to
position superpositions as time evolves, such that localization effects also
destroy coherences in momentum representation. This asymmetry between position
and momentum is clearly due to the presence of the kinetic term. Indeed,
Eq.~(\ref{eq:x}) and Eq.~(\ref{eq:p0}) acquire the same functional form in the
limit $M \rightarrow \infty$, which removes the Hamiltonian term from the
master equation.

\subsubsection{Relevance of the quantum correction term}

We now estimate the relative weight of the two contributions at the right-hand
side of Eq.~(\ref{eq:p0}), so as to gain an insight on the relevance of the
quantum correction term with coefficient $D_{xx}$. In order to do that we
recall the expressions (\ref{eq:pth}) and (\ref{eq:xth}) for the diffusion
coefficients in terms of the characteristic thermal scales. One thus obtains
an expression involving the friction constant $\eta$, as well as the
temperature and the mass of the test particle,
\begin{eqnarray}
  \langle \tmmathbf{P}| \rho_t |\tmmathbf{P}' \rangle & = & \exp \left( -
  \frac{\beta}{16} \frac{\left( \tmmathbf{P}-\tmmathbf{P}' \right)^2}{M}
  \left[ 1 + \frac{4}{3} \left( \frac{t}{\beta \hbar} \right)^2 \right] \eta t
  \right) \exp \left( - \frac{i}{\hbar} \frac{P^2 - P'^2}{2 M} t \right)
  \nonumber\\
  &  & \times \left( \frac{\beta}{4 \pi M \eta t} \right)^{3 / 2} \int \mathd
  \tmmathbf{Q} \bignone \exp \left( - \frac{\beta Q^2}{4 M \eta t} \right)
  \langle \tmmathbf{P}-\tmmathbf{Q}| \rho_0 |\tmmathbf{P}' -\tmmathbf{Q}
  \rangle .  \label{eq:xp}
\end{eqnarray}
This equation tells us that momentum coherences are exponentially suppressed
for $\tmmathbf{P}-\tmmathbf{P}'$ exceeding the thermal momentum spread
$\langle P^2 \rangle_{\beta} = 3 M / \beta$. The first factor corresponding to
the quantum term, however, is the dominating contribution only for times $t
\lesssim \beta \hbar$, while for longer times the effect due to the spatial
dispersion dominates and grows very quickly with elapsing time.

\section{The dynamic structure factor}\label{sec:dsf}

We now assume a slightly different point of view on the classical and the
quantum linear Boltzmann equation, by rewriting it in a way which puts into
evidence the statistical properties of the gas, as well as the role of the
energy and momentum transfer. This is achieved by expressing the collision
kernel in terms of the dynamic structure factor of the background gas, a
two-point correlation function of its density fluctuations. This quantity,
which is popular in the analysis of experiments where microscopic probes
scatter off macroscopic samples, is defined as the Fourier transform of the
time dependent density autocorrelation function of the gas with respect to
both the energy transfer and the momentum transfer of a single collision. Its
appearance in the quantum linear Boltzmann equation suggests a natural
interpretation of the ensuing dynamics as driven by the density fluctuations
in the medium, in accordance with Einstein's understanding of Brownian motion
{\cite{Einstein1905b}}. We emphasize that the results discussed in this
section are valid for a Maxwell-Boltzmann distribution for the gas particles,
while they generally rely on further assumptions for more general
distributions.

\subsection{An alternative formulation of the linear Boltzmann
equation}\label{sec:clbedsf}

\subsubsection{The classical expression}

Let us go back to the classical linear Boltzmann equation written in terms of
an explicit energy conservation as in Eq.~(\ref{eq:clbeMB}). In
Sect.~\ref{sec:holevo} it serves as a starting point for the heuristic
motivation of the quantum Boltzmann equation, where a decisive step is the
decomposition of the momenta of gas and test particle into components parallel
and perpendicular to the momentum transfer $\tmmathbf{Q}$. This permits one to
express the energy conserving delta function in the classical linear Boltzmann
equation in terms of the parallel momentum components only. In the scattering
cross-section, in contrast, only momentum components perpendicular to the
momentum transfer $\tmmathbf{Q}$ appear, as can be seen in Eq.~(\ref{eq:A}).

We can now observe that the Maxwell-Boltzmann distribution (\ref{eq:muMB}) is
invariant under rotations and that it factorizes in Cartesian coordinates.
This implies that for any fixed direction $\tmmathbf{Q}$ one can write
\begin{eqnarray}
  \mu_{\beta} \left( \tmmathbf{p} \right) & = & \mu_{\beta}^{\left( 2 d
  \right)} \left( \tmmathbf{p}_{\bot \tmmathbf{Q}} \right) \mu_{\beta}^{\left(
  1 d \right)} \left( \tmmathbf{p}_{\|\tmmathbf{Q}} \right),  \label{eq:fatt}
\end{eqnarray}
where
\begin{eqnarray}
  \mu_{\beta}^{\left( 2 d \right)} \left( \tmmathbf{p} \right) & = &
  \frac{\beta}{2 \pi m} \exp \left( - \frac{\beta \tmmathbf{p}^2}{2 m} \right)
  \label{eq:parper}\\
  \mu_{\beta}^{\left( 1 d \right)} \left( \tmmathbf{p} \right) & = &
  \sqrt{\frac{\beta}{2 \pi m}} \exp \left( - \frac{\beta \tmmathbf{p}^2}{2 m}
  \right)  \label{eq:1dnorm}
\end{eqnarray}
denote probability densities in two and one dimensions, respectively. It
follows that also the momentum distribution can be split into factors
depending either on the parallel or the perpendicular components of the gas
momenta with respect to the momentum transfer $\tmmathbf{Q}$.

Exploiting these facts, one can perform the integral over the parallel
component of the gas momentum in the linear Boltzmann equation
Eq.~(\ref{eq:A}). This can be done without specifying the scattering
cross-section since the latter depends only on the orthogonal components of
the gas momenta. One thus obtains the compact expression
\begin{eqnarray}
  \text{$\partial^{\tmop{coll}}_t f$} (\tmmathbf{P}) & = &
  \frac{n_{\tmop{gas}} }{m^2_{\ast}} \bigintlim \mathd \tmmathbf{Q}
  \int_{\tmmathbf{Q}^{\bot}} \mathd \tmmathbf{k}_{\bot} \, \mu_{\beta}^{\left(
  2 d \right)} \left( \tmmathbf{k}_{\bot} \right) \sigma \left( \tmop{rel}
  \left( \tmmathbf{k}_{\bot}, \tmmathbf{P}_{\perp \tmmathbf{Q}} \right) -
  \frac{\tmmathbf{Q}}{2}, \tmop{rel} \left( \tmmathbf{k}_{\bot},
  \tmmathbf{P}_{\perp \tmmathbf{Q}} \right) + \frac{\tmmathbf{Q}}{2} \right) 
  \label{eq:cdsf}\\
  &  & \times \left[ S_{\text{MB}} \left( \tmmathbf{Q},
  \tmmathbf{P}-\tmmathbf{Q} \right) f \left( \tmmathbf{P}-\tmmathbf{Q} \right)
  - S_{\text{MB}} \left( \tmmathbf{Q}, \tmmathbf{P} \right) f \left(
  \tmmathbf{P} \right) \right], \nonumber
\end{eqnarray}
upon defining
\begin{eqnarray}
  S_{\text{MB}} \left( \tmmathbf{Q}, \tmmathbf{P} \right) & \equiv & \int
  \mathd \tmmathbf{k} \, \mu_{\beta} \left( \tmmathbf{k} \right) \, \delta
  \left( \frac{\left( \tmmathbf{P} + \tmmathbf{Q} \right)^2}{2 M} +
  \frac{(\tmmathbf{k}- \tmmathbf{Q})^2}{2 m} - \frac{P^2}{2 M} - \frac{k^2}{2
  m} \right) \nonumber\\
  & = & \int_{\tmmathbf{Q}^{\|}} \mathd \bignone \tmmathbf{k}_{\|} \,
  \mu_{\beta}^{\left( 1 d \right)} \left( \tmmathbf{k}_{\|} \right)  \, \delta
  \left( \frac{Q^2}{2 m_{\ast}} - \frac{1}{m_{\ast}} \mathbf{\tmmathbf{Q}}
  \cdot \tmop{rel} \left( \mathbf{\tmmathbf{k}}_{\|},
  \tmmathbf{P}_{\|\tmmathbf{Q}} \right) \right) .  \label{eq:smb}
\end{eqnarray}
The second line follows from the identity Eq.~(\ref{eq:para}) and the
normalization is fixed in Eq.~(\ref{eq:1dnorm}). As will be discussed in
Sect.~\ref{sec:propdsf}, the function $S_{\text{MB}} \left( \tmmathbf{Q},
\tmmathbf{P} \right)$ is the dynamic structure factor of a gas of
Maxwell-Boltzmann particles.

By further introducing an averaged scattering cross-section, involving the
two-dimensional distribution (\ref{eq:parper}) over the perpendicular gas
momenta,
\begin{eqnarray}
  \sigma_{\tmop{av}} \left( \tmmathbf{P}_{\perp \tmmathbf{Q}}, \tmmathbf{Q}
  \right) & \equiv & \int_{\tmmathbf{Q}^{\bot}} \mathd \tmmathbf{k}_{\bot} \,
  \mu_{\beta}^{\left( 2 d \right)} \left( \tmmathbf{k}_{\bot} \right) \sigma
  \left( \tmop{rel} \left( \tmmathbf{k}_{\bot}, \tmmathbf{P}_{\perp
  \tmmathbf{Q}} \right) - \frac{\tmmathbf{Q}}{2}, \tmop{rel} \left(
  \tmmathbf{k}_{\bot}, \tmmathbf{P}_{\perp \tmmathbf{Q}} \right) +
  \frac{\tmmathbf{Q}}{2} \right),  \label{eq:tilda}
\end{eqnarray}
Equation~(\ref{eq:cdsf}) assumes the compact form
\begin{eqnarray}
  \text{$\partial^{\tmop{coll}}_t f$} (\tmmathbf{P}) & = &
  \frac{n_{\tmop{gas}} }{m^2_{\ast}} \bigintlim \mathd \tmmathbf{Q} \,
  \sigma_{\tmop{av}} \left( \tmmathbf{P}_{\perp \tmmathbf{Q}}, \tmmathbf{Q}
  \right)  \left[ S_{\text{MB}} \left( \tmmathbf{Q}, \tmmathbf{P}-\tmmathbf{Q}
  \right) f \left( \tmmathbf{P}-\tmmathbf{Q} \right) - S_{\text{MB}} \left(
  \tmmathbf{Q}, \tmmathbf{P} \right) f \left( \tmmathbf{P} \right) \right] . 
  \label{eq:soloq}
\end{eqnarray}
By exploiting the factorized expression of the gas distribution function
Eq.~(\ref{eq:fatt}) we thus obtained an expression of the classical linear
Boltzmann equation which involves only an integral over the momentum transfer
$\tmmathbf{Q}$. Note that if the scattering cross-section only depends on the
momentum transfer, as is the case in the Born approximation, the
$\tmmathbf{k}_{\bot}$-integration can be done even in the absence of a
factorization property like Eq.~(\ref{eq:fatt}). The averaged scattering
cross-section (\ref{eq:tilda}) then coincides with the proper cross-section.

The integral in Eq.~(\ref{eq:smb}) can be calculated explicitly, leading to an
expression which is best stated in terms of the momentum $\tmmathbf{Q}$ and
the energy $E$ transferred to the test particle, whose momentum prior to the
collision is $\tmmathbf{P}$. We take these quantities as positive if the test
particle gains momentum or energy. According to Eq.~(\ref{eq:etransfer}) the
energy transfer is $E \left( \tmmathbf{Q}, \tmmathbf{P} \right) = Q^2 / \left(
2 M \right) +\tmmathbf{Q} \cdot \tmmathbf{P}/ M$, such that it depends only on
$\tmmathbf{P}_{\|\tmmathbf{Q}}$. The function $S_{\text{MB}}$ can therefore be
expressed equivalently as a function of $\tmmathbf{Q}$ and
$\tmmathbf{P}_{\|\tmmathbf{Q}}$, or of $\tmmathbf{Q}$ and $E \left(
\tmmathbf{Q}, \tmmathbf{P} \right)$,
\begin{eqnarray}
  S_{\tmop{MB}} \left( \mathbf{\tmmathbf{Q}}, \tmmathbf{P} \right) =
  S_{\tmop{MB}} \left( \mathbf{\tmmathbf{Q}}, E \left( \tmmathbf{Q},
  \tmmathbf{P} \right) \right) & = & \sqrt{\frac{\beta m}{2 \pi}} \frac{1}{Q}
  \exp \left( - \frac{\beta}{8 m} \frac{(Q^2 + 2 mE \left( \tmmathbf{Q},
  \tmmathbf{P} \right))^2}{Q^2} \right) .  \label{eq:sqp}
\end{eqnarray}
This exact expression implies in particular
\begin{eqnarray}
  \frac{m}{Q} \mu_{\beta} \left( \tmmathbf{p} \mathbf{}_{\perp \tmmathbf{Q}}
  \text{$+ \frac{m}{m_{\ast}} \frac{\tmmathbf{Q}}{2} + \frac{m}{M}
  \tmmathbf{P}_{\| \tmmathbf{Q}}$} \right) & = & \mu_{\beta}^{\left( 2 d
  \right)} \left( \tmmathbf{p}_{\bot \tmmathbf{Q}} \right) S_{\text{MB}}
  \left( \tmmathbf{Q}, \tmmathbf{P}_{\| \tmmathbf{Q}} \right), 
  \label{eq:equiv}
\end{eqnarray}
where we stress explicitly that the function $S_{\text{MB}}$ depends on
$\tmmathbf{P}$ only through $\tmmathbf{P}_{\| \tmmathbf{Q}}$. Using
Eq.~(\ref{eq:equiv}) one immediately finds that Eq.~(\ref{eq:cdsf}) and
Eq.~(\ref{eq:clbe}) are two equivalent forms of the same equation.

\subsubsection{Generalization to the quantum case}

The particular form~(\ref{eq:cdsf}) of the classical linear Boltzmann equation
suggests an analogous way of writing the quantum linear Boltzmann equation
{\cite{Altenmuller1997a,Vacchini2001a,Vacchini2001b,Gocke2007a,Gocke2008a}},
\begin{eqnarray}
  \mathcal{L} \rho & = & \frac{n_{\tmop{gas}} }{m^2_{\ast}} \bigintlim \mathd
  \tmmathbf{Q} \int_{\tmmathbf{Q}^{\bot}} \mathd \tmmathbf{k}_{\bot} \,
  \mu_{\beta}^{\left( 2 d \right)} \left( \tmmathbf{k}_{\bot} \right) 
  \nonumber\\
  &  & \times \left[ f \left( \tmop{rel} \left( \tmmathbf{k}_{\bot},
  \mathsf{P}_{\bot \tmmathbf{Q}} \right) - \frac{\tmmathbf{Q}}{2}, \tmop{rel}
  \left( \tmmathbf{k}_{\bot}, \mathsf{P}_{\bot \tmmathbf{Q}} \right) +
  \frac{\tmmathbf{Q}}{2} \right) \mathe^{i\tmmathbf{Q} \cdot \mathsf{X} /
  \hbar} \sqrt{S_{\text{MB}} \left( \tmmathbf{Q}, \mathsf{P} \right)} \rho
  \right. \nonumber\\
  &  & \times \sqrt{S_{\text{MB}} \left( \tmmathbf{Q}, \mathsf{P} \right)}
  \mathe^{- i\tmmathbf{Q} \cdot \mathsf{X} / \hbar} f^{\dag} \left( \tmop{rel}
  \left( \tmmathbf{k}_{\bot}, \mathsf{P}_{\bot \tmmathbf{Q}} \right) -
  \frac{\tmmathbf{Q}}{2}, \tmop{rel} \left( \tmmathbf{k}_{\bot},
  \mathsf{P}_{\bot \tmmathbf{Q}} \right) + \frac{\tmmathbf{Q}}{2} \right)
  \nonumber\\
  &  & \left. - \frac{1}{2} \left\{ S_{\text{MB}} \left( \tmmathbf{Q},
  \mathsf{P} \right) \left| f \left( \tmop{rel} \left( \tmmathbf{k}_{\bot},
  \mathsf{P}_{\bot \tmmathbf{Q}} \right) - \frac{\tmmathbf{Q}}{2}, \tmop{rel}
  \left( \tmmathbf{k}_{\bot}, \mathsf{P}_{\bot \tmmathbf{Q}} \right) +
  \frac{\tmmathbf{Q}}{2} \right) \right|^2, \rho \right\} \right] . 
  \label{eq:qdsf}
\end{eqnarray}
This is exactly equivalent to Eq.~(\ref{eq:qlbe}) for a gas of
Maxwell-Boltzmann particles, described by Eq.~(\ref{eq:muMB}), as can easily
be seen using Eq.~(\ref{eq:equiv}). Note that the square root in
Eq.~(\ref{eq:qdsf}) is well defined since $S_{\text{MB}}$ is positive, which
is a general property of the dynamic structure factor, as discussed in the
following section.

\subsection{Properties of the dynamic structure factor}\label{sec:propdsf}

\subsubsection{Expression in terms of the density correlation function}

Let us consider a quantum many-body system, e.g. the gas of free particles
considered in the previous paragraph, described in second quantization by a
field operator $\hat{\psi} \left( \tmmathbf{x} \right)$, satisfying canonical
commutation or anti-commutation rules. One can then consider the Fourier
transformation of the operator density $\hat{n} \left( \tmmathbf{x} \right) =
\hat{\psi}^{^{\dag}} \left( \tmmathbf{x} \right) \hat{\psi} \left(
\tmmathbf{x} \right)$ given by
\begin{eqnarray}
  \hat{\rho}_{\tmmathbf{Q}} & = & \int \mathd \tmmathbf{x} \, \bignone
  \mathe^{- i \mathsf{\tmmathbf{Q} \cdot \tmmathbf{x}/ \hbar}}  \hat{n} \left(
  \tmmathbf{x} \right) .  \label{eq:r}
\end{eqnarray}
The associated spectral function
\begin{eqnarray}
  S \left( \tmmathbf{Q}, E \right) & = & \frac{1}{2 \pi \hbar}  \frac{1}{N}
  \int \mathd t \, \hspace{0.25em} \mathe^{iEt / \hbar} \langle
  \hat{\rho}_{\tmmathbf{Q}}^{^{\dag}} \hat{\rho}_{\tmmathbf{Q}} (t) \rangle, 
  \label{eq:rr}
\end{eqnarray}
with $N$ the total number of particles, is known as {\tmem{dynamic structure
factor}} {\cite{Lovesey1984,Schwabl2003,Pitaevskii2003}}, though other names
such as ``spectral density function'', ``scattering function'' and
``scattering law'' are also used in the literature. It can be expressed in
general as the Fourier transform with respect to energy and momentum transfer
of the time dependent density autocorrelation function
\begin{eqnarray}
  G (\tmmathbf{x}, t) & = & \frac{1}{N} \int \mathd \tmmathbf{y} \, \langle
  \hat{n} \left( \tmmathbf{y} \right) \bignone \hat{n} \left(
  \tmmathbf{x}+\tmmathbf{y}, t \right) \rangle  \label{eq:G}
\end{eqnarray}
as
\begin{eqnarray}
  S \left( \tmmathbf{Q}, E \right) & = & \frac{1}{2 \pi \hbar} \int \mathd t
  \int \mathd \tmmathbf{x} \, \bignone \hspace{0.25em} e^{i \left( Et
  -\tmmathbf{Q} \cdot \tmmathbf{x} \right) / \hbar} G (\tmmathbf{x}, t) . 
  \label{eq:FT}
\end{eqnarray}
By evaluating the dynamic structure factor for the case of a system of free
distinguishable particles one obtains, recalling Eq.~(\ref{eq:etransfer}) for
the energy transfer,
\begin{eqnarray}
  S \left( \mathbf{\tmmathbf{Q}}, E \left( \tmmathbf{Q}, \tmmathbf{P} \right)
  \right) & = & \frac{1}{n_{\tmop{gas}}} \int \frac{\mathd \tmmathbf{p}
  \bignone}{\left( 2 \pi \hbar \right)^3} \langle \hat{n}_{\tmmathbf{p}}
  \rangle \delta \left( E \left( \tmmathbf{Q}, \tmmathbf{P} \right) + \frac{(
  \tmmathbf{p} - \tmmathbf{Q})^2}{2 m} - \frac{p^2}{2 m} \right) . 
  \label{eq:MB}
\end{eqnarray}
Here $\langle \hat{n}_{\tmmathbf{p}} \rangle$ denotes the expectation value of
$\hat{n}_{\tmmathbf{p}}$, i.e. the mean number of particles with momentum
$\tmmathbf{p}$. If we consider a Maxwell-Boltzmann distribution, such that
\begin{eqnarray}
  \langle \hat{n}_{\tmmathbf{p}} \rangle \rightarrow n_{\tmop{gas}} \left( 2
  \pi \hbar \right)^3 \mu_{\beta} \left( \tmmathbf{p} \right), &  & 
  \label{eq:q5}
\end{eqnarray}
one comes back to the expression introduced in Eq.~(\ref{eq:sqp}) (see e.g.
{\cite{Vacchini2001b,Lanz2002a}}).

It thus appears that the dynamic structure factor $S_{\tmop{MB}}$, which can
be used to express the scattering kernel of the linear Boltzmann equation
(\ref{eq:qdsf}), has an important and transparent physical meaning. It is the
Fourier transform of the correlation function of the density fluctuations in
the medium, reflecting the molecular nature of the gas, which determines the
collisional interaction between test particle and gas particles. It thus
embodies in a suitable correlation function the original intuition by Einstein
which led to the correct understanding of Brownian motion: the dynamics of the
test particle is driven by collisions taking place due to the discrete nature
of the medium, whose density fluctuations provide the coupling mechanism.

\subsubsection{Connection to the laboratory-frame scattering cross-section}

The dynamic structure factor is particularly useful in describing the
scattering of microscopic probes off macroscopic samples, since it allows one
to express the energy dependent cross-section in terms of a correlation
function of the many-body system {\cite{Brenig1989,Schwabl2003}}. It was in
fact first introduced by van Hove in order to describe neutron scattering
{\cite{VanHove1954}}.

Using Fermi's golden rule and the Born approximation one can show that the
laboratory-frame cross-section $\Sigma$ for the scattering of a test particle
off a many-body system is characterized by the dynamic structure factor $S
\left( \tmmathbf{Q}, E \right)$ of the many-body system {\cite{Sears1989a}}
\begin{eqnarray}
  \frac{\mathd^2 \Sigma}{\mathd \Omega_{P'} \mathd E_{P'}} & = & \frac{M^2}{4
  \pi^2 \hbar^4} ^{} \frac{P'}{P} | \tilde{V} \left( \tmmathbf{Q} \right) |^2
  S (\tmmathbf{Q}, E) = \frac{M^2 }{m_{\ast}^2} \frac{P'}{P} \left| f_B \left(
  \tmmathbf{Q} \right) \right|^2 S (\tmmathbf{Q}, E) .  \label{eq:scs}
\end{eqnarray}
Here a scattering event is considered in which a particle of mass $M$ changes
its momentum from $\tmmathbf{P}$ to $\tmmathbf{P}'
=\tmmathbf{P}+\tmmathbf{Q}$, absorbing an energy $E$ according to
(\ref{eq:etransfer}), and $ \tilde{V} (\tmmathbf{Q})$ is the Fourier transform
of the two-body interaction potential
\begin{eqnarray}
  \tilde{V} (\tmmathbf{Q}) & = & \int \mathd \tmmathbf{x} \, \mathe^{i
  \mathsf{\tmmathbf{Q} \cdot \tmmathbf{x}/ \hbar}} V \left( \tmmathbf{x}
  \right),  \label{eq:q6}
\end{eqnarray}
related through Eq.~(\ref{eq:fBorn}) to the scattering amplitude in Born
approximation.

More generally, provided multiple scattering effects can be neglected, one has
{\cite{Brenig1989}}
\begin{eqnarray}
  \frac{\mathd^2 \Sigma}{\mathd \Omega_{P'} \mathd E_{P'}}  & = & \frac{\mathd
  \Sigma}{\mathd \Omega_{P'} } S (\tmmathbf{Q}, E),  \label{eq:q7}
\end{eqnarray}
where $\mathd \Sigma / \mathd \Omega_{P'}$ describes the single scattering
event in the laboratory frame.

\subsubsection{Detailed balance and the stationary solution}

For a strongly interacting many-body quantum system the dynamic structure
factor is no longer analytically tractable as in Eq.~(\ref{eq:sqp}), but it
can only be measured in suitable scattering experiments. However, due to its
very definition the dynamic structure factor acquires many interesting general
properties. First of all, it is always positive as a consequence of Bochner's
theorem {\cite{Feller1971}}, because the density autocorrelation function
Eq.~(\ref{eq:G}) is a positive definite function. This property is to be
expected on physical grounds, since the dynamic structure factor is directly
proportional to a scattering cross-section according to Eq.~(\ref{eq:scs}).
The dynamic structure factor further obeys the so-called {\tmem{detailed
balance condition}}
\begin{eqnarray}
  S \left( \tmmathbf{Q}, E \right) & = & \mathe^{- \beta E} S \left(
  -\tmmathbf{Q}, - E \right),  \label{eq:dbc}
\end{eqnarray}
which is the crucial property for ensuring the existence of a stationary
solution of both the classical and the quantum linear Boltzmann equation. In
order to confirm this property one has to consider Eq.~(\ref{eq:rr}) and to
observe that $\hat{\rho}^{\dag}_{\tmmathbf{Q}} = \hat{\rho}_{-\tmmathbf{Q}
\bignone}$, as follows directly from the definition Eq.~(\ref{eq:r}), and to
exploit the fact that the expectation value is obtained with respect to the
equilibrium state of the many-body system $\rho_{\tmop{EQ}} \propto \mathe^{-
\beta H}$, where $H$ is the full Hamiltonian. The property Eq.~(\ref{eq:dbc})
therefore holds generally, for any many-body system in a thermal equilibrium
state.

In our case, where the energy transfer can be expressed by
Eq.~(\ref{eq:etransfer}), the detailed balance condition can also be written
as
\begin{eqnarray}
  S \left( \tmmathbf{Q}, \tmmathbf{P} \right) & = & \mathe^{- \beta E \left(
  \tmmathbf{Q}, \tmmathbf{P} \right)} S \left( -\tmmathbf{Q},
  \tmmathbf{P}+\tmmathbf{Q} \right) .  \label{eq:dbcp}
\end{eqnarray}
It is now simple to prove that Eq.~(\ref{eq:dbcp}) ensures the existence of a
stationary solution of the canonical form $\rho_{\tmop{EQ}} (\tmmathbf{P})
\propto \exp \left( - \beta P^2 / \left( 2 M \right) \right)$ for both the
classical and the quantum linear Boltzmann equation. For $f \left(
\tmmathbf{P} \right) \rightarrow \rho_{\tmop{EQ}} (\tmmathbf{P})$ the
expression
\begin{eqnarray}
  S_{\text{MB}} \left( \tmmathbf{Q}, \tmmathbf{P}-\tmmathbf{Q} \right)
  \rho_{\tmop{EQ}} \left( \tmmathbf{P}-\tmmathbf{Q} \right) - S_{\text{MB}}
  \left( \tmmathbf{Q}, \tmmathbf{P} \right) \rho_{\tmop{EQ}} \left(
  \tmmathbf{P} \right) &  &  \label{eq:q8}
\end{eqnarray}
is odd in $\tmmathbf{Q}$, such that Eq.~(\ref{eq:cdsf}) implies
$\text{$\partial^{\tmop{coll}}_t \rho_{\tmop{EQ}}$} (\tmmathbf{P}) = 0$.
Similarly, in the quantum case the expression Eq.~(\ref{eq:qdsf}) for an
operator $\rho = \nu ( \mathsf{P})$, which is only a function of the momentum
operator $\mathsf{P}$, simply reads
\begin{eqnarray}
  \mathcal{L} \nu ( \mathsf{P}) & = & \frac{n_{\tmop{gas}} }{m^2_{\ast}}
  \bigintlim \mathd \tmmathbf{Q} \int_{\tmmathbf{Q}^{\bot}} \mathd
  \tmmathbf{k}_{\bot} \, \mu_{\beta}^{\left( 2 d \right)} \left(
  \tmmathbf{k}_{\bot} \right)  \left| f \left( \tmop{rel} \left(
  \tmmathbf{k}_{\bot}, \mathsf{P}_{\bot \tmmathbf{Q}} \right) -
  \frac{\tmmathbf{Q}}{2}, \tmop{rel} \left( \tmmathbf{k}_{\bot},
  \mathsf{P}_{\bot \tmmathbf{Q}} \right) + \frac{\tmmathbf{Q}}{2} \right)
  \right|^2  \label{eq:q9}\\
  &  & \times \left[ S_{\text{MB}} \left( \tmmathbf{Q}, \mathsf{P}
  -\tmmathbf{Q} \right) \nu ( \mathsf{P} -\tmmathbf{Q}) - S_{\text{MB}} \left(
  \tmmathbf{Q}, \mathsf{P} \right) \nu ( \mathsf{P}) \right] . \nonumber
\end{eqnarray}
For the case $\nu ( \mathsf{P}) \rightarrow \nu_{\tmop{EQ}} ( \mathsf{P})$ the
condition Eq.~(\ref{eq:dbcp}) again implies that the integrand is an odd
function of $\tmmathbf{Q}$, such that $\mathcal{L} \nu_{\tmop{EQ}} (
\mathsf{P}) = 0$. Note that the explicit expression Eq.~(\ref{eq:sqp}) of
$S_{\text{MB}}$ is not relevant for the proof, which relies only on the
property Eq.~(\ref{eq:dbcp}), valid for a generic medium.

As expected, also in the quantum case the existence and the form of the
stationary solution do not depend on the scattering amplitude, even though it
appears operator-valued. This strong correspondence with the classical case is
due to the fact that the covariance under translations implies that the
generator $\mathcal{L}$ leaves the algebra generated by the momentum operator
invariant, as discussed in Sect.~\ref{sec:emrelax}.

\subsubsection{Fluctuation-dissipation relationship}

Let us now discuss the origin of the detailed balance property, beyond its
direct calculation based on the definition of dynamic structure factor in
Eq.~(\ref{eq:rr}). To this end, we have to touch upon the
fluctuation-dissipation theorem.

The dynamic structure factor can be directly related to the dynamic
susceptibility studied in linear response theory. Denoting the imaginary part
of the dynamic response function as $\chi'' \left( \tmmathbf{Q}, E \right)$,
one finds {\cite{Pitaevskii2003}}
\begin{eqnarray}
  S \left( \tmmathbf{Q}, E \right) & = & \frac{1}{\pi} \frac{1}{1 - e^{\beta
  E}} \chi'' \left( \tmmathbf{Q}, E \right) .  \label{eq:drf}
\end{eqnarray}
Since the imaginary part of the dynamic susceptibility, which describes the
dissipative part of the response, is by definition an odd function of energy,
this relationship ensures that the dynamic structure factor obeys the detailed
balance condition Eq.~(\ref{eq:dbc}). We recall that, contrary to the usual
perspective in linear response theory, we are here concerned with the reduced
dynamics of the test particle, thus taking the momentum and energy transferred
to the particle as positive.

Introducing the real correlation function of the gas medium
\begin{eqnarray}
  &  & \phi^+ \left( \tmmathbf{Q}, t \right) = \frac{1}{\hbar N} \langle
  \left\{ \hat{\rho}_{\tmmathbf{Q}}^{} \left( t \right),
  \hat{\rho}_{\tmmathbf{Q}}^{^{\dag}} \right\} \rangle,  \label{eq:51}
\end{eqnarray}
for the fluctuations of the operator $\hat{\rho}_{\tmmathbf{Q}}$ given by
Eq.~(\ref{eq:r}), the fluctuation-dissipation relationship can be formulated
in terms of the dynamic structure factor as follows
{\cite{Lovesey1984,Petruccione2005a}}
\begin{eqnarray}
  \phi^+ \left( \tmmathbf{Q}, t \right) & = & - \frac{1}{\hbar} \int^{+
  \infty}_{- \infty} \mathd E \hspace{0.25em} \exp \left( i \frac{E}{\hbar} t
  \right) \coth \left(  \frac{\beta}{2} E \right) \left( 1 - e^{\beta E}
  \right) S \left( \tmmathbf{Q}, E \right) .  \label{eq:52}
\end{eqnarray}
This is a consequence of the definition of the dynamic structure factor
Eq.~(\ref{eq:rr}) as well as Eq.~(\ref{eq:51}). When expressed in terms of the
imaginary part of the dynamic susceptibility according to Eq.~(\ref{eq:drf})
this identity gives back the usual formulation of the fluctuation-dissipation
theorem, once again taking as positive the energy when transferred to the test
particle.

In this spirit, the expression~(\ref{eq:scs}) for the laboratory-frame
cross-section can be viewed as a formulation of the fluctuation-dissipation
relationship. The test particle experiences dissipation due to energy and
momentum transfer processes described by the cross-section on the left-hand
side of Eq.~(\ref{eq:scs}), which is related on the right-hand side to the
equilibrium density fluctuations, characterized by the dynamic structure
factor.

\subsection{Extension to different reservoirs}

In the previous sections, we considered the explicit expression of the dynamic
structure factor for an ideal gas described by the Maxwell-Boltzmann
distribution. The latter implies that the gas particles are either
distinguishable or that the gas temperature is sufficiently high such that
effects due to their indistinguishability can be neglected. We note that the
result~(\ref{eq:sqp}) is the same in the classical and the quantum calculation
{\cite{Lovesey1984}}.

For the case of an ideal gas, one can calculate the exact dynamic structure
factor also by taking into account the effect of quantum statistics. Starting
from the density autocorrelation function Eq.~(\ref{eq:G}) for a gas of
identical bosons or fermions it takes the form
\begin{eqnarray}
  S_{\tmop{BF}} \left( \mathbf{\tmmathbf{Q}}, E \left( \tmmathbf{Q},
  \tmmathbf{P} \right) \right) & = & \frac{1}{n_{\tmop{gas}}} \int
  \frac{\mathd \tmmathbf{p} \bignone}{\left( 2 \pi \hbar \right)^3} \langle
  \hat{n}_{\tmmathbf{p}} \rangle_{\tmop{BF}} \left( 1 \pm \langle
  \hat{n}_{\tmmathbf{p}} \rangle_{\tmop{BF}} \right) \delta \left( E \left(
  \tmmathbf{Q}, \tmmathbf{P} \right) + \frac{( \tmmathbf{p} -
  \tmmathbf{Q})^2}{2 m} - \frac{p^2}{2 m} \right),  \label{eq:FD}
\end{eqnarray}
which should be compared to Eq.~(\ref{eq:MB}). Here the upper sign stands for
Bose-Einstein and the lower one for Fermi-Dirac statistics, respectively. The
corresponding distribution functions
\begin{eqnarray}
  \langle \hat{n}_{\tmmathbf{p}} \rangle_{\tmop{BF}} & = & \frac{1}{z^{- 1}
  \mathe^{\beta \frac{\tmmathbf{p}^2}{2 M}} \mp 1}  \label{eq:q0}
\end{eqnarray}
are characterized by the fugacity $z = \exp \left( \beta \mu \right)$. The
integral~(\ref{eq:FD}) can be done, leading to {\cite{Vacchini2001b}}
\begin{eqnarray}
  S_{\tmop{BF}} \left( \mathbf{\tmmathbf{Q}}, E \left( \tmmathbf{Q},
  \tmmathbf{P} \right) \right) & = & \frac{1}{n_{\tmop{gas}}}  \frac{2 \pi
  m^2}{\beta \left( 2 \pi \hbar \right)^3}  \frac{1}{Q}  \frac{\pm 1}{1 -
  \mathe^{\beta E \left( \tmmathbf{Q}, \tmmathbf{P} \right)}} \log \left[
  \frac{1 \mp z \exp \left( - \frac{\beta}{8 m} \frac{(Q^2 + 2 mE \left(
  \tmmathbf{Q}, \tmmathbf{P} \right))^2}{Q^2} \right)}{1 \mp z \exp \left( -
  \frac{\beta}{8 m} \frac{(Q^2 - 2 mE \left( \tmmathbf{Q}, \tmmathbf{P}
  \right))^2}{Q^2} \right)} \right] .  \label{eq:freeFD}
\end{eqnarray}
In the limit of small fugacity $z \rightarrow n_{\tmop{gas}} \left( 2 \pi
\hbar^2 \beta / m \right)^{3 / 2}$, corresponding to negligible degeneracy,
the latter expression goes back to Eq.~(\ref{eq:MB}). As one can also check
directly, $S_{\tmop{BF}} \left( \mathbf{\tmmathbf{Q}}, E \right)$ obeys the
detailed balance condition~(\ref{eq:dbc}) for both statistics.

Derivations of the quantum linear Boltzmann equation in the weak-coupling
limit, in which a thermal reservoir of identical quantum particles is
considered, have been obtained in {\cite{Altenmuller1997a}} for the case of a
bosonic reservoir, and more generally in {\cite{Vacchini2001a}}, leading to
\begin{eqnarray}
  \mathcal{L} \rho & = & \frac{n_{\tmop{gas}} }{m^2_{\ast}} \bigintlim \mathd
  \tmmathbf{Q} \left| f_B \left( \tmmathbf{Q} \right) \right|^2 \left[
  \mathe^{i\tmmathbf{Q} \cdot \mathsf{X} / \hbar} \sqrt{S_{\tmop{BF}} \left(
  \tmmathbf{Q}, \mathsf{P} \right)} \rho \sqrt{S_{\tmop{BF}} \left(
  \tmmathbf{Q}, \mathsf{P} \right)} \mathe^{- i\tmmathbf{Q} \cdot \mathsf{X} /
  \hbar} - \frac{1}{2} \left\{ S_{\tmop{BF}} \left( \tmmathbf{Q}, \mathsf{P}
  \right), \rho \right\} \right] .  \label{eq:qdsffd}
\end{eqnarray}
This result is the quantum counterpart of a classical linear Boltzmann
equation, which is modified in order to account for statistical corrections.
The latter can be obtained upon setting $\mu_{\tmop{BF}} \left( \tmmathbf{p}
\right) = \langle \hat{n}_{\tmmathbf{p}} \rangle_{\tmop{BF}} / \left(
n_{\tmop{gas}} \left( 2 \pi \hbar \right)^3 \right)$ and replacing
$\mu_{\beta} \left( \tmmathbf{p} \right)$ in Eq.~(\ref{eq:clbeMB}) by
$\mu_{\tmop{BF}} \left( \tmmathbf{p} \right) (1 \pm n_{\tmop{gas}} \left( 2
\pi \hbar \right)^3 \mu_{\tmop{BF}} \left( \tmmathbf{p}' \right))$, which
leads to
\begin{eqnarray}
  \text{$\partial^{\tmop{coll}}_t f$} (\tmmathbf{P}) & = &
  \frac{n_{\tmop{gas}}}{m^2_{\ast}} \bigintlim \mathd \tmmathbf{P}' \mathd
  \tmmathbf{p}' \mathd \tmmathbf{p} \, \delta \left( \frac{P'^2}{2 M} +
  \frac{p'^2}{2 m} - \frac{P^2}{2 M} - \frac{p^2}{2 m} \right) \delta^3 \left(
  \tmmathbf{P}' +\tmmathbf{p}' -\tmmathbf{P}-\tmmathbf{p} \right) \text{}
  \nonumber\\
  &  & \times \sigma \left( \tmop{rel} \left( \tmmathbf{p}, \tmmathbf{P}
  \right), \tmop{rel} \left( \tmmathbf{p}', \tmmathbf{P}' \right) \right)
  \left[ \mu_{\tmop{BF}} \left( \tmmathbf{p}' \right) \left( 1 \pm
  n_{\tmop{gas}} \left( 2 \pi \hbar \right)^3 \mu_{\tmop{BF}} \left(
  \tmmathbf{p} \right) \right) f \left( \tmmathbf{P}' \right) \right.
  \nonumber\\
  &  & \left. - \mu_{\tmop{BF}} \left( \tmmathbf{p} \right) \left( 1 \pm
  n_{\tmop{gas}} \left( 2 \pi \hbar \right)^3 \mu_{\tmop{BF}} \left(
  \tmmathbf{p}' \right) \right) f \left( \tmmathbf{P} \right) \right] . 
  \label{eq:clbeFD}
\end{eqnarray}
The statistical corrections in this modified classical linear Boltzmann
equation are the analog of the modifications introduced by Uehling and
Uhlenbeck to the original classical equation {\cite{Uehling1933}}. They
express the fact that quantum correlations in the gas can be relevant even in
the absence of any self-interactions, just because the occupancy of the
different momentum states may enhance or suppress the related scattering rates
due to statistics.

We mention that a simplified form of the master equation (\ref{eq:qdsf}) was
used in {\cite{Altenmuller1997a}} to study how the effects of quantum
statistics in a Boltzmann gas affect the reduction of visibility in the
interference fringes due to collisional decoherence.

Finally, it is interesting to note that the comparison between
Eq.~(\ref{eq:qdsffd}) and Eq.~(\ref{eq:qdsf}) suggests a natural heuristic
approach to extend the quantum linear Boltzmann equation to real, interacting
gases. In place of the dynamic structure factor of the ideal gas one just uses
the relevant expression of the interacting gas. Even if an exact analytic
evaluation of the dynamic structure factor is in general impossible for a
truly interacting many-body system, one may rely either on experimental data
from scattering experiments or on a phenomenological ansatz, constrained by
the requirement of the detailed balance condition. This strategy is
straightforward for the weak-coupling form~(\ref{eq:qdsffd}) of the master
equation. However, complications arise if the interaction with the gas is to
be treated beyond the Born approximation, as in Eq.~(\ref{eq:qdsf}), once the
momentum distribution in the gas no longer factorizes as in
Eq.~(\ref{eq:fatt}).

\subsection{Inclusion of internal degrees of freedom}\label{sec:qbbe}

The quantum linear Boltzmann equation provides a realistic description for the
dynamics of a massive test particle traveling through a dilute gas of free
particles, and its predictions in the quantum framework have been tested in
recent experiments relevant for atom interferometry
{\cite{Schmiedmayer1995a,Hornberger2003a,Uys2005a,Jacquey2007a}}. It is
however natural to consider extensions of the equation, allowing one to cope
with a wider experimental scenario.

In the first instance both test particle and gas particles have been
considered as pointlike particles, whose only relevant degrees of freedom are
those connected to the center of mass. This is obviously not necessarily the
case and one should consider a proper quantum treatment of both internal and
center of mass degrees of freedom.

A hybrid equation, describing the internal degrees of freedom within quantum
mechanics, and the center of mass degrees of freedom as classical variables,
was recently introduced under the name ``Bloch-Boltzmann equation''
{\cite{Alicki2003a,Kryszewski2006a}}. It combines the master equation for the
gas-induced incoherent dynamics of a $N$-level system of internal states with
the classical linear Boltzmann equation. This Bloch-Boltzmann equation thus
accounts for the possibility of inelastic collisions connecting the various
possible scattering channels, which can be of relevance e.g. for spectroscopic
experiments {\cite{Rautian1991a}}, transport phenomena, or more refined
interferometric experiments testing decoherence.

A microscopic derivation of the effect of the collisional dynamics on the
internal degrees of freedom of a very massive test particle at rest was
considered in {\cite{Dumcke1985a,Hornberger2007b}}. A fully quantum
description of both internal and center of mass degrees of freedom of a
quantum test particle interacting through collisions with a gas has been
considered in {\cite{Vacchini2008a}}, leading to an equation called ``quantum
Bloch-Boltzmann equation''. From this equation all the others can be obtained
as limiting situations in which one of the two kinds of degrees of freedom is
traced over or treated classically. In this framework, a rich phenomenology of
dynamical situations can be considered, including non-Markovian effects
described by means of the so-called generalized Lindblad structure
{\cite{Budini2005a,Budini2006a,Breuer2006a,Breuer2007a}}.

The ``quantum Bloch-Boltzmann equation'' can be written in Lindblad form in
close analogy to the linear Boltzmann equation~(\ref{eq:lvonn}). However, the
statistical operator $\rho$ now acts on the Hilbert space describing both
internal and center of mass degrees of freedom. The correction to the free
Hamiltonian, which is the analog of Eq.~(\ref{eq:effectivebis}), now reads
\begin{eqnarray}
  \mathsf{H}_{\text{n}} & = & - 2 \pi \hbar^2 \frac{n_{\tmop{gas}}}{m_{\ast}} 
  \sum_{\tmscript{\begin{array}{c}
    ij\\
    \mathcal{E}_{ij} = 0
  \end{array}}} \int \mathd \tmmathbf{p} \, \mu \left( \tmmathbf{p} \right)
  \tmop{Re} \left[ f_{ij} \left( \tmop{rel} \left( \tmmathbf{p}, \mathsf{P}
  \right), \tmop{rel} \left( \tmmathbf{p}, \mathsf{P} \right) \right) \right]
  \otimes \mathsf{E}_{ij},  \label{eq:qbbe1}
\end{eqnarray}
where $\mathsf{E}_{ij} = |i \rangle \langle j|$ provides a basis of operators
in the space of the internal degrees of freedom, and $\mathcal{E}_{ij} = \hbar
\omega_i - \hbar \omega_j$ are the possible transition energies. The
multi-channel scattering amplitudes, denoted by $f_{ij} \left( \tmmathbf{p}_f,
\tmmathbf{p}_i \right) \equiv f \left( \tmmathbf{p}_i, j \rightarrow
\tmmathbf{p}_f, i \right)$, describe a transition from an {\tmem{in}}-state
with labels $\tmmathbf{p}_i, j$ to an {\tmem{out}}-state with labels
$\tmmathbf{p}_f, i$.

The incoherent part of the master equation can be written analogous to
Eq.~(\ref{eq:qlbe}). It takes the form
\begin{eqnarray}
  \mathcal{L} \rho & = & \sum_{\mathcal{E}} \bigintlim \mathd \tmmathbf{Q}
  \int_{\tmmathbf{Q}^{\bot}} \mathd \tmmathbf{k}_{\bot} \left[
  \mathe^{i\tmmathbf{Q} \cdot \mathsf{X} / \hbar} L \left(
  \tmmathbf{k}_{\bot}, \mathsf{P} ; \tmmathbf{Q}, \mathcal{E} \right) \rho
  L^{\dag} \left( \tmmathbf{k}_{\bot}, \mathsf{P} ; \tmmathbf{Q}, \mathcal{E}
  \right) \mathe^{- i\tmmathbf{Q} \cdot \mathsf{X} / \hbar} \right.
  \nonumber\\
  &  & \left. - \frac{1}{2} \left\{ \rho, L^{\dag} \left(
  \tmmathbf{k}_{\bot}, \mathsf{P} ; \tmmathbf{Q}, \mathcal{E} \right) L \left(
  \tmmathbf{k}_{\bot}, \mathsf{P} ; \tmmathbf{Q}, \mathcal{E} \right) \right\}
  \right],  \label{eq:qbbe}
\end{eqnarray}
where, apart from the momentum integrations, there is an additional sum over a
discrete index $\mathcal{E}$, labeling the energy transfer to the internal
degrees of freedom in case of inelastic scattering.

The function $L$, which describes the effect of the single collisions, is
given by
\begin{eqnarray}
  L \left( \tmmathbf{p}, \tmmathbf{P}; \tmmathbf{Q}, \mathcal{E} \right) & = &
  \sum_{\tmscript{\begin{array}{c}
    ij\\
    \mathcal{E}_{ij} = \mathcal{E}
  \end{array}}} \bignone f_{ij} \left( \tmop{rel} \left(
  \mathbf{\tmmathbf{p}}_{\perp \tmmathbf{Q}}, \tmmathbf{P}_{\perp
  \tmmathbf{Q}} \right) - \frac{\tmmathbf{Q}}{2} + \frac{\mathcal{E}_{ij}}{Q^2
  / m_{\ast}} \tmmathbf{Q}, \tmop{rel} \left( \mathbf{\tmmathbf{p}}_{\perp
  \tmmathbf{Q}}, \tmmathbf{P}_{\perp \tmmathbf{Q}} \right) +
  \frac{\tmmathbf{Q}}{2} + \frac{\mathcal{E}_{ij}}{Q^2 / m_{\ast}}
  \tmmathbf{Q} \right) \nonumber\\
  &  & \times \sqrt{\frac{n_{\tmop{gas}} m}{m_{\ast}^2 Q} \mu \left(
  \tmmathbf{p} \mathbf{}_{\perp \tmmathbf{Q}} \text{$+ \frac{m}{m_{\ast}}
  \frac{\tmmathbf{Q}}{2} + \frac{m}{M} \tmmathbf{P}_{\| \tmmathbf{Q}}$} +
  \frac{\mathcal{E}_{ij}}{Q^2 / m} \tmmathbf{Q} \right)} \otimes
  \mathsf{E}_{ij},  \label{eq:lltris}
\end{eqnarray}
which should be compared to Eq.~(\ref{eq:L}). This implies that the
multi-channel scattering amplitudes $f_{ij}$ and the distribution function
$\mu$ of the gas momenta appear again operator-valued in (\ref{eq:qbbe}).

Equations~(\ref{eq:qbbe1})-(\ref{eq:lltris}) represent the natural
generalization obtained by combining the two master equations for the quantum
motion of a point particle and for the internal dynamics of a system at rest,
respectively. It should be emphasized, however, that a physically stringent
derivation of Eqs.~(\ref{eq:qbbe1})-(\ref{eq:lltris}) is still missing. It is
a subject of current research whether this can be achieved by means of the
monitoring approach described in {\cite{Hornberger2007b}}, which was also
instrumental in deriving the quantum linear Boltzmann equation.

\section{Conclusion}\label{sec:ceo}

In the present report we discussed a Lindblad master equation for the quantum
motion of a test particle in an ambient gas, arguing that it is the natural
quantum counterpart of the classical linear Boltzmann equation. This means
that the gas-induced phenomena described by the equation are accounted for not
just in a phenomenological sense, but are incorporated in a non-perturbative
fashion by means of the exact scattering amplitudes. This microscopically
realistic description of the individual scattering processes allows one to
predict experimentally observable properties such as dissipation and
decoherence rates directly from the microscopic interaction laws. At the same
time, it explains why the structure of the equation is more complicated than
most master equations used for the description of open quantum systems.

We presented a fairly straightforward, heuristic motivation for the particular
form of the equation, referring the reader to the literature for a more
stringent derivation {\cite{Hornberger2008a,Hornberger2006b}}. It is
remarkable that already two basic requirements serve to fix the master
equation almost completely, namely the translation-covariance of its Lindblad
structure and the compatibility with the classical linear Boltzmann equation
for quantum states which are indistinguishable from classical ones. The
remaining freedom in this heuristic approach can be settled by a plausibility
argument, which is corroborated by the fact that the various limits that can
be taken turn the equation into established master equations. In particular, a
generalized Caldeira-Leggett master equation of quantum Brownian motion is
obtained in the diffusive limit, whose Lindblad structure is as close as
possible to the classical Kramers equation with microscopically defined
friction and diffusion constants.

Some attention was devoted to analyzing the general structural features and
symmetry properties of the equation, as well as the predicted approach to an
equilibrium state. We saw that many of the relaxation and dissipation effects
are shared with those of the classical linear Boltzmann equation, while pure
quantum phenomena, such as collisional decoherence and the gas induced index
of refraction for matter waves, are genuine results of the quantum version. A
remarkable feature of the equation is that it accounts for the interplay of
all relevant gas-induced phenomena, from the short-time decoherence dynamics
to the approach towards equilibrium for asymptotically large times. At the
same time, it is clearly impossible to state the general solution in an
explicit form, even in the simplest case of s-wave scattering with a constant
scattering length. However, we saw that the Lindblad structure of the master
equation admits a stochastic unravelling of the master equation in terms of
quantum trajectories, which is particularly transparent in the momentum basis.

The validity of the quantum linear Boltzmann equation is expected to cease in
situations where the medium cannot be considered an ideal, homogeneous
Maxwell-Boltzmann gas, or where the scattering interaction is inelastic. We
discussed an alternative formulation of the master equation in terms of the
dynamic structure factor of the gas. Apart from providing physical insights,
it suggests a natural way of incorporating corrections due to quantum
degeneracies and self-interactions in the gas, at least in the framework of
the weak-coupling approximation. As for inelastic scattering, we described a
natural multi-channel scattering extension of the equation, capable to account
for the presence of internal degrees of freedom in the test particle.

A further important extension of the equation concerns the treatment of an
inhomogeneous background gas, e.g. due to an external potential. In view of
the corresponding classical equation, it seems natural to incorporate this by
means of an operator-valued gas density $n_{\tmop{gas}} \left( \mathsf{X}
\right)$, rendering the transition rate $M_{\tmop{in}}$ a function of the
position operator of the test particle. However, the non-commutativity of the
position and momentum operators complicates this task considerably.

Finally, the limit of a dense and strongly self-interacting gas turns the
background medium into a liquid. Clearly, the interactions with the test
particle are then no longer described by statistically independent
two-particle collisions, leading to a breakdown of the Markov assumption. In
spite of many years of research, the field of non-Markovian open quantum
dynamics is still in its infancy. However, it will certainly be of central
relevance to cope with such non-Markovian effects in order to describe a truly
quantum microscopic dynamics.

The long path leading to the present version of the quantum linear Boltzmann
equation suggests that most of these improvements are still a long way ahead,
and possibly will require a different mathematical framework. Nonetheless,
they will be of great importance for a quantitative assessment of quantum
transport and decoherence phenomena.

\section*{Acknowledgments}

The work was partially supported by the Italian MIUR under PRIN2005 (BV) and
by the Emmy Noether programme of Deutsche Forschungsgemeinschaft (KH). We
would like to thank A.~Barchielli, H.-P.~Breuer, L.~Lanz, and J.~E.~Sipe for
many useful discussions on the subject of this article.

\appendix\section{Classical formulation}

\subsection{The classical linear Boltzmann equation}\label{sec:a1}

The classical linear Boltzmann equation can be written in quite different
ways, which reflects its complicated structure and the many variables
appearing in it {\cite{Cercignani1975a}}. The most compact and perhaps
well-known form is given by
\begin{eqnarray}
  \text{$\partial^{\tmop{coll}}_t f$} (\tmmathbf{P}) & = & \bigintlim \mathd
  \tmmathbf{p} \int \bignone \mathd \Omega \, \sigma \left( \tmop{rel} \left(
  \tmmathbf{p}, \tmmathbf{P} \right), \tmop{rel} \left( \tmmathbf{p}',
  \tmmathbf{P}' \right) \right) j_{\tmop{rel}} \left( \tmmathbf{p},
  \tmmathbf{P} \right) \nonumber\\
  &  & \times \left[ \mu \left( \tmmathbf{p}' \right) f \left( \tmmathbf{P}'
  \right) - \mu \left( \tmmathbf{p} \right) f \left( \tmmathbf{P} \right)
  \right],  \label{eq:classica}
\end{eqnarray}
where $j_{\tmop{rel}} \left( \tmmathbf{p}, \tmmathbf{P} \right) \equiv
n_{\tmop{gas}} | \tmop{rel} \left( \tmmathbf{p}, \tmmathbf{P} \right) | /
m_{\ast}$ denotes the current density in the relative motion. In this
equation, the values of $\tmmathbf{p}'$ and $\tmmathbf{P}'$ are determined
implicitly by the conservation of momentum, $\tmmathbf{p}' +\tmmathbf{P}'
=\tmmathbf{p}+\tmmathbf{P}$, and energy, $\left| \tmop{rel} \left(
\tmmathbf{p}', \tmmathbf{P}' \right) | = \left| \tmop{rel} \left(
\tmmathbf{p}, \tmmathbf{P} \right) \right| \right.$, as well as by the angles
of rotation $\Omega$ which connect the relative momenta appearing in the
differential cross-section $\sigma$. Moreover, we have only considered the
collisional term.

However, it is much more convenient for our purposes to use a more explicit
expression. In particular, is permits to clearly demonstrate the connection
between the classical and the quantum version of the equation.

Let us start from the basic expression obtained by turning the original,
non-linear Boltzmann equation into the linear version. By replacing one of the
distribution functions $f \left( \tmmathbf{p} \right)$ appearing in the
bilinear collision term with the equilibrium distribution $\mu \left(
\tmmathbf{p} \right)$ of the gas particles, one gets an equation
\begin{eqnarray}
  \text{$\partial^{\tmop{coll}}_t f$} (\tmmathbf{P}) & = &
  \frac{n_{\tmop{gas}}}{m^2_{\ast}} \bigintlim \mathd \tmmathbf{P}' \int
  \mathd \tmmathbf{p}' \int \bignone \mathd \tmmathbf{p} \, \delta \left(
  \frac{P'^2}{2 M} + \frac{p'^2}{2 m} - \frac{P^2}{2 M} - \frac{p^2}{2 m}
  \right) \delta^3 \left( \tmmathbf{P}' +\tmmathbf{p}'
  -\tmmathbf{P}-\tmmathbf{p} \right) \nonumber\\
  &  & \times \sigma \left( \tmop{rel} \left( \tmmathbf{p}, \tmmathbf{P}
  \right), \tmop{rel} \left( \tmmathbf{p}', \tmmathbf{P}' \right) \right)
  \left[ \mu \left( \tmmathbf{p}' \right) f \left( \tmmathbf{P}' \right) - \mu
  \left( \tmmathbf{p} \right) f \left( \tmmathbf{P} \right) \right], 
  \label{eq:clbeMBapp}
\end{eqnarray}
which can be brought into the form Eq.~(\ref{eq:classica}). Here the dynamics
determined by collisions between test particle and gas particles is put into
evidence. The collisions are characterized by a differential scattering
cross-section $\sigma \left( \tmmathbf{p}_f, \tmmathbf{p}_i \right)$, together
with the constraints of energy and momentum conservation expressed by the
$\delta$-functions.

Exploiting the momentum conservation and introducing the momentum transfer
$\tmmathbf{Q}= \tmmathbf{P}' -\tmmathbf{P}$, which corresponds to the momentum
gained by the test particle in a collision, we obtain
\begin{eqnarray}
  \text{$\partial^{\tmop{coll}}_t f$} (\tmmathbf{P}) & = &
  \frac{n_{\tmop{gas}}}{m^2_{\ast}} \bigintlim \mathd \tmmathbf{Q} \int
  \bignone \mathd \tmmathbf{p} \, \delta \left( \frac{\left(
  \tmmathbf{P}+\tmmathbf{Q} \right)^2}{2 M} +
  \frac{(\tmmathbf{p}-\tmmathbf{Q})^2}{2 m} - \frac{P^2}{2 M} - \frac{p^2}{2
  m} \right) \nonumber\\
  &  & \times \sigma \left( \tmop{rel} \left( \tmmathbf{p}, \tmmathbf{P}
  \right), \tmop{rel} \left( \tmmathbf{p}-\tmmathbf{Q},
  \tmmathbf{P}+\tmmathbf{Q} \right) \right) \left[ \mu \left(
  \tmmathbf{p}-\tmmathbf{Q} \right) f \left( \tmmathbf{P}+\tmmathbf{Q} \right)
  - \mu \left( \tmmathbf{p} \right) f \left( \tmmathbf{P} \right) \right] . 
  \label{eq:clbeMBbis}
\end{eqnarray}
It is now convenient to express the $\delta$-function of energy conservation
in terms of the momentum transfer and of the momentum components parallel and
perpendicular to it. Introducing $\tmmathbf{P}_{\|\tmmathbf{Q}} = \left(
\tmmathbf{P} \cdot \tmmathbf{Q} \right) \tmmathbf{Q}/ Q^2$ and
$\tmmathbf{P}_{\bot \tmmathbf{Q}} =\tmmathbf{P}-\tmmathbf{P}_{\|\tmmathbf{Q}}$
and further exploiting the definition of relative momenta given in
Eq.~(\ref{eq:reldef}) one has
\begin{eqnarray}
  \delta \left( \frac{\left( \tmmathbf{P} + \tmmathbf{Q} \right)^2}{2 M} +
  \frac{( \tmmathbf{p} - \tmmathbf{Q})^2}{2 m} - \frac{P^2}{2 M} -
  \frac{p^2}{2 m} \right) & = & \delta \left( \frac{Q^2}{2 m_{\ast}} -
  \frac{1}{m_{\ast}} \mathbf{\tmmathbf{Q}} \cdot \tmop{rel} \left(
  \mathbf{\tmmathbf{p}}_{\| \tmmathbf{Q}}, \tmmathbf{P}_{\| \tmmathbf{Q}}
  \right) \right),  \label{eq:para}
\end{eqnarray}
which implies
\begin{eqnarray}
  \tmop{rel} \left( \mathbf{\tmmathbf{p}}_{\| \tmmathbf{Q}}, \tmmathbf{P}_{\|
  \tmmathbf{Q}} \right) & = & \frac{\tmmathbf{Q}}{2} \mathbf{\tmmathbf{}} . 
  \label{eq:delta}
\end{eqnarray}
Since the reduced mass $m_{\ast}$ obeys $m_{\ast} / m + m_{\ast} / M = 1$, one
has the useful relation
\begin{eqnarray}
  \tmop{rel} \left( \tmmathbf{p}, \tmmathbf{P} \right) +\tmmathbf{Q} & = &
  \tmop{rel} \left( \tmmathbf{p}+\tmmathbf{Q}, \tmmathbf{P}-\tmmathbf{Q}
  \right),  \label{eq:rel}
\end{eqnarray}
which, together with Eq.~(\ref{eq:delta}), leads to
\begin{eqnarray}
  \text{$\sigma \left( \tmop{rel} \left( \tmmathbf{p}, \tmmathbf{P} \right),
  \tmop{rel} \left( \tmmathbf{p}, \tmmathbf{P} \right) -\tmmathbf{Q} \right)$}
  & = & \sigma \left( \tmop{rel} \left(
  \mathbf{\tmmathbf{p}}_{\|\tmmathbf{Q}}, \tmmathbf{P}_{\|\tmmathbf{Q}}
  \right) + \tmop{rel} \left( \mathbf{\tmmathbf{p}}_{\perp \tmmathbf{Q}},
  \tmmathbf{P}_{\perp \tmmathbf{Q}} \right) -\tmmathbf{Q}, \right. \nonumber\\
  &  & \left. \phantom{\sigma \left( \right.} \tmop{rel} \left(
  \mathbf{\tmmathbf{p}}_{\|\tmmathbf{Q}}, \tmmathbf{P}_{\|\tmmathbf{Q}}
  \right) + \tmop{rel} \left( \mathbf{\tmmathbf{p}}_{\perp \tmmathbf{Q}},
  \tmmathbf{P}_{\perp \tmmathbf{Q}} \right) \right) \nonumber\\
  & = & \sigma \left( \tmop{rel} \left( \mathbf{\tmmathbf{p}}_{\perp
  \tmmathbf{Q}}, \tmmathbf{P}_{\perp \tmmathbf{Q}} \right) - \frac{1}{2}
  \tmmathbf{Q}, \tmop{rel} \left( \mathbf{\tmmathbf{p}}_{\perp \tmmathbf{Q}},
  \tmmathbf{P}_{\perp \tmmathbf{Q}} \right) + \frac{1}{2} \tmmathbf{Q} \right)
  .  \label{eq:sigma}
\end{eqnarray}
Eq.~(\ref{eq:clbeMBbis}) now becomes
\begin{eqnarray}
  \text{$\partial^{\tmop{coll}}_t f$} (\tmmathbf{P}) & = &
  \frac{n_{\tmop{gas}}}{m^2_{\ast}} \bigintlim \mathd \tmmathbf{Q} \int
  \bignone \mathd \tmmathbf{p} \, \sigma \left( \tmop{rel} \left(
  \tmmathbf{p}_{\bot}, \tmmathbf{P}_{\perp \tmmathbf{Q}} \right) - \frac{1}{2}
  \tmmathbf{Q}, \tmop{rel} \left( \tmmathbf{p}_{\bot}, \tmmathbf{P}_{\perp
  \tmmathbf{Q}} \right) + \frac{1}{2} \tmmathbf{Q} \right) \nonumber\\
  &  & \times \delta \left( \frac{Q^2}{2 m_{\ast}} - \frac{1}{m_{\ast}}
  \mathbf{\tmmathbf{Q}} \cdot \tmop{rel} \left(
  \mathbf{\tmmathbf{p}}_{\|\tmmathbf{Q}}, \tmmathbf{P}_{\|\tmmathbf{Q}}
  \right) \right) \left[ \mu \left( \tmmathbf{p}-\tmmathbf{Q} \right) f \left(
  \tmmathbf{P}+\tmmathbf{Q} \right) - \mu \left( \tmmathbf{p} \right) f \left(
  \tmmathbf{P} \right) \right] .  \label{eq:A}
\end{eqnarray}
By means of the shift of coordinates
\begin{eqnarray}
  \tmmathbf{p} & \rightarrow & \text{$\tmmathbf{p}+ \frac{m}{m_{\ast}}
  \frac{\tmmathbf{Q}}{2} + \frac{m}{M} \tmmathbf{P}_{\| \tmmathbf{Q}}$}
  \text{},  \label{eq:shift}
\end{eqnarray}
which leaves $\tmmathbf{p}_{\bot}$ invariant, the argument of the delta
function simplifies to $\mathbf{\tmmathbf{Q}} \cdot \tmmathbf{p}$, such that
it can be evaluated explicitly leading to
\begin{eqnarray}
  \text{$\partial^{\tmop{coll}}_t f$} (\tmmathbf{P}) & = &
  \frac{n_{\tmop{gas}} m}{m^2_{\ast}} \bigintlim \frac{\mathd \tmmathbf{Q}}{Q}
  \int_{\tmmathbf{Q}^{\bot}} \mathd \tmmathbf{k}_{\bot} \, \sigma \left(
  \tmop{rel} \left( \tmmathbf{k}_{\bot}, \tmmathbf{P}_{\perp \tmmathbf{Q}}
  \right) - \frac{\tmmathbf{Q}}{2}, \tmop{rel} \left( \tmmathbf{k}_{\bot},
  \tmmathbf{P}_{\perp \tmmathbf{Q}} \right) + \frac{\tmmathbf{Q}}{2} \right) 
  \label{eq:clbebis}\\
  &  & \times \left[ \mu \left( \tmmathbf{k}_{\bot} \text{$\left. +
  \text{$\frac{m}{m_{\ast}} \frac{\tmmathbf{Q}}{2} + \frac{m}{M} \left(
  \tmmathbf{P}_{\| \tmmathbf{Q}} -\tmmathbf{Q} \right)$} \right)$} f \left(
  \tmmathbf{P}-\tmmathbf{Q} \right) - \mu \left( \tmmathbf{k}_{\bot} \text{$+
  \frac{m}{m_{\ast}} \frac{\tmmathbf{Q}}{2} + \frac{m}{M} \tmmathbf{P}_{\|
  \tmmathbf{Q}}$} \right) f \left( \tmmathbf{P} \right) \right] . \right.
  \nonumber
\end{eqnarray}
This is the explicit expression allowing one to straightforwardly see the
connection between the classical and the quantum linear Boltzmann equation.
Another useful form is given by Eq.~(\ref{eq:522}).

\subsection{Approach to equilibrium}\label{sec:a2}

We now prove that the relative entropy of a solution of the classical linear
Boltzmann equation with respect to its stationary solution is a monotonically
decreasing function, equal to zero if and only if the solution is at
equilibrium. Let us first consider the explicit derivative of the relative
entropy (\ref{eq:521}),
\begin{eqnarray}
  \frac{\mathd H \left( f|f_{\tmop{EQ}} \right)}{\mathd t} & = & \int \mathd
  \tmmathbf{P} \, \text{ $\partial^{\tmop{coll}}_t f$} (\tmmathbf{P}) \bignone
  \left[ 1 + \log \frac{f \left( \tmmathbf{P} \right)}{f_{\tmop{EQ}} \left(
  \tmmathbf{P} \right)} \right] .  \label{eq:w1}
\end{eqnarray}
Exploiting the relations Eq.~(\ref{eq:rateout}) and Eq.~(\ref{eq:rate}) for
the loss and gain rates, the classical linear Boltzmann
equation~(\ref{eq:inout}) reads
\begin{eqnarray}
  \text{$\partial^{\tmop{coll}}_t f$} (\tmmathbf{P}) & = & \int \mathd
  \tmmathbf{P}'  \bignone \left[ M^{\tmop{cl}} \left( \tmmathbf{P}'
  \rightarrow \tmmathbf{P} \right) f \left( \tmmathbf{P}'  \right) -
  M^{\tmop{cl}} \left( \tmmathbf{P} \rightarrow \tmmathbf{P}'  \right) f
  \left( \tmmathbf{P} \right) \right] .  \label{eq:w2}
\end{eqnarray}
The fact that the transition rates obey the detailed balance condition
Eq.~(\ref{eq:dbcm}) implies that the quantity
\begin{eqnarray}
  W \left( \tmmathbf{P}, \tmmathbf{P}' \right) & = & M^{\tmop{cl}} \left(
  \tmmathbf{P}' \rightarrow \tmmathbf{P} \right) \mathe^{\beta \frac{P^2}{2
  M}}  \label{eq:w3}
\end{eqnarray}
is symmetric in its arguments. We thus obtain an expression
\begin{eqnarray}
  \text{$\partial^{\tmop{coll}}_t f$} (\tmmathbf{P}) & = & \int \mathd
  \tmmathbf{P}' \, W \left( \tmmathbf{P}, \tmmathbf{P}' \right) \bignone
  \left[ f_{\tmop{EQ}} \left( \tmmathbf{P} \right) f \left( \tmmathbf{P}' 
  \right) - f_{\tmop{EQ}} \left( \tmmathbf{P}'  \right) f \left( \tmmathbf{P}
  \right) \right],  \label{eq:522}
\end{eqnarray}
that puts into evidence that the integrand is a product of functions which are
symmetric and antisymmetric under the exchange of $\tmmathbf{P}$ and
$\tmmathbf{P}'$. Using this form, the time derivative of the relative entropy
therefore reads as
\begin{eqnarray}
  \frac{\mathd H \left( f|f_{\tmop{EQ}} \right)}{\mathd t} & = & \frac{1}{2}
  \int \mathd \tmmathbf{P} \int \mathd \tmmathbf{P}' \, W \left( \tmmathbf{P}'
  \rightarrow \tmmathbf{P} \right) \bignone \left[ f_{\tmop{EQ}} \left(
  \tmmathbf{P} \right) f \left( \tmmathbf{P}'  \right) - f_{\tmop{EQ}} \left(
  \tmmathbf{P}'  \right) f \left( \tmmathbf{P} \right) \right] \nonumber\\
  &  & \times \left[ \log \left( f_{\tmop{EQ}} \left( \tmmathbf{P}'  \right)
  f \left( \tmmathbf{P} \right) \right) - \log \left( f_{\tmop{EQ}} \left(
  \tmmathbf{P} \right) f \left( \tmmathbf{P}'  \right) \right) \right] . 
  \label{eq:w4}
\end{eqnarray}
By noting that
\begin{eqnarray}
  \left( X - Y \right) \left( \log Y - \log X \right) & \leqslant & 0, 
  \label{eq:w5}
\end{eqnarray}
where the equal sign holds if and only if $X = Y$, this proves that the time
derivative of the relative entropy is equal to zero if and only if one
considers the equilibrium solution.

Moreover, since $\partial^{\tmop{coll}}_t f (\tmmathbf{P}) = 0$ implies
$\mathd H \left( f|f_{\tmop{EQ}} \right) / \mathd t = 0$, it follows that
$\mathd H \left( f|f_{\tmop{EQ}} \right) / \mathd t = 0$, or equivalently $f =
f_{\tmop{EQ}}$, is a necessary condition for a stationary solution. But we
already know that $f = f_{\tmop{EQ}}$ is a sufficient condition for
stationarity, such that the stationary solution is unique.

\subsection{The friction coefficient of classical Brownian
motion}\label{sec:a3}

We now explicitly derive the microscopic expression for the friction
coefficient appearing in the classical Kramers equation~(\ref{eq:fp}). Let us
start from the classical linear Boltzmann equation written as in
Eq.~(\ref{eq:classica}), with the constraint $\left| \tmop{rel} \left(
\tmmathbf{p}, \tmmathbf{P} \right) | = \left| \tmop{rel} \left( \tmmathbf{p}',
\tmmathbf{P}' \right) \right| \right.$ keeping track of energy conservation in
each collision. We put $f \left( \tmmathbf{P} \right) = \nu_{\tmop{EQ}} \left(
\tmmathbf{P} \right) \chi \left( \tmmathbf{P} \right)$, where $\nu_{\tmop{EQ}}
\left( \tmmathbf{P} \right)$ has the form Eq.~(\ref{eq:muMB2}) of a
Maxwell-Boltzmann distribution of the {\tmem{test}} particle. Exploiting
energy conservation Eq.~(\ref{eq:classica}) now takes the more explicit form
\begin{eqnarray}
  \text{$\partial_t f$} (\tmmathbf{P}) & = & \frac{n_{\tmop{gas}}}{m_{\ast} } 
  \text{$\nu_{\tmop{EQ}} \left( \tmmathbf{P} \right)$} \int^{\infty}_0 \mathd
  p \, p^2 \mu_{\beta} \left( \tmmathbf{p} \right) \int^{2 \pi}_0 \bignone
  \mathd \Phi \int^{\pi}_0 \mathd \bignone \Theta \sin \Theta \int^{2 \pi}_0
  \bignone \mathd \varphi \int^{\pi}_0 \mathd \bignone \vartheta \sin
  \vartheta \bignone   \label{eq:angoli}\\
  &  & \times \bignone | \tmop{rel} \left( \tmmathbf{p}, \tmmathbf{P} \right)
  | \sigma \left( \vartheta ; | \tmop{rel} \left( \tmmathbf{p}, \tmmathbf{P}
  \right) | \right) \left[ \chi \left( \tmmathbf{P}' \right) - \chi \left(
  \tmmathbf{P} \right) \right], \nonumber
\end{eqnarray}
where $\left( \vartheta, \varphi \right)$ denote the polar angles of
$\tmop{rel} \left( \tmmathbf{p}', \tmmathbf{P}' \right)$ with respect to
$\tmop{rel} \left( \tmmathbf{p}, \tmmathbf{P} \right)$, and $\left( \Theta,
\Phi \right)$ the polar angles of $\tmmathbf{p}$ with respect to
$\tmmathbf{P}$.

As discussed in Sect.~\ref{sec:qbm}, in the Brownian motion limit one
considers a test particle much heavier than the gas particles, such that $m /
M \ll 1$, and the test particle is assumed to be close to thermal equilibrium.
Under these assumptions the momentum transfer in a single collision, given by
\begin{eqnarray}
  \tmmathbf{Q}=\tmmathbf{P}' -\tmmathbf{P}= - \left[ \tmop{rel} \left(
  \tmmathbf{p}', \tmmathbf{P}' \right) - \tmop{rel} \left( \tmmathbf{p},
  \tmmathbf{P} \right) \right], &  &  \label{eq:basic}
\end{eqnarray}
can be considered small. We therefore expand the function $\chi \left(
\tmmathbf{P}' \right)$ in a Taylor series, thus obtaining up to second order
\begin{eqnarray}
  \text{$\partial_t f$} (\tmmathbf{P}) & = & \frac{n_{\tmop{gas}}}{m_{\ast} } 
  \text{$\nu_{\tmop{EQ}} \left( \tmmathbf{P} \right)$} \int^{\infty}_0 \mathd
  p \, p^2 \mu_{\beta} \left( \tmmathbf{p} \right) \int^{2 \pi}_0 \bignone
  \mathd \Phi \int^{\pi}_0 \mathd \bignone \Theta \sin \Theta \int^{2 \pi}_0
  \bignone \mathd \varphi \int^{\pi}_0 \mathd \bignone \vartheta \sin
  \vartheta \bignone  \nonumber\\
  &  & \times \bignone | \tmop{rel} \left( \tmmathbf{p}, \tmmathbf{P} \right)
  | \sigma \left( \vartheta ; | \tmop{rel} \left( \tmmathbf{p}, \tmmathbf{P}
  \right) | \right)  \label{eq:angoli2}\\
  &  & \times \left[ \sum^3_{i = 1} \left( \tmmathbf{P}' -\tmmathbf{P}
  \right)_i \frac{\partial}{\partial P_i} \chi \left( \tmmathbf{P} \right) +
  \frac{1}{2} \sum^3_{i, j = 1} \left( \tmmathbf{P}' -\tmmathbf{P} \right)_j
  \left( \tmmathbf{P}' -\tmmathbf{P} \right)_i \frac{\partial^2}{\partial P_j
  \partial P_i} \chi \left( \tmmathbf{P} \right) \right] . \nonumber
\end{eqnarray}
Thanks to the identity Eq.~(\ref{eq:basic}) and to the fact that the
scattering cross-section only depends on the angles $\vartheta$ and $\Theta$,
one can perform the integration over the azimuthal angle $\varphi$ appearing
only in $\tmmathbf{P}' -\tmmathbf{P}$. It is done by exploiting the integrals
\begin{eqnarray}
  \int^{2 \pi}_0 \bignone \mathd \varphi \left( \tmmathbf{K}' -\tmmathbf{K}
  \right)_i & = & 2 \pi \left( \frac{K'}{K} \cos \vartheta - 1 \right) K_i 
  \label{eq:relaz}\\
  \int^{2 \pi}_0 \bignone \mathd \varphi \left( \tmmathbf{K}' -\tmmathbf{K}
  \right)_j \left( \tmmathbf{K}' -\tmmathbf{K} \right)_i & = & 2 \pi \left[
  \frac{1}{2} \sin^2 \vartheta K^2 \delta_{ij} - \frac{1}{2} \left( 3 \cos
  \vartheta - 1 \right) \left( 1 - \cos \vartheta \right) K_j K_i \right], 
  \label{eq:relaz2}
\end{eqnarray}
which are valid whenever $\left( \vartheta, \varphi \right)$ denote the polar
angles of $\tmmathbf{K}'$ with respect to $\tmmathbf{K}$.

To obtain Eq.~(\ref{eq:relaz}) let us denote by $\tmmathbf{e}_1$,
$\tmmathbf{e}_2$, and $\tmmathbf{e}_3$ the unit basis vectors for the
Cartesian coordinates, such that $K_i =\tmmathbf{e}_i \cdot \tmmathbf{K}$, and
consider another set of basis vectors $\tmmathbf{e}'_1, \tmmathbf{e}'_2
\tmop{and} \tmmathbf{e}'_3$ such that $\tmmathbf{e}'_3$ is in the direction of
$\tmmathbf{K}$. One then has
\begin{eqnarray}
  \left( \tmmathbf{K}' -\tmmathbf{K} \right)_i & = & \tmmathbf{e}_i \cdot
  \left( \tmmathbf{K}' -\tmmathbf{K} \right)  \label{eq:sbobba}\\
  & = & \tmmathbf{e}_i \cdot \left[ K' \sin \vartheta \cos \varphi
  \tmmathbf{e}'_1 + K' \sin \vartheta \sin \varphi \tmmathbf{e}'_2 + \left( K'
  \cos \vartheta - K \right) \tmmathbf{e}'_3 \right] . \nonumber
\end{eqnarray}
Equation~(\ref{eq:relaz}) is obtained by integrating the azimuthal angle
$\varphi$ and recalling $\tmmathbf{K}= K\tmmathbf{e}'_3$. A similar, but much
longer calculation leads to Eq.~(\ref{eq:relaz2}).

We are left with contributions in the integrand proportional to $\tmop{rel}
\left( \tmmathbf{p}, \tmmathbf{P} \right)_i$ and $\tmop{rel} \left(
\tmmathbf{p}, \tmmathbf{P} \right)_j \text{$\tmop{rel} \left( \tmmathbf{p},
\tmmathbf{P} \right)_i$}$, such that the integrals~(\ref{eq:relaz}) and
(\ref{eq:relaz2}) can be applied once more to perform the integrations over
$\Phi$. The result of the integration over the azimuthal angles reads
\begin{eqnarray}
  \text{$\partial_t f$} (\tmmathbf{P}) & = & 4 \pi^2
  \frac{n_{\tmop{gas}}}{m_{\ast} }  \text{$\nu_{\tmop{EQ}} \left( \tmmathbf{P}
  \right)$} \int^{\infty}_0 \mathd p \, p^2 \mu_{\beta} \left( \tmmathbf{p}
  \right)  \int^{\pi}_0 \mathd \bignone \Theta \sin \Theta \int^{\pi}_0 \mathd
  \bignone \vartheta \sin \vartheta \bignone  \left( 1 - \cos \vartheta
  \right) | \tmop{rel} \left( \tmmathbf{p}, \tmmathbf{P} \right) | \nonumber\\
  &  & \times \sigma \left( \vartheta ; | \tmop{rel} \left( \tmmathbf{p},
  \tmmathbf{P} \right) | \right) \left\{ \left( \frac{M}{m}  \frac{p}{P} \cos
  \Theta - 1 \right) \frac{m_{\ast}}{M} \sum^3_{i = 1} P_i
  \frac{\partial}{\partial P_i} \chi \left( \tmmathbf{P} \right) 
  \label{eq:angoli3} \right.\\
  &  & + \frac{1}{4} \left[ \left( 1 + \cos \vartheta \right) | \tmop{rel}
  \left( \tmmathbf{p}, \tmmathbf{P} \right) |^2 + \frac{1}{2} \left( 1 - 3
  \cos \vartheta \right)  \frac{m^2_{\ast}}{m^2} p^2 \sin^2 \Theta \right]
  \sum^3_{i = 1} \frac{\partial^2}{\partial P^2_i} \chi \left( \tmmathbf{P}
  \right) \nonumber\\
  &  & \left. + \frac{1}{4} \left( 1 - 3 \cos \vartheta \right) \left[ \left(
  \frac{M}{m}  \frac{p}{P} \cos \Theta - 1 \right)^2 - \frac{1}{2} \left(
  \frac{M}{m}  \frac{p}{P} \sin \Theta \right)^2 \right]
  \frac{m^2_{\ast}}{M^2} \sum^3_{i, j = 1} P_j P_i \frac{\partial^2}{\partial
  P_j \partial P_i} \chi \left( \tmmathbf{P} \right) \right\} . \nonumber
\end{eqnarray}
To proceed we exploit the fact that the test particle is assumed to be close
to equilibrium, such that $p / P \approx \sqrt{m / M}$, and $m / M \ll 1$. We
expand the expression of $| \tmop{rel} \left( \tmmathbf{p}, \tmmathbf{P}
\right) |$ in this small ratio, thus obtaining up to order $\sqrt{m / M}$
\begin{eqnarray}
  | \tmop{rel} \left( \tmmathbf{p}, \tmmathbf{P} \right) | & \approx & p
  \left[ 1 - \frac{m}{M} \frac{P}{p} \cos \Theta \right],  \label{eq:drel}
\end{eqnarray}
and therefore in particular
\begin{eqnarray}
  | \tmop{rel} \left( \tmmathbf{p}, \tmmathbf{P} \right) | \sigma \left(
  \vartheta ; | \tmop{rel} \left( \tmmathbf{p}, \tmmathbf{P} \right) | \right)
  & \approx & p \left[ \sigma \left( \vartheta ; p \right) - \frac{m}{M}
  \frac{P}{p} \cos \Theta \left( \sigma \left( \vartheta ; p \right) + p
  \frac{\partial}{\partial p} \sigma \left( \vartheta ; p \right) \right)
  \right] .  \label{eq:dsigma}
\end{eqnarray}
As a result of this approximation one can perform the integral over $\Theta$,
without explicit knowledge of the scattering cross-section, which only depends
on $p$ and $\vartheta$ for an isotropic interaction potential. Inserting
Eq.~(\ref{eq:dsigma}) and Eq.~(\ref{eq:drel}) in Eq.~(\ref{eq:angoli3}) and
performing the angular integral one obtains after lengthy but straightforward
calculations
\begin{eqnarray}
  \text{$\partial_t f$} (\tmmathbf{P}) & = & - \frac{8}{3} \pi^2
  \frac{n_{\tmop{gas}}}{m}  \text{$\nu_{\tmop{EQ}} \left( \tmmathbf{P}
  \right)$} \frac{\beta}{M} \int^{\infty}_0 \mathd p \, p^5 \mu_{\beta} \left(
  \tmmathbf{p} \right)  \int^{\pi}_0 \mathd \bignone \vartheta \sin \vartheta
  \bignone  \left( 1 - \cos \vartheta \right) \sigma \left( \vartheta ; p
  \right)  \label{eq:angoli4}\\
  &  & \times \sum^3_{i = 1} \left[ P_i \frac{\partial}{\partial P_i} \chi
  \left( \tmmathbf{P} \right) - \frac{M}{\beta} \frac{\partial^2}{\partial
  P^2_i} \chi \left( \tmmathbf{P} \right) \right] . \nonumber
\end{eqnarray}
Here, an integration by parts has been exploited,
\begin{eqnarray}
  \int^{\infty}_0 \mathd p \, p^4 \mu_{\beta} \left( \tmmathbf{p} \right) 
  \frac{\partial}{\partial p} \sigma \left( \vartheta ; p \right) & = & -
  \int^{\infty}_0 \mathd p \left( 4 p^3 - \frac{2}{p^2_{\beta}} p^5 \right)
  \mu_{\beta} \left( \tmmathbf{p} \right) \sigma \left( \vartheta ; p \right)
  .  \label{eq:parts}
\end{eqnarray}
As a last step, we return to an equation for the original distribution $f
\left( \tmmathbf{P} \right) = \nu_{\tmop{EQ}} \left( \tmmathbf{P} \right) \chi
\left( \tmmathbf{P} \right)$. We recall that the equilibrium solution
$\nu_{\tmop{EQ}} \left( \tmmathbf{P} \right)$, as defined in
Eq.~(\ref{eq:muMB2}), is a null eigenvector of the Fokker-Planck operator
\begin{eqnarray}
  D \left[ g \left( \tmmathbf{P} \right) \right] & = & \sum^3_{i = 1} \left[
  \frac{\partial}{\partial P_i} \left( P_i g \left( \tmmathbf{P} \right)
  \right) + \frac{M}{\beta} \frac{\partial^2}{\partial P^2_i} g \left(
  \tmmathbf{P} \right) \right],  \label{eq:fpop}
\end{eqnarray}
such that in particular
\begin{eqnarray}
  D \left[ \nu_{\tmop{EQ}} \left( \tmmathbf{P} \right) \chi \left(
  \tmmathbf{P} \right) \right] & = & - \nu_{\tmop{EQ}} \left( \tmmathbf{P}
  \right) \sum^3_{i = 1} \left[ P_i \frac{\partial}{\partial P_i} \chi \left(
  \tmmathbf{P} \right) - \frac{M}{\beta} \frac{\partial^2}{\partial P^2_i}
  \chi \left( \tmmathbf{P} \right) \right] .  \label{eq:fpop2}
\end{eqnarray}
We are thus left with
\begin{eqnarray}
  \text{$\partial_t f$} (\tmmathbf{P}) & = & \frac{8}{3} \pi^2
  \frac{n_{\tmop{gas}}}{m}  \frac{\beta}{M} \int^{\infty}_0 \mathd p \, p^5
  \mu_{\beta} \left( \tmmathbf{p} \right)  \int^{\pi}_0 \mathd \bignone
  \vartheta \sin \vartheta \bignone  \left( 1 - \cos \vartheta \right) \sigma
  \left( \vartheta ; p \right)  \label{eq:angoli45}\\
  &  & \times \sum^3_{i = 1} \left[ \frac{\partial}{\partial P_i} \left( P_i
  f \left( \tmmathbf{P} \right) \right) + \frac{M}{\beta}
  \frac{\partial^2}{\partial P^2_i} f \left( \tmmathbf{P} \right) \right] .
  \nonumber
\end{eqnarray}
Switching to dimensionless variables $u = p / p_{\beta}$ and using
Eq.~(\ref{eq:muMB}) for $\mu_{\beta} \left( \tmmathbf{p} \right)$, one finally
has
\begin{eqnarray}
  \text{$\partial_t f$} (\tmmathbf{P}) & = & \eta \sum^3_{i = 1} \left[
  \frac{\partial}{\partial P_i} \left( P_i f \left( \tmmathbf{P} \right)
  \right) + \frac{M}{\beta} \frac{\partial^2}{\partial P^2_i} f \left(
  \tmmathbf{P} \right) \right]  \label{eq:fpapp}
\end{eqnarray}
with
\begin{eqnarray}
  \eta & = & \frac{16}{3}  \sqrt{\pi}  \frac{m}{M} n_{\tmop{gas}}
  \sqrt{\frac{2}{\beta m}}  \int^{\infty}_0 \mathd u \, u^5 \mathe^{- u^2}
  \int^{\pi}_0 \mathd \bignone \vartheta \sin \vartheta \bignone  \left( 1 -
  \cos \vartheta \right) \sigma \left( \vartheta ; up_{\beta} \right), 
  \label{eq:a33}
\end{eqnarray}
as in Eq.~(\ref{eq:etac}).

\section{List of symbols}

\begin{longtable}{ll}
  $\beta$ & inverse temperature, $\beta = 1 / \left( k_{\text{B}} T \right)$\\
  $D_{pp}$ & diffusion coefficient, see (\ref{eq:dpp})\\
  $D_{xx}$ & diffusion coefficient, see (\ref{eq:dxx})\\
  $\eta$ & friction coefficient, see (\ref{eq:friction}), (\ref{eq:e3})\\
  $\text{erf} (x)$ & error function {\cite{Abramowitz1965a}}\\
  $E \left( \tmmathbf{Q}, \tmmathbf{P} \right)$ & energy transfer to test
  particle, see (\ref{eq:etransfer})\\
  $\Phi_{\mathcal{P}}$ & characteristic function of momentum transfer
  distribution, see (\ref{eq:cf0}), (\ref{eq:715})\\
  $f \left( \tmmathbf{P} \right)$ & distribution function of the test particle
  momentum\\
  $f \left( \tmmathbf{p}_f, \tmmathbf{p}_i \right)$ & elastic scattering
  amplitude\\
  $f_0 \left( \tmmathbf{p} \right)$ & forward scattering amplitude, $f_0
  \left( \tmmathbf{p} \right) = f \left( \tmmathbf{p}, \tmmathbf{p} \right)$\\
  $f_B \left( \tmmathbf{Q} \right)$ & scattering amplitude in Born
  approximation, see (\ref{eq:fBorn})\\
  $\text{}_1 F_1$ & confluent hypergeometric function
  {\cite{Abramowitz1965a}}\\
  $\Gamma \left( x \right)$ & gamma function {\cite{Abramowitz1965a}}\\
  $\Gamma_{\tmop{tot}}$ & total scattering rate, see (\ref{eq:norma})\\
  $\Gamma_{\beta}$ & thermal scattering rate, see (\ref{eq:gamma0})\\
  $\tilde{\Gamma} \left( U \right)$ & dimensionless loss rate, see
  (\ref{eq:x14})\\
  $\mathsf{H}_0$ & kinetic energy operator of test particle, $\mathsf{H}_0 =
  \mathsf{P}^2 / \left( 2 M \right)$\\
  $H_{\text{n}}$ & gas induced energy shift, see (\ref{eq:effectivebis}),
  (\ref{eq:for})\\
  $\tmmathbf{K}$ & scaled momentum transfer, see (\ref{eq:K})\\
  $\lambda_{\tmop{th}}$ & thermal de Broglie wave length of test particle,
  $\lambda_{\tmop{th}} = \sqrt{2 \pi \hbar^2 \beta / M}$\\
  $L \left( \tmmathbf{p}, \tmmathbf{P}; \tmmathbf{Q} \right)$ & non-unitary
  part of Lindblad operator, see (\ref{eq:L})\\
  $L_B \left( \tmmathbf{P}; \tmmathbf{Q} \right)$ & non-unitary part of
  Lindblad operator in Born approximation, see (\ref{eq:Lb})\\
  $\mathcal{L}$ & superoperator of the dissipative part of the quantum linear
  Boltzmann equation, see (\ref{eq:lvonn})\\
  $\mu \left( \tmmathbf{p} \right)$ & distribution function of gas momenta\\
  $\mu_{\beta} \left( \tmmathbf{p} \right)$ & Maxwell-Boltzmann distribution,
  see (\ref{eq:muMB}); for $\mu_{\beta}^{\left( 1 d, 2 d \right)}$ see
  (\ref{eq:parper}), (\ref{eq:1dnorm})\\
  $m$ & mass of gas particle\\
  $m_{\ast}$ & reduced mass, $m_{\ast} = mM / \left( m + M \right)$\\
  $M$ & mass of test particle\\
  $M_{\tmop{in}}^{\tmop{cl}} \left( \tmmathbf{P}; \tmmathbf{Q} \right)$ &
  classical gain rate, see (\ref{eq:rate})\\
  $M_{\tmop{out}}^{\tmop{cl}} \left( \tmmathbf{P} \right)$ & classical loss
  rate, see (\ref{eq:rateout})\\
  $\mathcal{M}$ & superoperator of the quantum linear Boltzmann equation, see
  (\ref{eq:lvonn})\\
  $n_{\tmop{gas}}$ & density of gas particles\\
  $n \equiv n_1 + in_2$ & index of refraction, see (\ref{eq:nindex})\\
  $\text{$\nu_{\tmop{EQ}} \left( \tmmathbf{P} \right)$}$ & Maxwell-Boltzmann
  distribution for the test particle, see (\ref{eq:muMB2})\\
  $p_{\beta} \text{}$ & most probable momentum in thermal gas distribution,
  $p_{\beta} = \sqrt{2 m / \beta} \text{}$\\
  $\mathsf{P} \equiv \left( \mathsf{P}_1, \mathsf{P}_2, \mathsf{P}_3 \right)$
  & momentum operator of test particle\\
  $\tmmathbf{Q}$ & momentum transfer to test particle\\
  $\tmmathbf{Q}^{\bot}$ & set of all momenta perpendicular to $\tmmathbf{Q}$,
  i.e., $\tmmathbf{Q}^{\bot} = \left\{ \tmmathbf{p} \in \mathbbm{R}^3 :
  \tmmathbf{p} \cdot \tmmathbf{Q}= 0 \right\}$\\
  $\rho$ & quantum state of motion of test particle \\
  $\tmop{rel} \left( \tmmathbf{p}, \tmmathbf{P} \right)$ & relative momentum,
  see (\ref{eq:reldef})\\
  $\sigma \left( \tmmathbf{p}_f, \tmmathbf{p}_i \right)$ & differential
  cross-section, $\sigma \left( \tmmathbf{p}_f, \tmmathbf{p}_i \right) =
  \left| f \left( \tmmathbf{p}_f, \tmmathbf{p}_i \right) \right|^2$\\
  $\sigma \left( \vartheta ; p_i \right)$ & differential cross-section
  (isotropic potential), $\cos \vartheta =\tmmathbf{p}_i \cdot \tmmathbf{p}_f
  / p_i^2$, $p_i = \left| \tmmathbf{p}_i \right|$\\
  $\sigma_{\tmop{tot}}$ & total elastic scattering cross-section, see
  (\ref{eq:totcross}), (\ref{eq:c6})\\
  $\sigma_B \left( \tmmathbf{Q} \right)$ & differential cross-section in Born
  approximation, $\sigma_B \left( \tmmathbf{Q} \right) = \left| f_B \left(
  \tmmathbf{Q} \right) \right|^2$\\
  $S \left( \tmmathbf{Q}, E \right)$ & dynamic structure factor, see
  (\ref{eq:rr})\\
  $\tmmathbf{U}$ & scaled momentum, see (\ref{eq:U})\\
  $v_{\beta}$ & most probable velocity in thermal gas distribution, $v_{\beta}
  = p_{\beta} / m \text{}$\\
  $V \left( \tmmathbf{X}-\tmmathbf{x} \right)$ & interaction potential between
  test and gas particle\\
  $V_{\tmop{opt}}$ & optical potential, see (\ref{eq:opt})\\
  $\mathsf{X} \equiv \left( \mathsf{X}_1, \mathsf{X}_2, \mathsf{X}_3 \right)$
  \ \ \  & position operator of test particle\\
  $\tmmathbf{X}_{\|\tmmathbf{Q}}$ & vector component parallel to
  $\tmmathbf{Q}$, i.e., $\tmmathbf{X}_{\|\tmmathbf{Q}} = \left( \tmmathbf{X}
  \cdot \tmmathbf{Q} \right) \tmmathbf{Q}/ Q^2$\\
  $\tmmathbf{X}_{\bot \tmmathbf{Q}}$ & vector component orthogonal to
  $\tmmathbf{Q}$, i.e., $\tmmathbf{X}_{\bot \tmmathbf{Q}}
  =\tmmathbf{X}-\tmmathbf{X}_{\|\tmmathbf{Q}}$
\end{longtable}


\begin{thebibliography}{100}
  \bibitem[1]{Boltzmann1898a}L.~Boltzmann, Vorlesungen \"uber Gastheorie,
  Barth, Leipzig, 1898.
  
  \bibitem[2]{Cercignani1975a}C.~Cercignani, Theory and application of the
  Boltzmann equation, Scottisch Academic Press, Edinburgh, 1975.
  
  \bibitem[3]{Spohn1991}H.~Spohn, Large scale dynamics of interacting
  particles, Springer, Berlin, 1991.
  
  \bibitem[4]{Harris2004a}S.~Harris, An Introduction to the Theory of the
  Boltzmann Equation, Courier Dover Publications, 2004.
  
  \bibitem[5]{Balian2007b}R.~Balian, From microphysics to macrophysics:
  methods and applications of statistical physics. Vol. II, Springer, Berlin,
  2007.
  
  \bibitem[6]{Davison1957a}B.~Davison, B.~Sykes, Neutron transport theory,
  Clarendon Press, Oxford, 1957.
  
  \bibitem[7]{Williams1966}M.~M.~R. Williams, The Slowing Down and
  Thermalization of Neutrons, North-Holland, Amsterdam, 1966.
  
  \bibitem[8]{Nordheim1928}L.~W. Nordheim, On the kinetic method in the new
  statistics and its application in the electron theory of conductivity,
  Proc.~Roy.~Soc. 119 (1928) 689--698.
  
  \bibitem[9]{Uehling1933}E.~A. Uehling, G.~E. Uhlenbeck, Transport phenomena
  in Einstein-Bose and Fermi-Dirac gases. I, Phys. Rev. 43 (1933) 552--561.
  
  \bibitem[10]{Ross1954a}J.~Ross, J.~G. Kirkwood, The statistical-mechanical
  theory of transport processes. VIII. Quantum theory of transport in gases,
  J.~Chem. Phys. 22 (1954) 1094--1103.
  
  \bibitem[11]{QKI}C.~W. Gardiner, P.~Zoller, Quantum kinetic theory: A
  quantum kinetic master equation for condensation of a weakly interacting
  Bose gas without a trapping potential, Phys. Rev.~A 55 (1997) 2902--2921.
  
  \bibitem[12]{QKII}D.~Jaksch, C.~W. Gardiner, P.~Zoller, Quantum kinetic
  theory. II. Simulation of the quantum Boltzmann master equation, Phys.
  Rev.~A 56 (1997) 575--586.
  
  \bibitem[13]{QK-PRLI}C.~W. Gardiner, P.~Zoller, R.~J. Ballagh, M.~J. Davis,
  Kinetics of Bose-Einstein condensation in a trap, Phys. Rev. Lett. 79 (1997)
  1793--1796.
  
  \bibitem[14]{QKIII}C.~W. Gardiner, P.~Zoller, Quantum kinetic theory. III.
  Quantum kinetic master equation for strongly condensed trapped systems,
  Phys. Rev.~A 58 (1998) 536--556.
  
  \bibitem[15]{QKIV}D.~Jaksch, C.~W. Gardiner, K.~M. Gheri, P.~Zoller, Quantum
  kinetic theory. IV. Intensity and amplitude fluctuations of a Bose-Einstein
  condensate at finite temperature including trap loss, Phys. Rev.~A 58 (1998)
  1450--1464.
  
  \bibitem[16]{QK-PRLII}C.~W. Gardiner, M.~D. Lee, R.~J. Ballagh, M.~J. Davis,
  P.~Zoller, Quantum kinetic theory of condensate growth: Comparison of
  experiment and theory, Phys. Rev. Lett. 81 (1998) 5266--5269.
  
  \bibitem[17]{QKV}C.~W. Gardiner, P.~Zoller, Quantum kinetic theory. V.
  Quantum kinetic master equation for mutual interaction of condensate and
  noncondensate, Phys. Rev.~A 61 (2000) 033601.
  
  \bibitem[18]{Spohn1980a}H.~Spohn, Kinetic equations from Hamiltonian
  dynamics: Markovian limits, Rev. Mod. Phys. 52 (1980) 569--615.
  
  \bibitem[19]{Snider1998a}R.~F. Snider, Relaxation and transport of molecular
  systems in the gas phase, Int. Rev. Phys. Chem. 17 (1998) 185--225.
  
  \bibitem[20]{Erdos2004a}L.~Erd®s, M.~Salmhofer, H.~Yau, On the quantum
  Boltzmann equation, J.~Stat. Phys. 116 (2004) 367--380.
  
  \bibitem[21]{Benedetto2007a}D.~Benedetto, F.~Castella, R.~Esposito,
  M.~Pulvirenti, A short review on the derivation of the nonlinear quantum
  Boltzmann equations, Commun. Math. Sci. (2007) 55--71.
  
  \bibitem[22]{Lukkarinen2009a}J.~Lukkarinen, H.~Spohn, Not to normal order -
  notes on the kinetic limit for weakly interacting quantum fluids, J.~Stat.
  Phys. 134 (2009) 1133--1172.
  
  \bibitem[23]{Spohn2007a}H.~Spohn, Kinetic equations for quantum
  many-particle systems, in: Modern Encyclopedia of Mathematical Physics,
  Springer, to appear, preprint arXiv:0706.0807.
  
  \bibitem[24]{Castella2001a}F.~Castella, From the von-Neumann equation to the
  quantum Boltzmann equation in a deterministic framework, J.~Stat. Phys. 104
  (2001) 387--447.
  
  \bibitem[25]{Castella2002a}F.~Castella, From the von Neumann equation to the
  quantum Boltzmann equation II: identifying the Born series, J.~Stat. Phys.
  106 (2002) 1197--1220.
  
  \bibitem[26]{Eng2007a}D.~Eng, L.~Erd®s, The linear Boltzmann equation as the
  low density limit of a random Schr\"odinger equation, Rev. Math. Phys. 17
  (2005) 669--743.
  
  \bibitem[27]{Alicki2007}R.~Alicki, K.~Lendi, Quantum Dynamical Semigroups
  and Applications, 2nd Edition, Vol. 717 of Lecture Notes in Physics,
  Springer, Berlin, 2007.
  
  \bibitem[28]{Holevo2001}A.~S. Holevo, Statistical Structure of Quantum
  Theory, Vol. m 67 of Lecture Notes in Physics, Springer, Berlin, 2001.
  
  \bibitem[29]{Breuer2007}H.-P. Breuer, F.~Petruccione, The Theory of Open
  Quantum Systems, Oxford University Press, Oxford, 2007.
  
  \bibitem[30]{Gorini1976a}V.~Gorini, A.~Kossakowski, E.~C.~G. Sudarshan,
  Completely positive dynamical semigroups of $N$-level systems, J.~Math.
  Phys. 17 (1976) 821--825.
  
  \bibitem[31]{Lindblad1976a}G.~Lindblad, On the generators of quantum
  dynamical semigroups, Comm. Math. Phys. 48 (1976) 119--130.
  
  \bibitem[32]{Hornberger2008a}K.~Hornberger, B.~Vacchini, Monitoring
  derivation of the quantum linear Boltzmann equation, Phys. Rev.~A 77 (2008)
  022112.
  
  \bibitem[33]{Holevo1998a}A.~S. Holevo, Covariant quantum dynamical
  semigroups: unbounded generators, in: A.~B\"ohm, H.~D. Doebner,
  P.~Kielanowski (Eds.), Irreversibility and causality, Vol. 504 of Lecture
  Notes in Physics, Springer, Berlin, 1998, pp. 67--81.
  
  \bibitem[34]{Arndt2005a}M.~Arndt, K.~Hornberger, A.~Zeilinger, Probing the
  limits of the quantum world, Physics World 18 (2005) 35--40.
  
  \bibitem[35]{Cronin2009a}A.~Cronin, J.~Schmiedmayer, D.~Pritchard, Optics
  and interferometry with atoms and molecules, Rev. Mod. Phys. (2009) in press
  eprint arXiv: 0712.3703.
  
  \bibitem[36]{Gerlich2007a}S.~Gerlich, L.~Hackerm\"uller, K.~Hornberger,
  A.~Stibor, H.~Ulbricht, M.~Gring, F.~Goldfarb, T.~Savas, M.~M\"uri,
  M.~Mayor, M.~Arndt, A Kapitza-Dirac-Talbot-Lau interferometer for highly
  polarizable molecules, Nature Phys. 3 (2007) 711--715.
  
  \bibitem[37]{Schmiedmayer1995a}J.~Schmiedmayer, M.~S. Chapman, C.~R.
  Ekstrom, T.~D. Hammond, S.~Wehinger, D.~E. Pritchard, Index of refraction of
  various gases for sodium matter waves, Phys. Rev. Lett. 74 (1995)
  1043--1047.
  
  \bibitem[38]{Jacquey2007a}M.~Jacquey, M.~Buchner, G.~Trenec, J.~Vigue, First
  measurements of the index of refraction of gases for lithium atomic waves,
  Phys. Rev. Lett. 98 (2007) 240405.
  
  \bibitem[39]{Hornberger2003a}K.~Hornberger, S.~Uttenthaler, B.~Brezger,
  L.~Hackerm\"uller, M.~Arndt, A.~Zeilinger, Collisional decoherence observed
  in matter wave interferometry, Phys. Rev. Lett. 90 (2003) 160401.
  
  \bibitem[40]{Joos1985a}E.~Joos, H.~D. Zeh, The emergence of classical
  properties through interaction with the environment, Z. Phys. B: Condens.
  Matter 59 (1985) 223--243.
  
  \bibitem[41]{Hornberger2006b}K.~Hornberger, Master equation for a quantum
  particle in a gas, Phys. Rev. Lett. 97 (2006) 060601.
  
  \bibitem[42]{Vacchini2001a}B.~Vacchini, Test particle in a quantum gas,
  Phys. Rev.~E 63 (2001) 066115.
  
  \bibitem[43]{Gallis1990a}M.~R. Gallis, G.~N. Fleming, Environmental and
  spontaneous localization, Phys. Rev.~A 42 (1990) 38--48.
  
  \bibitem[44]{Diosi1995a}L.~Di\'osi, Quantum master equation of a particle in
  a gas environment, Europhys. Lett. 30 (1995) 63--68.
  
  \bibitem[45]{Lanz1997a}L.~Lanz, B.~Vacchini, Incoherent dynamics in
  neutron-matter interaction, Phys. Rev.~A 56 (1997) 4826--4838.
  
  \bibitem[46]{Vacchini2000a}B.~Vacchini, Completely positive quantum
  dissipation, Phys. Rev. Lett. 84 (2000) 1374--1377.
  
  \bibitem[47]{Vacchini2001b}B.~Vacchini, Translation-covariant Markovian
  master equation for a test particle in a quantum fluid, J.~Math. Phys. 42
  (2001) 4291--4312.
  
  \bibitem[48]{Alicki2002a}R.~Alicki, Search for a border between classical
  and quantum worlds, Phys. Rev.~A 65 (2002) 034104.
  
  \bibitem[49]{Dodd2003a}P.~J. Dodd, J.~J. Halliwell, Decoherence and records
  for the case of a scattering environment, Phys. Rev.~D 67 (2003) 105018.
  
  \bibitem[50]{Clark2008a}J.~Clark, Decoherence rates for Galilean covariant
  dynamics, J.~Math. Phys. 49 (2008) 052103.
  
  \bibitem[51]{Clark2009a}J.~Clark, An infinite-temperature limit for a
  quantum scattering process, Rep. Math. Phys. 63 (2009) 131--152.
  
  \bibitem[52]{Clark2009b}J.~Clark, The reduced effect of a single scattering
  with a low-mass particle via a point interaction, J. Funct. Anal. 256 (2009)
  2894--2916.
  
  \bibitem[53]{Clark-xxx}J.~Clark, W.~De Roeck, C.~Maes, Diffusive behavior
  from a quantum master equation, eprint arXiv:0812.2858v1.
  
  \bibitem[54]{Hornberger2003b}K.~Hornberger, J.~E. Sipe, Collisional
  decoherence reexamined, Phys. Rev.~A 68 (2003) 012105.
  
  \bibitem[55]{Hackermuller2003b}L.~Hackerm\"uller, K.~Hornberger, B.~Brezger,
  A.~Zeilinger, M.~Arndt, Decoherence in a Talbot Lau interferometer: The
  influence of molecular scattering, Appl. Phys.~B 77 (2003) 781--787.
  
  \bibitem[56]{Petruccione2005a}F.~Petruccione, B.~Vacchini, Quantum
  description of Einstein's Brownian motion, Phys. Rev.~E 71 (2005) 046134.
  
  \bibitem[57]{Lanz1996a}L.~Lanz, B.~Vacchini, Dynamical semigroup description
  of coherent and incoherent particle-matter interaction, Int. J. Theor. Phys.
  36 (1997) 67--88.
  
  \bibitem[58]{Altenmuller1997a}T.~P. Altenm\"uller, R.~M\"uller, A.~Schenzle,
  Atom-interferomeric study of Bose-Einstein condensation, Phys. Rev.~A 56
  (1997) 2959--2971.
  
  \bibitem[59]{Teta2004a}A.~Teta, On a rigorous proof of the Joos-Zeh formula
  for decoherence in a two-body problem, in: Multiscale methods in quantum
  mechanics, Trends Math., Birkh\"auser, Boston, 2004, pp. 197--205.
  
  \bibitem[60]{Durr2004a}D.~D\"urr, R.~Figari, A.~Teta, Decoherence in a
  two-particle model, J.~Math. Phys. 45 (2004) 1291--1309.
  
  \bibitem[61]{Adami2004a}R.~Adami, R.~Figari, D.~Finco, A.~Teta, On the
  asymptotic behaviour of a quantum two-body system in the small mass ratio
  limit, J.~Phys.~A: Math. Gen. 37 (2004) 7567--7580.
  
  \bibitem[62]{Cacciapuoti2005a}C.~Cacciapuoti, R.~Carlone, R.~Figari,
  Decoherence induced by scattering: a three-dimensional model, J.~Phys.~A:
  Math. Gen. 38 (2005) 4933--4946.
  
  \bibitem[63]{Adami2006a}R.~Adami, R.~Figari, D.~Finco, A.~Teta, On the
  asymptotic dynamics of a quantum system composed by heavy and light
  particles, J.~Phys.~A: Math. Gen. 268 (2006) 819--852.
  
  \bibitem[64]{Stenholm1993a}S.~Stenholm, Occurrences, observations and
  measurements in quantum mechanics, Phys. Scr. 47 (1993) 724--731.
  
  \bibitem[65]{Hornberger2007b}K.~Hornberger, Monitoring approach to open
  quantum dynamics using scattering theory, EPL 77 (2007) 50007.
  
  \bibitem[66]{Hornberger2007a}K.~Hornberger, Open quantum dynamics via
  environmental monitoring, J. Phys.: Conf. Ser. 67 (2007) 012002.
  
  \bibitem[67]{Jacobs2006a}K.~Jacobs, D.~A. Steck, A straightforward
  introduction to continuous quantum measurement, Contemp. Phys. 47 (2006)
  279--303.
  
  \bibitem[68]{Barchielli2009}A.~Barchielli, M.~Gregoratti, Quantum
  Trajectories and Measurements in Continuous Time, Vol. 782 of Lecture Notes
  in Physics, Springer, Berlin, 2009.
  
  \bibitem[69]{Dumcke1985a}R.~D\"umcke, The low density limit for an N-level
  system interacting with a free Bose or Fermi gas, Commun. Math. Phys. 97
  (1985) 331--359.
  
  \bibitem[70]{Raffelt1993a}G.~Raffelt, G.~Sigl, L.~Stodolsky, Non-Abelian
  Boltzmann equation for mixing and decoherence, Phys. Rev. Lett. 70 (1993)
  2363--2366.
  
  \bibitem[71]{Tsonchev2000a}S.~Tsonchev, P.~Pechukas, Binary collision model
  for quantum Brownian motion, Phys. Rev.~E 61 (2000) 6171--6182.
  
  \bibitem[72]{Kleckner2001a}M.~Kleckner, A.~Ron, Decoherence of a pointer by
  a gas reservoir, Phys. Rev.~A 63 (2001) 022110.
  
  \bibitem[73]{Hellmich2004a}M.~Hellmich, Alicki's model of scattering-induced
  decoherence derived from Hamiltonian dynamics, J.~Phys.~A: Math. Gen. 37
  (2004) 8711--8719.
  
  \bibitem[74]{Pechen2004a}A.~N. Pechen, Quantum stochastic equation for a
  test particle interacting with a dilute Bose gas, J.~Math. Phys. 45 (2004)
  400--417.
  
  \bibitem[75]{Pechen2005a}A.~N. Pechen, White noise approach to the low
  density limit of a quantum particle in a gas, in: M.~Schuermann, U.~Franz
  (Eds.), QP-PQ:Quantum Probability and White Noise Analysis, Vol.~18, World
  Scientific, Singapore, 2005, p. 428.
  
  \bibitem[76]{Barnett2005a}S.~M. Barnett, J.~D. Cresser, Quantum theory of
  friction, Phys. Rev.~A 72 (2005) 022107.
  
  \bibitem[77]{Adler2006a}S.~L. Adler, Normalization of collisional
  decoherence: squaring the delta function, and an independent cross-check,
  J.~Phys.~A: Math. Gen. 39 (2006) 14067--14074.
  
  \bibitem[78]{Halliwell2007a}J.~J. Halliwell, Two derivations of the master
  equation of quantum Brownian motion, J.~Phys.~A: Math. Theor. 40 (2007)
  3067--3080.
  
  \bibitem[79]{Dominguez-Clarimon2007a}A.~Dominguez-Clarimon, A particle
  across a medium: How decoherence relates to the index of refraction., Ann.
  Phys. 322 (2007) 2085--2103.
  
  \bibitem[80]{Mintert200Xa}F.~Mintert, E.~J. Heller, Simulation of open
  quantum systems, arXiv:0803.3883.
  
  \bibitem[81]{Benedetto2004a}D.~Benedetto, F.~Castella, R.~Esposito,
  M.~Pulvirenti, Some considerations on the derivation of the nonlinear
  quantum Boltzmann equation, J.~Stat. Phys. 116 (2004) 381--410.
  
  \bibitem[82]{Benedetto2006a}D.~Benedetto, F.~Castella, R.~Esposito,
  M.~Pulvirenti, Some considerations on the derivation of the nonlinear
  quantum Boltzmann equation II: the low density regime, J.~Stat. Phys. 124
  (2006) 951--996.
  
  \bibitem[83]{Benedetto2008a}D.~Benedetto, F.~Castella, R.~Esposito,
  M.~Pulvirenti, From the $N$-body Schr\"odinger equation to the quantum
  Boltzmann equation: a term-by-term convergence result in the weak coupling
  regime, Commun. Math. Phys. 277 (2008) 1--44.
  
  \bibitem[84]{Steinigeweg2007a}R.~Steinigeweg, H.-P. Breuer, J.~Gemmer,
  Transition from diffusive to ballistic dynamics for a class of finite
  quantum models, Phys. Rev. Lett. 99 (2007) 150601.
  
  \bibitem[85]{Kadiroglu2007a}M.~Kadiroglu, J.~Gemmer, Boltzmann-equation
  approach to transport in finite modular quantum systems, Phys. Rev.~B 76
  (2007) 024306.
  
  \bibitem[86]{Kossakowski1972a}A.~Kossakowski, On quantum statistical
  mechanics of non-Hamiltonian systems, Rep. Math. Phys. 3 (1972) 247--274.
  
  \bibitem[87]{Manita1991a}A.~D. Manita, Properties of translation-invariant
  quantum-dynamical semigroups, Theor. Math. Phys. 89 (1991) 1271--1281.
  
  \bibitem[88]{Botvich1991a}D.~D. Botvich, V.~A. Malyshev, A.~D. Manita,
  Translation invariant quantum master equation, Helv. Phys. Acta 64 (1991)
  1072--1092.
  
  \bibitem[89]{Holevo1993a}A.~S. Holevo, A note on covariant dynamical
  semigroups, Rep. Math. Phys. 32 (1993) 211--216.
  
  \bibitem[90]{Holevo1993b}A.~S. Holevo, On conservativity of covariant
  dynamical semigroups, Rep. Math. Phys. 33 (1993) 95--110.
  
  \bibitem[91]{Holevo1995a}A.~S. Holevo, On translation-covariant quantum
  Markov equations., Izv. Math. 59 (1995) 427--443.
  
  \bibitem[92]{Holevo1996a}A.~S. Holevo, Covariant quantum Markovian
  evolutions, J.~Math. Phys. 37 (1996) 1812--1832.
  
  \bibitem[93]{Vacchini-xxx}B.~Vacchini, Covariant mappings for the
  description of measurement, dissipation and decoherence in quantum
  mechanics, in E.~Br\"unning, F.~Pettrucione (Eds.) Theoretical Foundations
  of Quantum Information Processing and Communication, Lecture Notes in
  Physics 878, Springer-Verlag, Berlin, 2010, to appear; ~eprint
  arXiv:quant-ph/0707.0603.
  
  \bibitem[94]{Huang1987}K.~Huang, Statistical Mechanics, John Wiley \& Sons,
  New York, 1987.
  
  \bibitem[95]{Spohn1978b}H.~Spohn, J.~L. Lebowitz, Irreversible
  thermodynamics for quantum systems weakly coupled to thermal reservoirs,
  Adv. Chem. Phys. 38 (1978) 109--142.
  
  \bibitem[96]{Lindblad1983}G.~Lindblad, Non-Equilibrium Entropy and
  Irreversibility, Reidel Publishing Company, Dordrecht, 1983.
  
  \bibitem[97]{Alicki2001}R.~Alicki, M.~Fannes, Quantum dynamical systems,
  Oxford University Press, Oxford, 2001.
  
  \bibitem[98]{Wehrl1978}A.~Wehrl, General properties of entropy, Rev. Mod.
  Phys. 50 (1978) 221--260.
  
  \bibitem[99]{Ingarden1997}R.~S. Ingarden, A.~Kossakowski, M.~Ohya,
  Information dynamics and open systems, Kluwer Academic Publishers,
  Dordrecht, 1997.
  
  \bibitem[100]{Vacchini2007d}B.~Vacchini, K.~Hornberger, Relaxation dynamics
  of a quantum Brownian particle in an ideal gas, Eur. Phys. J.~ST 151 (2007)
  59--72.
  
  \bibitem[101]{Breuer2007c}H.-P. Breuer, B.~Vacchini, Three-dimensional Monte
  Carlo simulations of the quantum linear Boltzmann equation, Phys. Rev.~E 76
  (2007) 036706.
  
  \bibitem[102]{Diosi1986a}L.~Di\'osi, Stochastic pure state representation
  for open quantum systems, Phys. Lett. 114A (1986) 451--454.
  
  \bibitem[103]{Gardiner1992a}C.~W. Gardiner, A.~S. Parkins, P.~Zoller,
  Wave-function quantum stochastic differential equations and quantum-jump
  simulation methods, Phys. Rev.~A 46 (1992) 4363--4381.
  
  \bibitem[104]{Carmichael1993a}H.~Carmichael, An Open Systems Approach to
  Quantum Optics, Springer, Berlin, 1993.
  
  \bibitem[105]{Molmer1993a}K.~Molmer, Y.~Castin, J.~Dalibard, Monte Carlo
  wave-function method in quantum optics, J.~Opt. Soc. Am.~B 10 (1993)
  524--538.
  
  \bibitem[106]{Vacchini2004a}B.~Vacchini, Quantum and classical features in
  the explanation of collisional decoherence, J.~Mod. Opt. 51 (2004)
  1025--1029.
  
  \bibitem[107]{Gillespie1992}D.~T. Gillespie, Markov Processes, Academic
  Press, Boston, 1992.
  
  \bibitem[108]{Grabert1988a}H.~Grabert, P.~Schramm, G.~Ingold, Quantum
  Brownian motion: The functional integral approach, Phys. Rep. 168 (1988)
  115--207.
  
  \bibitem[109]{Ingold2002a}G.~Ingold, Path integrals and their application to
  dissipative quantum systems, in: A.~Buchleitner, K.~Hornberger (Eds.),
  Coherent Evolution in Noisy Environments, Lecture Notes in Physics 611,
  Springer, Berlin, 2002, pp. 1--53.
  
  \bibitem[110]{Hanggi2005a}P.~H\"anggi, G.~Ingold, Fundamental aspects of
  quantum Brownian motion, Chaos 15 (2005) 026105.
  
  \bibitem[111]{Weiss2008}U.~Weiss, Quantum Dissipative Systems, 3rd Edition,
  World Scientific, Singapore, 2008.
  
  \bibitem[112]{Rayleigh1891a}J.~W.~S. Rayleigh, Dynamical problems in
  illustration of the theory of gases, Phil. Mag. 32 (1891) 424--445.
  
  \bibitem[113]{Green1951a}M.~S. Green, Brownian motion in a gas of
  noninteracting molecules, J.~Chem. Phys. 19 (1951) 1036--1046.
  
  \bibitem[114]{Uhlenbeck1970a}C.~S.~W. Chang, G.~E. Uhlenbeck, The kinetic
  theory of gases, in: J.~D. Boer (Ed.), Studies in statistical mechanics,
  Vol.~5, North-Holland, Amsterdam, 1970, Ch.~V.
  
  \bibitem[115]{Ferrari1987a}L.~Ferrari, An improved differential form of the
  Boltzmann collision operator for a Rayleigh gas (or Brownian particles),
  Physica A 142 (1987) 441--466.
  
  \bibitem[116]{Ambegaokar1991a}V.~Ambegaokar, Quantum Brownian motion and its
  classical limit, Ber. Bunsenges. Phys. Chem. 95 (1991) 400--404.
  
  \bibitem[117]{Pechukas1991a}P.~Pechukas, Quantum Brownian Motion, in:
  W.~Gans, A.~Blumen, A.~Amann (Eds.), Large Scale Molecular Systems, Vol. 258
  of NATO ASI Series, Plenum Press, New York, 1991, p. 123.
  
  \bibitem[118]{Bassi2005b}A.~Bassi, E.~Ippoliti, B.~Vacchini, On the energy
  increase in space-collapse models, J.~Phys.~A: Math. Gen. 38 (2005)
  8017--8038.
  
  \bibitem[119]{Caldeira1983a}A.~O. Caldeira, A.~J. Leggett, Path integral
  approach to quantum Brownian motion, Physica A 121 (1983) 587--616.
  
  \bibitem[120]{Caldeira1983b}A.~O. Caldeira, A.~J. Leggett, Quantum
  tunnelling in a dissipative system, apny 149 (1983) 374--456.
  
  \bibitem[121]{Leggett1987a}A.~J. Leggett, S.~Chakravarty, A.~T. Dorsey,
  M.~P.~A. Fisher, A.~Garg, W.~Zwerger, Dynamics of the dissipative two-state
  system, Rev. Mod. Phys. 59 (1987) 1--85.
  
  \bibitem[122]{Barchielli1983b}A.~Barchielli, Continual measurements for
  quantum open systems, Nuovo Cimento B 74 (1983) 113--138.
  
  \bibitem[123]{Sandulescu1987a}A.~Sandulescu, H.~Scutaru, Open quantum
  systems and the damping of collective modes in deep inelastic collisions,
  Ann. Physics 173 (1987) 277--317.
  
  \bibitem[124]{Isar1999a}A.~Isar, Uncertainty, entropy and decoherence of the
  damped harmonic oscillator in the Lindblad theory of open quantum systems,
  Fortschr. Phys. 47 (1999) 855--879.
  
  \bibitem[125]{Vacchini2002b}B.~Vacchini, Quantum optical versus quantum
  Brownian motion master equation in terms of covariance and equilibrium
  properties, J.~Math. Phys. 43 (2002) 5446--5458.
  
  \bibitem[126]{Isar1994a}A.~Isar, A.~Sandulescu, H.~Scutaru, E.~Stefanescu,
  W.~Scheid, Open quantum systems, Int. J. Mod. Phys. E 3 (1994) 635--714.
  
  \bibitem[127]{Jacobs2009a}K.~Jacobs, A Monte Carlo method for modeling
  thermal damping: Beyond the Brownian motion master equation, EPL 85 (2009)
  40002.
  
  \bibitem[128]{Diosi1993a}L.~Di\'osi, On high temperature Markovian equation
  for quantum Brownian motion, Europhys. Lett. 22 (1993) 1--3.
  
  \bibitem[129]{Diosi1993b}L.~Di\'osi, Caldeira-Leggett master equation and
  medium temperatures, Physica A 199 (1993) 517--526.
  
  \bibitem[130]{Vacchini2002a}B.~Vacchini, Non-abelian linear Boltzmann
  equation and quantum correction to Kramers and Smoluchowski equation, Phys.
  Rev.~E 66 (2002) 027107.
  
  \bibitem[131]{Sears1989a}V.~F. Sears, Neutron Optics, Oxford University
  Press, Oxford, 1989.
  
  \bibitem[132]{Werner2000a}S.~A. Werner, H.~Rauch, Neutron Interferometry,
  Oxford University Press, Oxford, 2000.
  
  \bibitem[133]{Adams1994a}C.~S. Adams, M.~Siegel, J.~Mlynek, Atom optics,
  Phys. Rep. 240 (1994) 143--210.
  
  \bibitem[134]{Miffre2006a}A.~Miffre, M.~Jacquey, M.~B\"uchner, G.~Tr\'enec,
  J.~Vigu\'e, Atom interferometry, Phys. Scr. 74 (2006) C15--C23.
  
  \bibitem[135]{Arndt2009a}M.~Arndt, K.~Hornberger, Quantum interferometry
  with complex molecules, in: Quantum Coherence in Solid State Systems,
  Proceedings of the International School of Physics ``Enrico Fermi'', Course
  CLXXI, Societ\`a Italiana di Fisica, 2009, (to appear), arXiv:0903.1614.
  
  \bibitem[136]{Vigue1995a}J.~Vigue, Index of refraction of dilute matter in
  atomic interferometry, Phys. Rev.~A 52 (1995) 3973--3975.
  
  \bibitem[137]{Rauch1974a}H.~Rauch, W.~Treimer, U.~Bonse, Test of a single
  crystal neutron interferometer, Phys. Lett.~A 47 (1974) 369--371.
  
  \bibitem[138]{Taylor1972a}J.~R. Taylor, Scattering Theory, John Wiley \&
  Sons, New York, 1972.
  
  \bibitem[139]{Champenois1999a}C.~Champenois, Interf\'erom\'etrie atomique
  avec l'atome de lithium: analyse th\'eorique et construction d'un
  interf\'erom\`etre, applications., Ph.D. thesis, Universit\'e Paul Sabatier,
  Toulouse, available at http://tel.archives-ouvertes.fr/tel-00003602 (1999).
  
  \bibitem[140]{Champenois2008a}C.~Champenois, M.~Jacquey, S.~Lepoutre,
  M.~Buchner, G.~Trenec, J.~Vigue, Index of refraction of gases for matter
  waves: Effect of the motion of the gas particles on the calculation of the
  index, Phys. Rev.~A 77 (2008) 013621.
  
  \bibitem[141]{Forrey1996a}R.~C. Forrey, L.~You, V.~Kharchenko, A.~Dalgarno,
  Index of refraction of noble gases for sodium matter waves, Phys. Rev.~A 54
  (1996) 2180--2184.
  
  \bibitem[142]{Kharchenko2001a}V.~Kharchenko, A.~Dalgarno, Refractive index
  for matter waves in ultracold gases, Phys. Rev.~A 63 (2001) 023615.
  
  \bibitem[143]{Roberts2002a}T.~D. Roberts, A.~D. Cronin, D.~A. Kokorowski,
  D.~E. Pritchard, Glory oscillations in the index of refraction for matter
  waves, Phys. Rev. Lett. 89 (2002) 200406.
  
  \bibitem[144]{Sanders2009a}S.~N. Sanders, F.~Mintert, E.~J. Heller, Coherent
  scattering from a free gas, Phys. Rev.~A 79 (2009) 023610.
  
  \bibitem[145]{Zurek1991a}W.~H. Zurek, Decoherence and the transition from
  quantum to classical, Phys. Today 44 (1991) 36--44.
  
  \bibitem[146]{Joos2003}E.~Joos, H.~D. Zeh, C.~Kiefer, D.~Giulini, J.~Kupsch,
  I.-O. Stamatescu, Decoherence and the Appearance of a Classical World in
  Quantum Theory, 2nd Edition, Springer, Berlin, 2003.
  
  \bibitem[147]{Zurek2003a}W.~H. Zurek, Decoherence, einselection, and the
  quantum origins of the classical, Rev. Mod. Phys. 75 (2003) 715--775.
  
  \bibitem[148]{Schlosshauer2007}M.~Schlosshauer, Decoherence and the
  Quantum-to-Classical Transition, Springer-Verlag, Berlin, 2007.
  
  \bibitem[149]{Hornberger2009a}K.~Hornberger, Introduction to decoherence
  theory, in: A.~Buchleitner, C.~Viviescas, M.~Tiersch (Eds.), Entanglement
  and Decoherence, Lecture Notes in Physics 768, Springer, Berlin, 2009, pp.
  221--276.
  
  \bibitem[150]{Vacchini2005b}B.~Vacchini, Master-equations for the study of
  decoherence, Int. J. Theor. Phys. 44 (2005) 1011--1021.
  
  \bibitem[151]{Hornberger2004a}K.~Hornberger, J.~E. Sipe, M.~Arndt, Theory of
  decoherence in a matter mave Talbot-Lau interferometer, Phys. Rev.~A 70
  (2004) 053608.
  
  \bibitem[152]{Feller1971}W.~Feller, An introduction to probability theory
  and its applications. Vol. II, John Wiley \& Sons Inc., New York, 1971.
  
  \bibitem[153]{Lukacs1966a}L.~Lukacs, Characteristic Functions, Griffin,
  London, 1966.
  
  \bibitem[154]{Vacchini2005a}B.~Vacchini, Theory of decoherence due to
  scattering events and L\'evy processes, Phys. Rev. Lett. 95 (2005) 230402.
  
  \bibitem[155]{Uys2005a}H.~Uys, J.~D. Perreault, A.~D. Cronin, Matter-wave
  decoherence due to a gas environment in an atom interferometer, Phys. Rev.
  Lett. 95 (2005) 150403.
  
  \bibitem[156]{Ruiz1997a}A.~Ruiz, J.~Breton, J.~Gomez Llorente, Scattering
  cross-sections for low-energy rare-gas+C$_{60}$ and C$_{60}$+C$_{60}$
  collisions, Chem. Phys. Lett. 270 (1997) 121--128.
  
  \bibitem[157]{Maitland1981a}G.~C. Maitland, M.~Rigby, E.~B. Smith, W.~A.
  Wakeham, Intermolecular Forces - Their Origin and Determination, Clarendon
  Press, Oxford, 1981.
  
  \bibitem[158]{Cheng1999a}C.-C. Cheng, M.~G. Raymer, Long-range saturation of
  spatial decoherence in wave-field transport in random multiple-scattering
  media, Phys. Rev. Lett. 82 (1999) 4807--4810.
  
  \bibitem[159]{Pfau1994a}T.~Pfau, S.~Sp\"alter, C.~Kurtsiefer, C.~Ekstrom,
  J.~Mlynek, Loss of spatial coherence by a single spontaneous emission, Phys.
  Rev. Lett. 73 (1994) 1223--1226.
  
  \bibitem[160]{Chapman1995a}M.~S. Chapman, T.~D. Hammond, A.~Lenef,
  J.~Schmiedmayer, R.~A. Rubenstein, E.~Smith, D.~E. Pritchard, Photon
  scattering from atoms in an atom interferometer: Coherence lost and
  regained, Phys. Rev. Lett. 75 (1995) 3783 -- 3787.
  
  \bibitem[161]{Kokorowski2001a}D.~A. Kokorowski, A.~D. Cronin, T.~D. Roberts,
  D.~E. Pritchard, From single- to multiple-photon decoherence in an atom
  interferometer, Phys. Rev. Lett. 86 (2001) 2191--2195.
  
  \bibitem[162]{Hackermuller2004a}L.~Hackerm\"uller, K.~Hornberger,
  B.~Brezger, A.~Zeilinger, M.~Arndt, Decoherence of matter waves by thermal
  emission of radiation, Nature 427 (2004) 711--714.
  
  \bibitem[163]{Hornberger2009b}K.~Hornberger, S.~Gerlich, H.~Ulbricht,
  L.~Hackerm\"uller, S.~Nimmrichter, I.~V. Goldt, O.~Boltalina, M.~Arndt,
  Theory and experimental verification of Kapitza-Dirac Talbot-Lau
  interferometry, New J.~Phys. 11 (2009) 043032.
  
  \bibitem[164]{Facchi2004a}P.~Facchi, Thermal decoherence in mesoscopic
  interference, J.~Mod. Opt. 51 (2004) 1049--1055.
  
  \bibitem[165]{Hornberger2005a}K.~Hornberger, L.~Hackerm\"uller, M.~Arndt,
  Influence of molecular temperature on the coherence of fullerenes in a
  near-field interferometer, Phys. Rev.~A 71 (2005) 023601.
  
  \bibitem[166]{Hornberger2006a}K.~Hornberger, Thermal limitation of far-field
  matter-wave interference, Phys. Rev.~A 73 (2006) 052102.
  
  \bibitem[167]{Nimmrichter2008b}S.~Nimmrichter, K.~Hornberger, H.~Ulbricht,
  M.~Arndt, Absolute absorption spectroscopy based on molecule interferometry,
  Phys. Rev.~A 78 (2008) 063607.
  
  \bibitem[168]{Viale2003a}A.~Viale, M.~Vicari, N.~Zanghi, Analysis of the
  loss of coherence in interferometry with macromolecules, Phys. Rev.~A 68
  (2003) 063610.
  
  \bibitem[169]{Lombardo2005a}F.~C. Lombardo, F.~D. Mazzitelli, P.~I. Villar,
  Decoherence induced by a fluctuating Aharonov-Casher phase, Phys. Rev.~A 72
  (2005) 042111.
  
  \bibitem[170]{Villar2007a}P.~Villar, F.~Lombardo, Visibility fringe
  reduction due to noise-induced effects: Microscopic approach to interference
  experiments, Int. J. Mod. Phys. B 21 (2007) 4659--4676.
  
  \bibitem[171]{Qureshi2008a}T.~Qureshi, A.~Venugopalan, Decoherence and
  matter wave interferometry, Int. J. Mod. Phys. B 22 (2008) 981--990.
  
  \bibitem[172]{Rubenstein1999a}R.~Rubenstein, D.~A. Kokorowski, A.~Dhirani,
  T.~Roberts, S.~Gupta, J.~Lehner, W.~Smith, E.~Smith, H.~Bernstein,
  D.~Pritchard, Measurement of the density matrix of a longitudinally
  modulated atomic beam, Phys. Rev. Lett. 83 (1999) 2285--2288.
  
  \bibitem[173]{Rubenstein1999b}R.~Rubenstein, A.~Dhirani, D.~Kokorowski,
  T.~Roberts, E.~Smith, W.~W. Smith, H.~Bernstein, J.~Lehner, S.~Gupta,
  D.~Pritchard, Search for off-diagonal density matrix elements for atoms in a
  supersonic beam, Phys. Rev. Lett. 82 (1999) 2018--2021.
  
  \bibitem[174]{Anglin1997a}J.~Anglin, J.~Paz, W.~Zurek, Deconstructing
  decoherence, Phys. Rev.~A 55 (1997) 4041--4053.
  
  \bibitem[175]{Kusnezov1999a}D.~Kusnezov, A.~Bulgac, G.~D. Dang, Quantum
  L\'evy processes and fractional kinetics, Phys. Rev. Lett. 82 (1999)
  1136--1139.
  
  \bibitem[176]{Lutz2002a}E.~Lutz, Anomalous L\'evy decoherence, Phys. Lett.~A
  293 (2002) 123--128.
  
  \bibitem[177]{Schomerus2007a}H.~Schomerus, E.~Lutz, Nonexponential
  decoherence and momentum subdiffusion in a quantum L\'evy kicked rotator,
  Phys. Rev. Lett. 98 (2007) 260401.
  
  \bibitem[178]{Bellomo2007a}B.~Bellomo, S.~M. Barnett, J.~Jeffers, Frictional
  quantum decoherence, J.~Phys.~A: Math. Theor. 40 (2007) 9437--9453.
  
  \bibitem[179]{Fehti-xxx}F.~M. Ramazanoglu, Approach to thermal equilibrium
  in the Caldeira-Leggett model, eprint arXiv:0812:2520v1.
  
  \bibitem[180]{Wigner1983a}E.~P. Wigner, Review of quantum mechanical
  measurement problem, in: P.~Meystre, M.~O. Scully (Eds.), Quantum Optics,
  Experimental Gravity and Measurement Theory, Plenum Press, New York, 1983,
  p.~43.
  
  \bibitem[181]{Einstein1905b}A.~Einstein, \"Uber die von der
  molekularkinetischen Theorie der W\"arme geforderte Bewegung von in ruhenden
  Fl\"ussigkeiten suspendierten Teilchen, Ann. Phys. (Leipzig) 322 (1905)
  549--560.
  
  \bibitem[182]{Gocke2007a}C.~Gocke, G.~R\"opke, Localization of coherent wave
  packets in plasmas due to decoherence, Contrib. Plasma Phys. 47 (2007)
  291--296.
  
  \bibitem[183]{Gocke2008a}C.~Gocke, G.~R\"opke, Master equation of the
  reduced statistical operator of an atom in a plasma, Theor. Math. Phys. 154
  (2008) 26--51.
  
  \bibitem[184]{Lovesey1984}S.~Lovesey, Theory of neutron scattering from
  condensed matter. Vol.1. Nuclear scattering, Clarendon Press, Oxford, UK,
  1984.
  
  \bibitem[185]{Schwabl2003}F.~Schwabl, Advanced quantum mechanics, Springer,
  New York, 2003.
  
  \bibitem[186]{Pitaevskii2003}L.~Pitaevskii, S.~Stringari, Bose-Einstein
  condensation, Oxford University Press, Oxford, 2003.
  
  \bibitem[187]{Lanz2002a}L.~Lanz, B.~Vacchini, Subdynamics of relevant
  observables: a field theoretical approach, Int. J. Modern Phys. A 17 (2002)
  435--463.
  
  \bibitem[188]{Brenig1989}W.~Brenig, Statistical theory of heat,
  Springer-Verlag, Berlin, 1989.
  
  \bibitem[189]{VanHove1954}L.~Van Hove, Correlations in space and time and
  Born approximation scattering in systems of interacting particles, Phys.
  Rev. 95 (1954) 249--262.
  
  \bibitem[190]{Alicki2003a}R.~Alicki, S.~Kryszewski, Completely positive
  Bloch-Boltzmann equations, Phys. Rev.~A 68 (2003) 013809.
  
  \bibitem[191]{Kryszewski2006a}S.~Kryszewski, J.~Czechowska, Positivity of
  Bloch-Boltzmann equations, Phys. Rev.~A 74 (2006) 022719.
  
  \bibitem[192]{Rautian1991a}S.~G. Rautian, A.~M. Shalagin, Kinetic Problems
  of Non-Linear Spectroscopy, North-Holland, Amsterdam, 1991.
  
  \bibitem[193]{Vacchini2008a}B.~Vacchini, Non-Markovian dynamics for
  bipartite systems, Phys. Rev.~A 78 (2008) 022112.
  
  \bibitem[194]{Budini2005a}A.~A. Budini, Random Lindblad equations from
  complex environments, Phys. Rev.~E 72 (2005) 056106.
  
  \bibitem[195]{Budini2006a}A.~A. Budini, Lindblad rate equations, Phys.
  Rev.~A 74 (2006) 053815.
  
  \bibitem[196]{Breuer2006a}H.-P. Breuer, J.~Gemmer, M.~Michel, Non-Markovian
  quantum dynamics: Correlated projection superoperators and Hilbert space
  averaging, Phys. Rev.~E 73 (2006) 016139.
  
  \bibitem[197]{Breuer2007a}H.-P. Breuer, Non-Markovian generalization of the
  Lindblad theory of open quantum systems, Phys. Rev.~A 75 (2007) 022103.
  
  \bibitem[198]{Abramowitz1965a}M.~Abramowitz, I.~Stegun, Handbook of
  Mathematical Functions, Dover Publications, New York, 1965.
\end{thebibliography}
\end{document}